\documentclass[%
superscriptaddress,
preprint,
 amsmath,amssymb,
 aps,
prx,
]{revtex4-2}

\usepackage{comment}
\usepackage{graphicx}% Include figure files
\usepackage{dcolumn}% Align table columns on decimal point
\usepackage{bm}% bold math
\usepackage[version=4]{mhchem}
\usepackage{chemformula}
\usepackage{siunitx}
\usepackage{hyperref}% add hypertext capabilities
\newcommand{\Qc}{\mathcal{Q}}
\begin{document}

\preprint{q-dependent phonons in RIXS}

%\title{The Dialogue Concerning Phonons and EPC in RIXS }
%\title{How closely is the intensity of phonon peaks in resonant inelastic x-ray scattering spectra of cuprates related to the electron-phonon coupling?}

%\title{The electron-phonon coupling in cuprates and its dependence on phonon momentum and electronic structure}

%\title{Momentum-dependent electron-phonon coupling in cuprates by RIXS: \\the roles of phonon symmetry and electronic structure}

\title{The influence of phonon symmetry and electronic structure on the electron-phonon coupling momentum dependence in cuprates}

\author{Maryia\,Zinouyeva}
\email[e-mail ]{maryia.zinouyeva@polimi.it}
\affiliation{Dipartimento di Fisica, Politecnico di Milano, piazza Leonardo da Vinci 32, I-20133 Milano, Italy}

\author{Rolf\, Heid}
\affiliation{Institut für QuantenMaterialien und Technologien, Karlsruher Institut für Technologie, D-76021 Karlsruhe, Germany}

\author{Giacomo\, Merzoni}
 \affiliation{Dipartimento di Fisica, Politecnico di Milano, piazza Leonardo da Vinci 32, I-20133 Milano, Italy}
 \affiliation{European XFEL, Holzkoppel 4, Schenefeld, D-22869 Schenefeld, Germany}
 
\author{Riccardo\, Arpaia}
\affiliation{
Quantum Device Physics Laboratory, Department of Microtechnology and Nanoscience, Chalmers University of Technology, SE-41296 Göteborg, Sweden
}
\affiliation{Department of Molecular Sciences and Nanosystems, Ca’ Foscari University of Venice, I-30172 Venezia, Italy}

\author{Nikolai\, Andreev}
\affiliation{Dipartimento di Ingegneria Civile e Ingegneria Informatica, Università di Roma Tor Vergata, I-00133 Roma, Italy}

\author{Marco\, Biagi}
\affiliation{Dipartimento di Fisica, Politecnico di Milano, piazza Leonardo da Vinci 32, I-20133 Milano, Italy}
\affiliation{
Quantum Device Physics Laboratory, Department of Microtechnology and Nanoscience, Chalmers University of Technology, SE-41296 Göteborg, Sweden
}

\author{Nicholas B.\, Brookes}
\affiliation{ESRF, The European Synchrotron, 71 Avenue des Martyrs, CS 40220, F-38043 Grenoble, France}

\author{Daniele\, Di Castro}
\affiliation{Dipartimento di Ingegneria Civile e Ingegneria Informatica, Università di Roma Tor Vergata, I-00133 Roma, Italy}
\affiliation{CNR-SPIN, Università di Roma Tor Vergata, I-00133 Roma, Italy}

\author{Alexei\, Kalaboukhov}
\affiliation{
Quantum Device Physics Laboratory, Department of Microtechnology and Nanoscience, Chalmers University of Technology, SE-41296 Göteborg, Sweden
}

\author{Kurt\, Kummer}
\affiliation{ESRF, The European Synchrotron, 71 Avenue des Martyrs, CS 40220, F-38043 Grenoble, France}

\author{Floriana\, Lombardi}
\affiliation{
Quantum Device Physics Laboratory, Department of Microtechnology and Nanoscience, Chalmers University of Technology, SE-41296 Göteborg, Sweden
}

\author{Leonardo\, Martinelli}
\altaffiliation[current address ]{Physik-Institut, Universität Zürich, Winterthurerstrasse 190, CH-8057 Zürich, Switzerland}
 \affiliation{Dipartimento di Fisica, Politecnico di Milano, piazza Leonardo da Vinci 32, I-20133 Milano, Italy}

\author{Francesco\, Rosa}
\affiliation{Dipartimento di Fisica, Politecnico di Milano, piazza Leonardo da Vinci 32, I-20133 Milano, Italy}

\author{Matteo\, Rossi}
\affiliation{Dipartimento di Fisica, Politecnico di Milano, piazza Leonardo da Vinci 32, I-20133 Milano, Italy}

\author{Flora\, Yakhou-Harris}
\affiliation{ESRF, The European Synchrotron, 71 Avenue des Martyrs, CS 40220, F-38043 Grenoble, France}

\author{Lucio\, Braicovich}
\affiliation{Dipartimento di Fisica, Politecnico di Milano, piazza Leonardo da Vinci 32, I-20133 Milano, Italy}
\affiliation{ESRF, The European Synchrotron, 71 Avenue des Martyrs, CS 40220, F-38043 Grenoble, France}

\author{Marco\, Moretti Sala}
\affiliation{Dipartimento di Fisica, Politecnico di Milano, piazza Leonardo da Vinci 32, I-20133 Milano, Italy}

\author{Paolo G.\, Radaelli}
\email[e-mail ]{paolo.radaelli@physics.ox.ac.uk}
\affiliation{Clarendon Laboratory, Department of Physics, University of Oxford, OX1 3PU Oxford, UK}

\author{Giacomo\, Ghiringhelli}
\email[e-mail ]{giacomo.ghiringhelli@polimi.it}
\affiliation{Dipartimento di Fisica, Politecnico di Milano, piazza Leonardo da Vinci 32, I-20133 Milano, Italy}
\affiliation{CNR-SPIN, Dipartimento di Fisica, Politecnico di Milano, I-20133 Milano, Italy}

\date{\today}% It is always \today, today,

\begin{abstract}
The experimental determination of the magnitude and momentum dependence of electron-phonon coupling (EPC) is an outstanding problem in condensed matter physics. The intensity of phonon peaks in Resonant Inelastic X-ray Scattering (RIXS) spectra can be related to the underlying EPC strength under significant approximations whose validity deserves careful verification. We measured the Cu L$_3$ RIXS phonon intensity as function of incident photon energy and of momentum transfer in several layered cuprates. For  CaCuO$_2$, La$_{2-x}$Sr$_{x}$CuO$_{4+\delta}$, and \ch{YBa_2Cu_3O_{6}}, using a generally accepted theoretical model, we estimate quantitatively the EPC for the bond-stretching mode along the high-symmetry directions ($\zeta$,0) and ($\zeta$,$\zeta$), and as a function of the azimuthal angle $\varphi$ at fixed $q_\parallel$. We compare our results with theoretical predictions and we find that the $\mathbf{q}_\parallel$-dependence of the phonon RIXS intensity can be largely ascribed to the phonon symmetry. However, a more satisfactory prediction of the experimental results requires an accurate  description of the electronic structure close to the Fermi level.  Our extensive investigation indicates that Cu L$_3$ RIXS can be reliably used to determine the  momentum dependence of EPC for the bond-stretching modes of cuprates. Moreover, the large experimental basis provided in this article can serve as stringent test for advanced theoretical predictions on the EPC.

\begin{description}
\item[DOI]

\end{description}
\end{abstract}

\maketitle

%\tableofcontents

\section{\label{sec:level1}Introduction}
Since the discovery of high-temperature superconductivity (HTS) in layered copper oxides (cuprates) \cite{Bednorz1986}, the role of phonons has been hotly debated. Although the early consensus was that phonons are insufficient to describe the physics of HTS, subsequent theoretical \cite{Savrasov_PRL,Sakai_PRB,Jepsen1998,Ishihara_PRB,Johnston_PRB} and experimental studies \cite{Pashkin_PRL,Fausti_Science,Kaiser_PRB,Hu_Nature,Liu_PRX,Fava_Nature} have concluded that the electron-phonon interaction cannot be entirely neglected.  These results have prompted a reassessment of the influence of phonons on the superconducting pairing of HTS, and have rekindled the interest in determining the electron-phonon coupling (EPC).\par

The generic EPC depends on the initial ($\mathbf{k}$) and final ($\mathbf{k'}$) electronic wave vectors and on that of the phonon  $\mathbf{q}=\mathbf{k'}-\mathbf{k}$. The EPC is notoriously difficult to calculate theoretically and to determine experimentally, especially in strongly correlated materials such as cuprates. Various theoretical approaches have been proposed, including quantum Monte Carlo \cite{Huang_PRB,Marsiglio_PRB,Esterlis_PRB,Nosarzewski_PRB}, dynamical mean field theory (DMFT) \cite{Capone_PRL,Sangiovanni_PRL,Werner_PRL,Macridin_PRL,Macridin2012}, and density functional theory (DFT) \cite{Heid2008,Bohnen2003,Giustino2008,Sterling_PRB}. Although DFT can accurately predict the EPC in conventional superconductors \cite{Bohnen_PRB_2001}, this is not always the case for cuprates, where DFT can underestimate the EPC strength by more than one order of magnitude \cite{Reznik2008}. \par

On the experimental side, prominent techniques to determine the EPC in correlated systems include inelastic neutron scattering (INS) \cite{Pintschovius2005,Reznik2010,Ahmadova2020}, inelastic x-ray scattering (IXS) \cite{LeTacon_Nature,Miao_PRB,Souliou2020,Souliou2021,Peng_PRM},
Raman and infrared spectroscopy \cite{Cardona1988,Motida1999,Zhang2013,Farina_PRB}, angle-resolved photoemission spectroscopy (ARPES) \cite{Damascelli_RMP,Cuk2005}, scanning tunneling spectroscopy (STS) \cite{JLee_Nature,Eickhoff_PRB} and resonant inelastic x-ray scattering (RIXS) \cite{Ament_EPL,Devereaux2016_PRX,Geondzhian_PRB,Gilmore2023,Braicovich_PRR,Tanaka_PRB,WSLee_PRL13,Johnston_Nature2016,Scott_PRB2024,DashwoodPRX2024}. Due to the distinctive characteristics of each technique, experimental findings are routinely accompanied by a supporting theoretical framework, which facilitates the interpretation of the data and the derivation of the EPC strength. For example, INS and IXS probe the EPC at well-defined phonon momenta in the Brillouin zone (BZ), averaging over all electron momenta near the Fermi surface. There, the information about the EPC is derived from the linewidth of the phonon peak ($\gamma$), which also strongly depends on the electronic density of states \cite{Allen_PRB}. 
\par
The use of RIXS to study phonons and the EPC is largely based on a theoretical work by Ament $et$ $al$. \cite{Ament_EPL}, who showed that the dimensionless EPC can be determined from RIXS \emph{directly}, i.e., without further, material-specific, theoretical or computational steps (e.g., to derive the EPC from phonon linewidths). Systematic studies of the momentum dependence of the EPC in Nd$_{1+x}$Ba$_{2-x}$Cu$_3$O$_{7-\delta}$ along the ($\zeta$,0) direction \cite{Rossi_PRL, Braicovich_PRR} led to results in good agreement with previous Raman and optical conductivity measurements \cite{De_Filippis_PRL,Farina_PRB} as well as with theoretical calculations \cite{Johnston_PRB,Rosch_PRB,Ishihara_PRB,Devereaux_PRL_2004}. These studies were an important evidence that RIXS is a reliable spectroscopic probe of EPC and that, thanks to its unique ability to track multiple degrees of freedom (charge \cite{ArpaiaScience,Merzoni_PRB,Hepting_PRB}, orbital \cite{Moretti_NJP,Martinelli_PRL2024}, spin \cite{Braicovich2010_PRL} and lattice \cite{JJLee_PRB,Chaix2017}) simultaneously, it can be used to determine the interplay of phonons with other elementary excitations.
%This result was exploited in several Cu L$_3$ RIXS studies of high-energy optical modes in cuprates \cite{Ament_EPL,Devereaux2016_PRX,Braicovich_PRR,Rossi_PRL,Peng_PRB22,DashwoodPRX2024}.
\par

Ament’s pioneering work \cite{Ament_EPL,AmentPhD} established a direct connection between the RIXS one-phonon peak intensity and the dimensionless EPC constant $\tilde{g}$ that can be deduced from detuning measurements (see below for details).  Initially obtained from a simple model containing a non-dispersive Einstein phonon, this result can be formally extended to realistic (dispersive) phonons, enabling the momentum ($\boldsymbol{\mathrm{q}}$) dependence of the EPC to be extracted from RIXS data \cite{Braicovich_PRR,BieniaszPRB2022}.  However, calculating $\boldsymbol{\mathrm{q}}$-dependent EPCs in practice, either from first principles or from simplified models, is complex, making Ament’s key result very appealing but also rather difficult to \emph{fully} validate.  This is compounded by the lack of a suitable experimental benchmark, since, to our knowledge, besides the initial work by Braicovich \textit{et al.} \cite{Braicovich_PRR} and some specific applications related to charge order \cite{Peng_PRL,ChaixPRR2022,Wang2021,Huang_PRX2021,Scott_SciAdv2023}, there are no systematic studies of the $\boldsymbol{\mathrm{q}}$-dependent one-phonon intensity in cuprates. \par

In this paper, we present measurements, performed on some representative cuprates (CaCuO$_2$, La$_{2-x}$Sr$_{x}$CuO$_{4+\delta}$, and \ch{YBa_2Cu_3O_{6}}) of the Cu L$_3$ edge RIXS intensity of the bond-stretching phonon peak as a function of detuning energy and then of momentum along two high-symmetry directions ($\zeta$,0), ($\zeta$,$\zeta$) and as a function of the azimuthal angle $\varphi$ at fixed modulus of the in-plane momentum $\mathbf{q_\parallel}$. We find that the EPC strength in three antiferromagnetic cuprates is between $\sim0.15$ and $0.17$ \,eV, consistent with previous results for \ch{NdBa_2Cu_3O_6} \cite{Rossi_PRL} at the same point in the Brillouin zone. For the momentum dependence, we compare our results with two different theoretical calculations: one based on Ament's model and either the tight-binding approximation or DFT, and the other one based on a frozen-phonon model that does not directly invoke the EPC.  Strikingly, for dispersionless phonons the $\textbf{q}$-dependent intensities from tight-binding and frozen phonon calculations are \emph{identical} and capture the main momentum dependence along the two high-symmetry directions. This clearly indicates that the $\boldsymbol{\mathrm{q}}$-dependence of the one-phonon RIXS intensity is primarily determined by the symmetry of the phonon, being mostly sensitive to the projection of the oxygen phonon displacements onto the directions of the $x^2-y^2$ orbitals, \emph{regardless} of the model used to interpret the data.  However, the $\varphi$ dependence is better reproduced by DFT model calculations, suggesting that, when calculating EPC, the electron wave vectors $\boldsymbol{\mathrm{k}}$ and $\boldsymbol{\mathrm{k'}}$ cannot be neglected, despite the strong electron-electron correlation characteristics of cuprates.

\section{Theoretical and experimental background}\label{Theoretical background}
\subsection{Probing EPC with RIXS}
\begin{figure*}
    %\centering
     \textbf{Features of phonon excitation in RIXS.}\\[0.5em]
    \includegraphics[width=\textwidth]{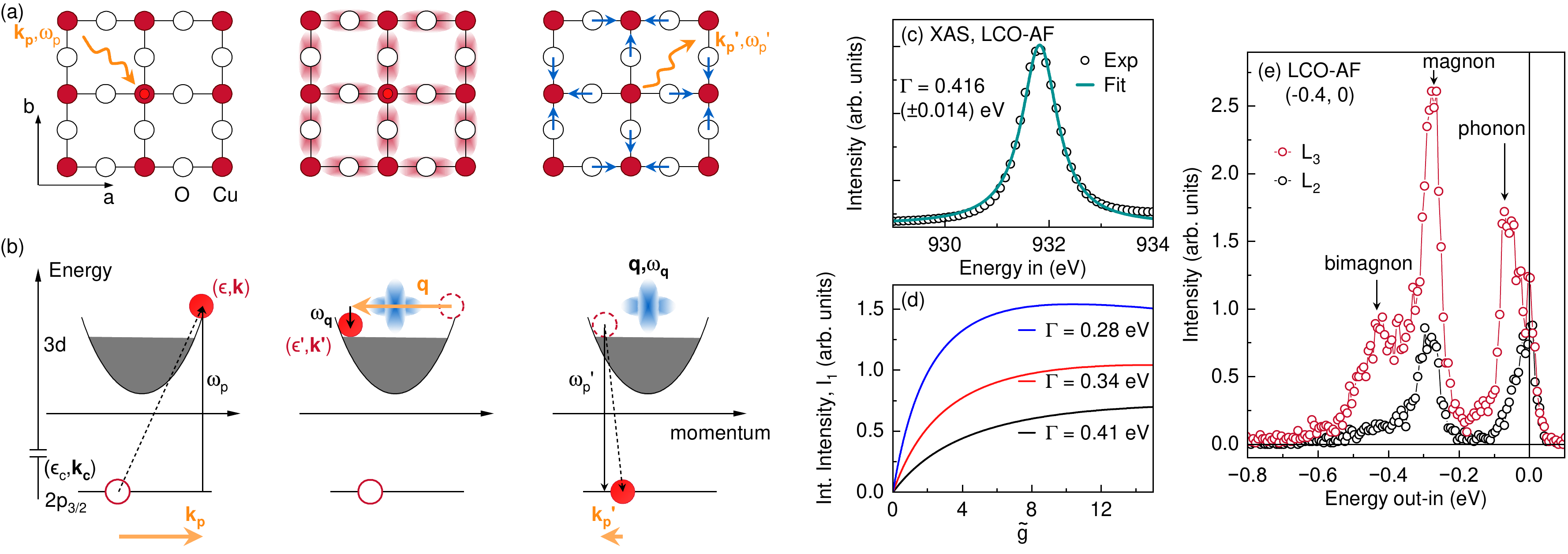}
    \caption{A scheme of phonon generation in RIXS process at the Cu L$_3$ edge in (a) real and (b) reciprocal spaces. Here, $\boldsymbol{\mathrm{k}}_{\mathrm{p}}$, $\omega_p$  ($\boldsymbol{\mathrm{k}}_{\mathrm{p}}'$, $\omega_p'$) are the wavevector and energy of the incoming (outgoing) photon; $\boldsymbol{\mathrm{k}}_{\mathrm{c}}$, $\epsilon_c$ are the wavector and energy of the core-hole; $\boldsymbol{\mathrm{k}}_{\mathrm{p}}$, $\epsilon$  ($\boldsymbol{\mathrm{k}}_{\mathrm{p}}'$, $\epsilon'$) are the wavevector and energy of the photo-excited electron before (after) the phonon generation; $\boldsymbol{\mathrm{q}}$, $\omega_{\boldsymbol{\mathrm{q}}}$ are the wavevector and energy of the created phonon. For the conservation of the momentum and energy: $\boldsymbol{\mathrm{k}}'=\boldsymbol{\mathrm{k}}+\boldsymbol{\mathrm{q}}$, $\boldsymbol{\mathrm{k}}_{\mathrm{p}}'=\boldsymbol{\mathrm{k}}_{\mathrm{p}}+\boldsymbol{\mathrm{q}}$, $\omega_{\boldsymbol{\mathrm{q}}}=\omega_{\mathrm{p}}-\omega_{\mathrm{p}}'=\epsilon-\epsilon'$, $\epsilon_{\mathrm{c}}=\epsilon-\omega_{\mathrm{p}}=\epsilon'-\omega_{\mathrm{p}}'$.  (c) Experimental XAS curve in total electron yield of the antiferromagnetic La$_2$CuO$_4$ (LCO-AF) around the Cu L$_3$ edge resonance and its Lorentzian fitting. (d) Calculated $\tilde{g}$-$I_1$ relationship from Eq.\ref{eq:RIXS_Int_1} for $\omega_{\textbf{q}}=75$\,meV, $\Delta=0$  and three different values of $\Gamma$. (e) Raw (not normalized) RIXS spectra of LCO-AF at the Cu L$_3$ (black) and L$_2$ (red) edges. The high energy phonon mode at $\sim 0.08$\,eV is indicated by arrow. The ratio between the phonon and elastic peak is considerably lower at the L$_2$ edge.}
       \label{fig:RIXS_phonons}
\end{figure*}
We remind the reader of the main features of lattice excitations in the RIXS process (see Fig.\,\ref{fig:RIXS_phonons}\,(a)-(b)) that are relevant for our discussion. The absorption of a x-ray photon at the Cu L$_3$ resonance creates an excitation of an electron from the $2p_{3/2}$ core level up to the highest $3d$ valence state, where it couples with the local charge distribution. In this RIXS `intermediate state', coupling with the lattice occurs primarily through coordinated (virtual) displacements of the ligand oxygens, which can be conceptualized as though the equilibrium lattice were perturbed by a coherent superposition of vibrational modes. The photo-excited electron radiatively decays back into the $2p_{3/2}$ state with a typical time constant of a few fs, leaving one or more excited phonons in the final state. The conservation laws impose that the total wave vector $\boldsymbol{\mathrm{q}}$ and energy $\omega_{\boldsymbol{\mathrm{q}}}$ of the final phonons are exactly the difference between wave vectors and energies of the incoming ($\boldsymbol{\mathrm{k}}_{\mathrm{p}},\omega$) and outgoing ($\boldsymbol{\mathrm{k}}_{\mathrm{p}}', \omega'$) photons, $\boldsymbol{\mathrm{q}} = \boldsymbol{\mathrm{k}}_{\mathrm{p}}'-\boldsymbol{\mathrm{k}}_{\mathrm{p}}$ and $\omega_{\boldsymbol{\mathrm{q}}} =  \omega-\omega'$.\par
As mentioned above, the model developed by Ament \textit{et al.} \cite{Ament_EPL} implies that quantitative information about the EPC can be directly
extracted from the RIXS spectra. The initial assumption of this model is to consider both electrons and phonons as \emph{localized} during the RIXS process. Hence, the electronic structure is approximated with a single-ion model, and the phonon is considered non-dispersive (Einstein phonon).  From this basis, Ament proceeds to establish a direct connection between the one-phonon peak intensity $I_1$ observed in RIXS and the dimensionless EPC strength $\tilde{g}$ ($\tilde{g} = M^2/{\omega_{\boldsymbol{\mathrm{q}}}}^2$), with no momentum dependence as the wave vectors of the vibrational mode and of the Cu $2p$ (core) and $3d$ (valence) electronic states are undefined:  
\begin{align}
    \label{eq:RIXS_Int_1}
    I_1 \propto \frac{e^{-2\Tilde{g}}}{\Tilde{g}}\left|\sum_{\substack{n_i=0}}^{\infty}\frac{\Tilde{g}^{n_i}(n_i-\Tilde{g})}{n_i![\Delta+i\Gamma+(\Tilde{g}-n_i)\omega_{\boldsymbol{\mathrm{q}}}]}\right|^2
    \end{align}
Here, $M$ is the EPC in eV; $\Delta$ is the detuning energy, i.e., the difference between the incident and resonance energy; $\Gamma$ is the natural width of the resonance, which is inversely proportional to the core-hole lifetime $\tau$ ($\tau \propto 1/\Gamma$); $\omega_{\boldsymbol{\mathrm{q}}}$ is the phonon frequency (in that model $\omega_{\boldsymbol{\mathrm{q}}}$ is a constant) and $n_i$ is the number of phonons in the intermediate state. This expression assumes a single vibrational mode that, in general, could be excited multiple times in the intermediate and in the final state of RIXS; the case of several modes possibly being excited simultaneously is considered by Gilmore \textit{et al.} in Ref.\,\cite{Gilmore2023, Geondzhian_PRB} and discussed by Scott \textit{et al.} in \cite{Scott_PRB2024}.\par
In applying Eq.\,\ref{eq:RIXS_Int_1}, it is important to consider that multiple values of $\Gamma$ have been reported in the literature for a given absorption edge \cite{keski1974total,krause1979natural,muller1982x,Fuggle_PRA}. Experimentally, $\Gamma$ corresponds to the half-width at half-maximum (HWHM) of the x-ray absorption (XAS) spectrum. In this work at the Cu L$_3$ edge we mainly use $\Gamma \sim 0.41$\,eV evaluated from XAS spectrum of antiferromagnetic La$_2$CuO$_4$ (see Fig.\,\ref{fig:RIXS_phonons}(c)) This number is in agreement with previously reported experimental values from Ref.\,\cite{Fuggle_PRA}.\par
A graphical representation of the relationship between $\tilde{g}$ and $I_1$ at $\Delta=0$ and three different $\Gamma$ calculated numerically with Eq.\,\ref{eq:RIXS_Int_1} for $n_i$ running up to 150 is shown in Fig.\,\ref{fig:RIXS_phonons}\,(d). It illustrates how $\Gamma$ influences the shape of the Ament curve: as $\Gamma$ increases, the relationship $I_1$-$\tilde{g}$ becomes more linear in a larger range of $\tilde{g}$, and the plateau shifts to higher $\tilde{g}$. For $\tilde{g}<10$ $I_1$ is a monotonically increasing function of $\tilde{g}$, quasi-linear for $\tilde{g}<2$. Using this relation one can map the relative value of $\Tilde{g}$ (or $I_1$) as a function of a parameter, e.g. the momentum transferred by the scattering photon, opening the possibility of measuring the EPC as a function of $\mathbf{q}$. \par
Eq.\,\ref{eq:RIXS_Int_1} points out the crucial dependence of $I_1$ on the lifetime of the intermediate state $\tau$ during which the phonon is created. For higher $\Gamma$ (lower $\tau$) the probability of the electron-phonon interaction significantly decreases, such that at the L$_3$ edge of $5d$ elements phonons are not observed. Figure\,\ref{fig:RIXS_phonons}\,(e) shows the comparison between raw RIXS spectra at the Cu L$_3$ and L$_2$ edges. Here we note that the ratio between the phonon and elastic peak is considerably lower at the L$_2$ edges, as the intermediate state lifetime is shorter than that at the L$_3$ edge. Similarly, the bimagnon to single magnon intensity ratio is much smaller at L$_2$ than at L$_3$, indicating that the fomer excitation, but not the latter, is influenced by the intermediate state life time \cite{Bisogni_bimagnon}\par 
Since the RIXS intensities are never measured on an absolute scale, one cannot use intensity data at a single energy in combination with Eq.\,\ref{eq:RIXS_Int_1} to determine the \emph{absolute} EPC \cite{Devereaux2016_PRX}, and alternative procedures must be employed \cite{Ament_EPL,WSLee_PRL13, Braicovich_PRR,Jinu_PRX}. A possible method is based on the ratio of the intensity of one-phonon peak $I_1$ and its first overtone $I_2$, since $I_2/I_1$ is scale-independent but explicitly contains $\tilde{g}$ \cite{Ament_EPL}. However, this approach is hardly applicable to Cu L$_3$ edge RIXS of cuprates, since the phonon satellites are usually difficult or even impossible to identify unambiguously. Moreover the two-phonon excitation has a momentum dependence different from the single phonon, because the total momentum is given by the sum of the momenta of the two phonons, leading to a rather \textbf{q}-independent $I_2$ \cite{Scott_PRB2024}. In such a case, an alternative is the so-called `detuning' method \cite{Braicovich_PRR,Rossi_PRL}, which consists in acquiring multiple RIXS spectra at incident energies progressively detuned by $\Delta$ away from the resonance. By tracking the decrease of $I_1$ with $\Delta$, the absolute value of $\tilde{g}$ can be determined using Eq.\,\ref{eq:RIXS_Int_1}. \par

In Ament's original model, the local character of the excited phonons forces $\tilde{g}$ to be independent on $\boldsymbol{\mathrm{q}}$, while in reality, $\tilde{g}$ (or $M$) is known to be momentum dependent, while the experimental $I_1$ are also $\boldsymbol{\mathrm{q}}$-dependent.  Eq.\,\ref{eq:RIXS_Int_1} has been employed in the past (with limited precautions) to extract the \emph{relative} $\boldsymbol{\mathrm{q}}$-dependence of the EPC constant $\tilde{g}$ ($M$) from RIXS data, which does not require any detuning \cite{Chaix2017,Wang2021,Huang_PRX2021,Li2020}.  \par

\subsection{Momentum-dependent EPC in cuprates: tight-binding calculations}
First theoretical attempts to capture the experimental results were made using a tight-binding (TB) model of the electronic structure \cite{Song_PRB1995, Johnston_PRB,Rosch_PRB,Ishihara_PRB,Devereaux_PRL_2004,horsch2005}. This model provides a simple analytical expression of the $\boldsymbol{\mathrm{q}}$-dependent EPC strength, and was first applied to the Cu-O in-plane bond-stretching phonon mode (also known as breathing mode). One of the reasons for this choice is the connection between bond-stretching phonons and charge-order correlations \cite{Chaix2017}, which are ubiquitous in all cuprate families and result in phonon energy softening and increases in intensity at the charge order wave-vector \cite{Peng_PRB22,ChaixPRR2022,Wang2021,Huang_PRX2021}. Moreover, from the theoretical perspective, the role of breathing phonons in HTS superconductivity is controversial, since different authors have argued that they provide either an attractive \cite{Ishihara_PRB,Shen2002} or a repulsive \cite{Johnston_PRB, Bulut_PRB,Sandvik_PRB} interaction.\par
For the breathing mode, using a TB model, Johnston \textit{et al.} \cite{Johnston_PRB} found that the EPC $M(\textbf{q},\textbf{k},\textbf{k'})$, once averaged over all the electronic states at the Fermi level, simplifies into
\begin{align}
\label{eq:M_br_fermi}
     M_{\mathrm{br}}(\boldsymbol{\mathrm{q}}) \propto  \frac{1}{\sqrt{\omega_{\textbf{q}}}}\sqrt{\mathrm{sin^2}(q_x\pi)+\mathrm{sin^2}(q_y\pi)},
    \end{align}
and the estimated EPC strength at $\mathbf{q_{\parallel}}=(0.5,0)$\,r.l.u. is of the order of 86\,meV \cite{JohnstonPhD}. It must be noted that for other modes the dependence on $\mathbf{k}$ cannot be eliminated. In Eq.\,\ref{eq:M_br_fermi} the EPC $M$ depends on the phonon wave-vector \textbf{q} through the sine function and the phonon frequency $\omega_{\textbf{q}}$. As discussed below, the frequency term is crucial, as the breathing modes exhibit a non-negligible dispersion in the BZ. Using Eq.\,\ref{eq:RIXS_Int_1} or its simplified expression $I_1(\textbf{q}) \propto \tilde{g}(\textbf{q})$, one can thus extract the experimental value of $M(\textbf{q})$ to be compared with the theory. \par

\subsection{Nomenclature of phonons in cuprates and previous measurements}

\begin{figure}
    %\centering
        \textbf{Bond-stretching modes in the BZ.}\\[0.5em]
    \includegraphics[width=\columnwidth]{ 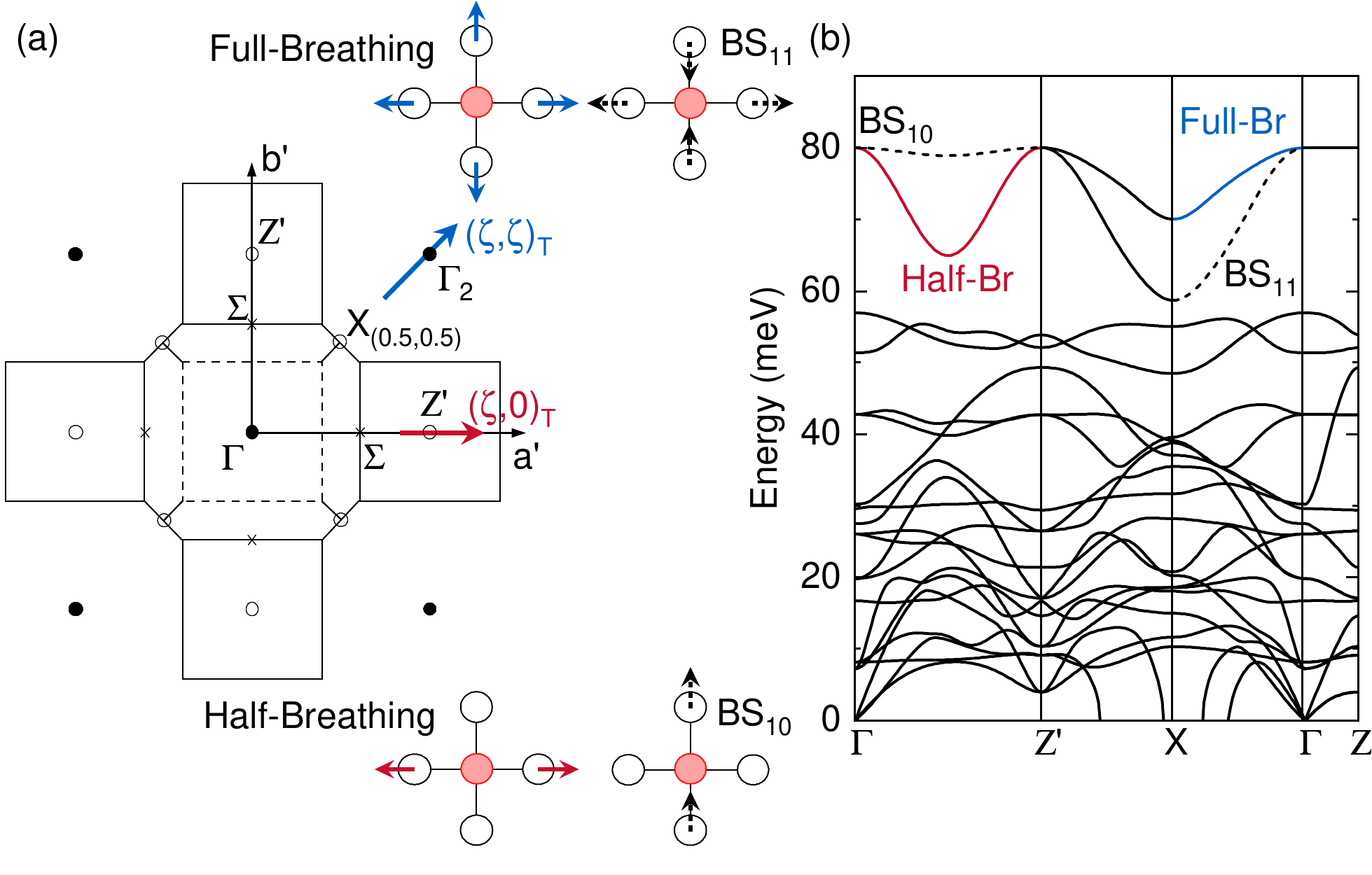}
    \caption{(a) A scheme of oxygen displacements of the breathing modes along the ($\zeta$, 0) and ($\zeta$,$\zeta$) directions in the BZ of the pseudo-tetragonal \ch{La_2CuO_4}. (b) Calculated phonon dispersion curves in \ch{La_2CuO_4}. The high-energy phonon branches that can be studied in RIXS are highlighted with colors while silent RIXS modes are shown with dotted lines.}
    \label{fig:nomenclature}
\end{figure}

Bond-stretching optical phonons in cuprates have traditionally been given different names, depending on the direction in the BZ. Along the Cu-O bond (the ($\zeta$,0) direction in our conventions --- see below), only two Cu-O bond lengths oscillate and the mode is called half-breathing; at 45$^\circ$ from the Cu-O bond (the ($\zeta$,$\zeta$) direction), all four Cu-O bond lengths oscillate and the mode is called full-breathing \cite{Zhang_PRB,horsch2005,sterling2021} (see Fig.\,\ref{fig:nomenclature}\,(a)). In addition to the breathing and half-breathing branches, along each of the two high-symmetry directions $(\zeta, 0)$ and $(\zeta, \zeta)$ there is another bond-stretching branch, which we label BS$_{10}$ and BS$_{11}$ in the two directions.  These branches are indicated with dashed lines in Fig.\,\ref{fig:nomenclature}\,(b), and merge with the breathing/half-breathing branches into a degenerate phonon at the $\Gamma$ point, as confirmed by our DFT calculations. However, the BS$_{10}$ and BS$_{11}$ modes have little effect on the average Cu-O bond lengths all along the dispersion, and are expected to be weakly coupled to RIXS (see Methods\,\ref{sec:Methods}). This is indeed confirmed by all the theoretical models we tested.\par

For some cuprate families, the phonon dispersion along high-symmetry directions in the BZ has been determined experimentally by means of INS measurements, and the energy of the breathing modes was found to be in the range between 65 and 85\,meV \cite{REICHARDT_PhC,PINTSCHOVIUS_PhC,Pintschovius_PRB,Braden_PRB}. However, their coupling to the electrons has not been determined, because INS/IXS phonon linewidths are resolution-limited.\par

More recently, RIXS investigations have shown that the intensity of the half-breathing mode scales as sin$^2(q_x\pi)$ in the ($\zeta$,0) direction for Nd$_{1+x}$Ba$_{2-x}$Cu$_3$O$_{7-\delta}$ \cite{Braicovich_PRR, Rossi_PRL}, La$_{1.8-x}$Eu$_{0.2}$Sr$_{x}$CuO$_{4+\delta}$ \cite{Peng_PRL,Wang2021}, La$_{2-x}$Sr$_x$CuO$_4$ \cite{Lin_PRL,Martinelli2024}, Bi$_2$Sr$_{2-x}$La$_x$CuO$_{6+\delta}$ \cite{Li2020}, Bi$_2$Sr$_2$CaCu$_2$O$_{8+\delta}$ \cite{Chaix2017,WSLee2021}.
The agreement between RIXS observations and Eq.\,\ref{eq:M_br_fermi} strengthened the idea that the RIXS intensity is proportional to $\tilde{g}$ and that the TB model is appropriate to describe the EPC in cuprates. However, so far, no systematic investigation of the phonon intensity has been carried out along other paths in the BZ, calling for a more robust experimental test of the Ament prediction (Eq.\,\ref{eq:RIXS_Int_1}) and of the TB model prediction of EPC (Eq.\,\ref{eq:M_br_fermi}).\par

\section{\label{sec:RIXS experiment} RIXS experiments}
\begin{figure*}
    \textbf{The detuning dependence of RIXS spectra.}\\[0.5em]
    %\centering
    \includegraphics[width=0.75\textwidth]{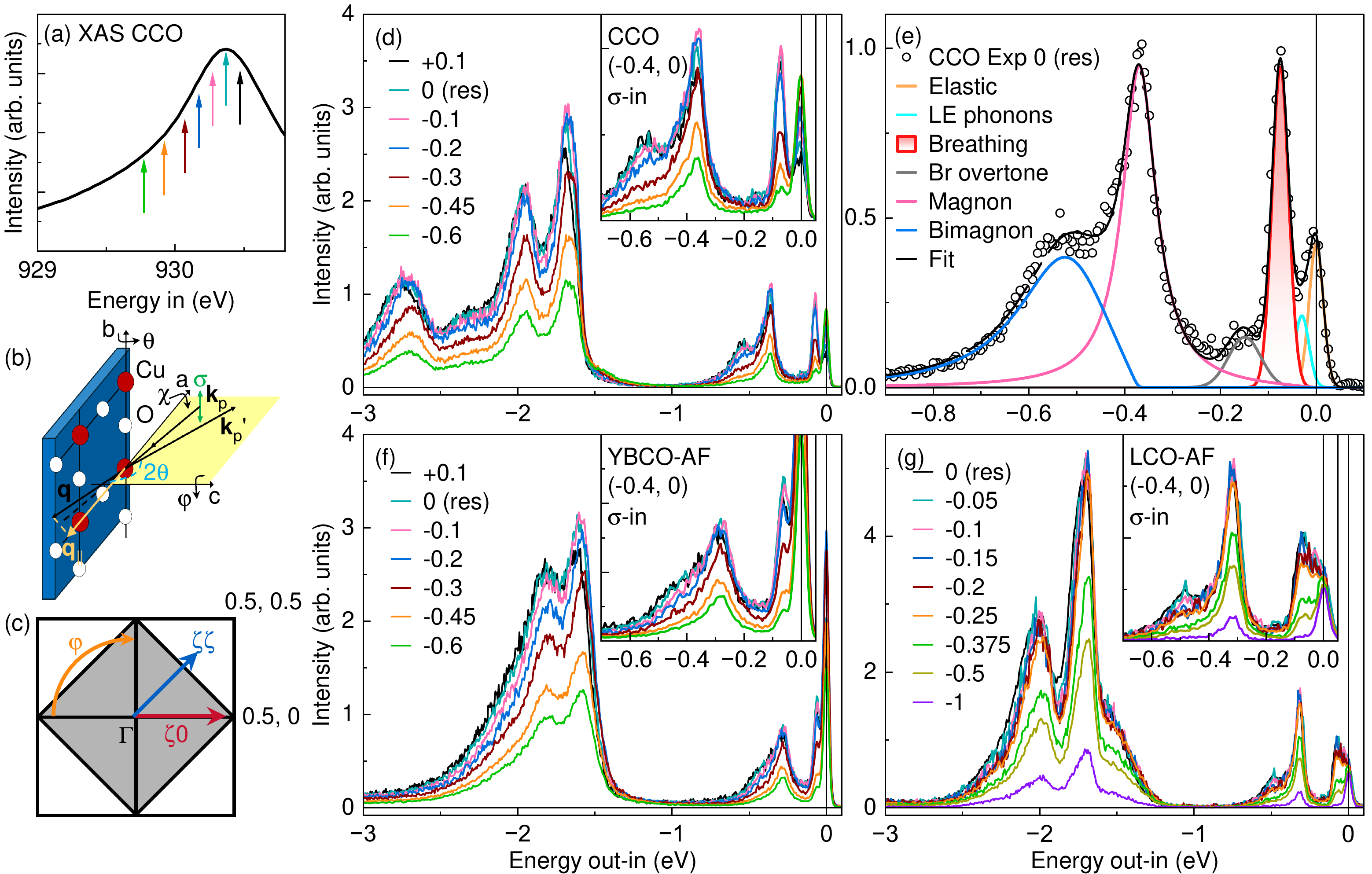}
    \caption{(a) XAS plot of CCO around the Cu L$_3$ resonance, taken by measuring total electron yield. 
    Arrows indicate incident photon energies in RIXS detuning study. (b) A sketch of the experimental geometry. (c) Trajectories in the BZ along which the $\textbf{q}$-dependence of the EPC was measured. (d) RIXS spectra of CCO collected at $\textbf{q}_{\parallel}=(-0.4,0)$\,r.l.u. as a function of the detuning energy $\Delta$. The inset highlights low-energy features, including phonon excitations. (e) Decomposition and fitting of a RIXS spectrum of CCO at resonance ($\Delta = 0$). The fitting procedure is described in the main text.
    (f), (g)  Same as panel (d), but for YBCO-AF and LCO-AF, respectively.
}
       \label{fig:Detuning1}
\end{figure*}

\subsection{\label{subsec:Experimental details} Experimental details}

We measured the phonon dispersion and intensities in three distinct cuprates families, all grown as thin films: \par
i) Infinite-layer \ch{CaCuO_2} (CCO), thickness $t = 40$\,nm, lattice constants $\textit{a} = \textit{b} = 3.85$\,\si{\angstrom}, $\textit{c} = 3.2$\,\si{\angstrom}.\par
ii) La$_{2-x}$Sr$_{x}$CuO$_{4+\delta}$ series. Antiferromagnetic (LCO-AF), $x\approx 0$, $\delta \approx  0$, $t \approx 30$\,nm. Strongly underdoped, non-superconducting (LCO-UD0),  $x \approx 0$, $\delta\approx 0.03$, $t\approx 30$\,nm. Underdoped with $T_c = 26$\,K (LCO-UD26), $x \approx 0$, $\delta \approx 0.12$, $t \approx 30$\,nm. Optimally doped with $T_c = 33$\,K (LSCO-OP33), $x \approx 0.16$, $\delta \approx 0$, $t \approx 18$\,nm. Although bulk L(S)CO would be orthorhombic at the temperatures where our experiment was conducted, our epitaxial L(S)CO films are all tetragonal with approximate lattice constants $\textit{a} = \textit{b} = 3.76$\,\si{\angstrom} for all L(S)CO samples, while $\textit{c} = 13.15, 13.16, 13.19$ and 13.24\,\si{\angstrom} for the four samples respectively. Additionally, in Appendix\,\ref{sec:Appendix A} we discuss the RIXS measurements on another underdoped LCO sample. \par 
iii) Antiferromagnetic \ch{YBa_2Cu_3O_{6}} (YBCO-AF), hole doping $p< 0.04$, $t = 50$\,nm, lattice constants $\textit{a} = \textit{b} = 3.87$, $\textit{c} = 11.83$\,\si{\angstrom}.\par
Details of sample growth and characterization are discussed in Ref.\,\cite{Martinelli_PRL2024, diCastro_CCO_2014,dicastro_2009_CCO} for CCO, in Ref.\,\cite{bozovic_2002_LSCO,Martinelli_PRL2024,BiagiPhD} for L(S)CO and in Ref.\,\cite{arpaia_2018_YBCO} for YBCO. All measurements were performed at the beamline ID32 of the European Synchrotron (ESRF, France) using the ERIXS spectrometer \cite{BROOKES_ID32}, with an overall energy resolution between 35 and 45\,meV at the Cu L$_3$ edge ($\approx$ 931\,eV).  All RIXS data are normalized only to the incident flux. To maximize the phonon signal, the incident polarization was chosen to be $\sigma$, which is perpendicular to the scattering plane and parallel to the $ab$ Cu-O planes, while there was no outgoing polarization analysis. \par
The detuning method was applied to estimate EPC in CCO, LCO-AF and YBCO-AF. The data were acquired at $\textbf{q}_{\parallel}=(-0.4,0)$\,r.l.u. with incident photon energy progressively detuned from the Cu L$_3$ resonance, $\Delta$ (see Fig.\,\ref{fig:Detuning1}\,(a)). The scattering angle $2\theta$ was set to 149.5$^\circ$ and the temperature was kept constant at about 20\,K. The overall experimental geometry is schematically illustrated in Figure\,\ref{fig:Detuning1}\,(b). All detuning spectra are corrected for geometrical and energy-dependent self-absorption effects.\par
The momentum (\textbf{q})-dependence of the EPC was measured in all described samples by moving along high-symmetry ($\zeta$, 0) and ($\zeta$,$\zeta$) directions and along the azimuthal direction with fixed $q_{\parallel} = 0.4$\,r.l.u. in the BZ (see Fig.\,\ref{fig:Detuning1}\,(c)). In this work, the momentum transfer $\textbf{q}$ is expressed in reciprocal lattice units (r.l.u.), and its projection on the basal $ab$ plane is indicated as $\textbf{q$_{\parallel}$}$. Special points in the BZ are labeled according to the conventions introduced by Pintschovius \textit{et al.} \cite{PINTSCHOVIUS_PhC}, which are appropriate for the body-centered cell of L(S)CO (see schematic drawing in Fig.\,\ref{fig:RIXS_phonons}).\par
 At a fixed scattering angle $2\theta$ the modulus of the \textbf{q$_{\parallel}$} ($q_{\parallel}$)  is defined by the angle $\theta$, while the in-plane orientation of \textbf{q$_{\parallel}$} is set by the azimuth angle $\varphi$. It must be noted that by $2\theta$ we indicate the angle formed by $\textbf{k}_\mathrm{p}$ and $\textbf{k'}_\mathrm{p}$, which is independent of the value of $\theta$; $2\theta=2 \times \theta$ only for $q_{\parallel}=0$.  With this convention, the ($\zeta$, 0) direction corresponds to $\varphi = 0^\circ$ and ($\zeta$,$\zeta$) to $\varphi = 45^\circ$. Since in $\varphi$ scans the angles 2$\theta$, $\theta$ and $\chi$ are fixed, the self-absorption effect for each $q_{\parallel}$ is exactly the same \cite{Minola_PRL}. For that reason, we did not apply the self-absorption correction to $\varphi$ scans, whereas we did it for all other momentum trajectories.\par

\section{\label{subsec:Experimental results}Experimental results and theoretical predictions}
\subsection{\label{subsec:Detuning}Detuning method}

\begin{figure}
    %\centering
    \textbf{The detuning dependence of RIXS excitations.}\\[0.5em] 
    \includegraphics[width=0.5\columnwidth]{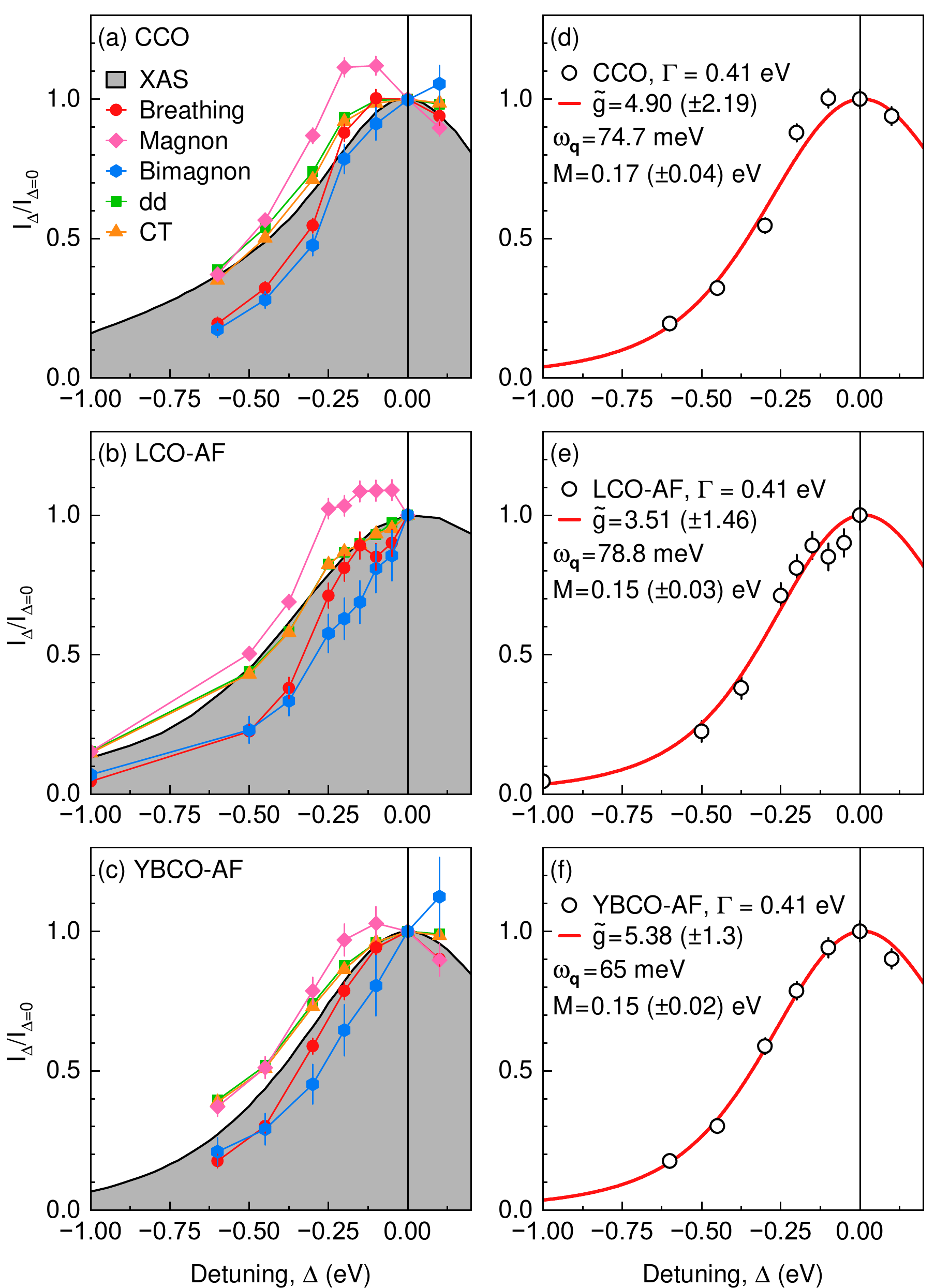}
    \caption{(a)-(c) The evolution of the integrated intensity of various excitations (symbols) upon detuning in CCO, LCO-AF and YBCO-AF on top of corresponding XAS plot (gray area). (d)-(f) Fitting curves (solid lines) calculated using Eq.\ref{eq:RIXS_Int_1} to extract EPC of the breathing mode. The empty circles show detuning dependence of the integrated phonon intensity. The value of $\tilde{g}$ is related to the width of the detuning curve: a wider curve corresponds to a larger $\tilde{g}$. The parameters for the Ament model and extracted EPC are summarized in each panel. Here and in the following, error bars indicate the 95\% confidence interval from the fit.}
       \label{fig:Detuning2}
\end{figure}

\begin{figure*}
    %\centering
            \textbf{RIXS spectra of CaCuO$_2$.}\\[0.5em] 
    \includegraphics[width=0.75\textwidth]{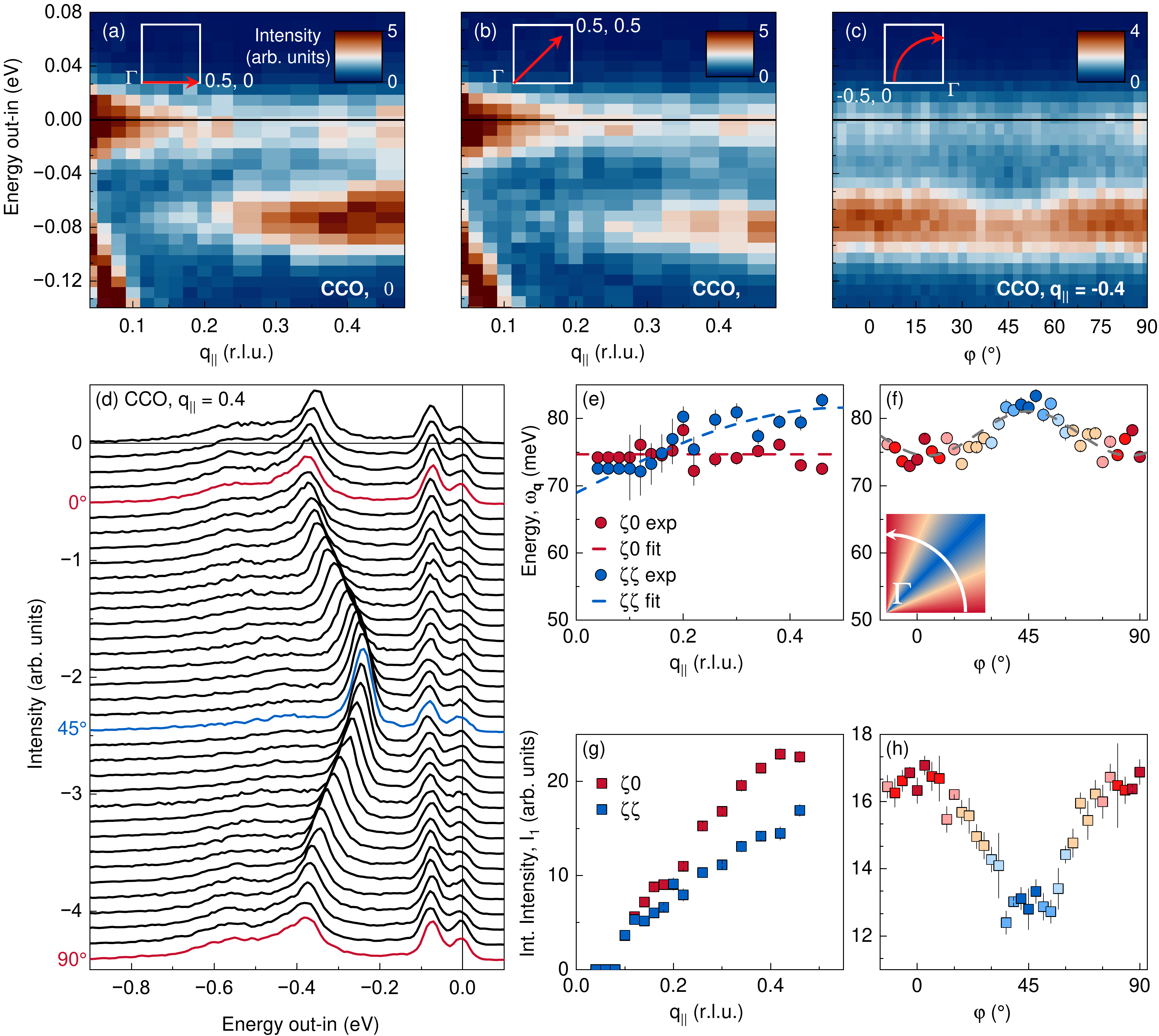}
    \caption{Experimental RIXS intensity maps along the (a) ($\zeta$,0), (b) ($\zeta$,$\zeta$) and (c) azimuth path. The trajectories in the reciprocal space are indicated in the insets. (d) A stack of RIXS spectra in the $\varphi$ scan. The red and blue curves indicate spectra at (-0.4,0), (0,0.4) and (-0.4/$\sqrt{2}$,0.4/$\sqrt{2}$) in r.l.u, respectively. (e), (f) Experimental breathing phonon energy (symbols) with \textbf{q$_{\parallel}$} along the same path as in (a), (b) and (c). The inset colormap in (f) indicates the adopted color-\textbf{q$_{\parallel}$} convention. The dashed lines are the fitting of the phonon dispersion which are subsequently used to reduce the number of free fit parameters as an input for theoretical calculations. (g), (h) Experimental evolution of the breathing phonon integrated intensity with $\mathbf{q}_{\parallel}$ along the same paths as in (e) and (f). All data are taken at 20\,K and 2$\theta=149.5^{\circ}$.
}
    \label{fig:RIXS_CCO1}
\end{figure*}

\begin{figure}
    %\centering
    \textbf{EPC in CaCuO$_2$. }\\[0.5em]
    \includegraphics[width=\columnwidth]{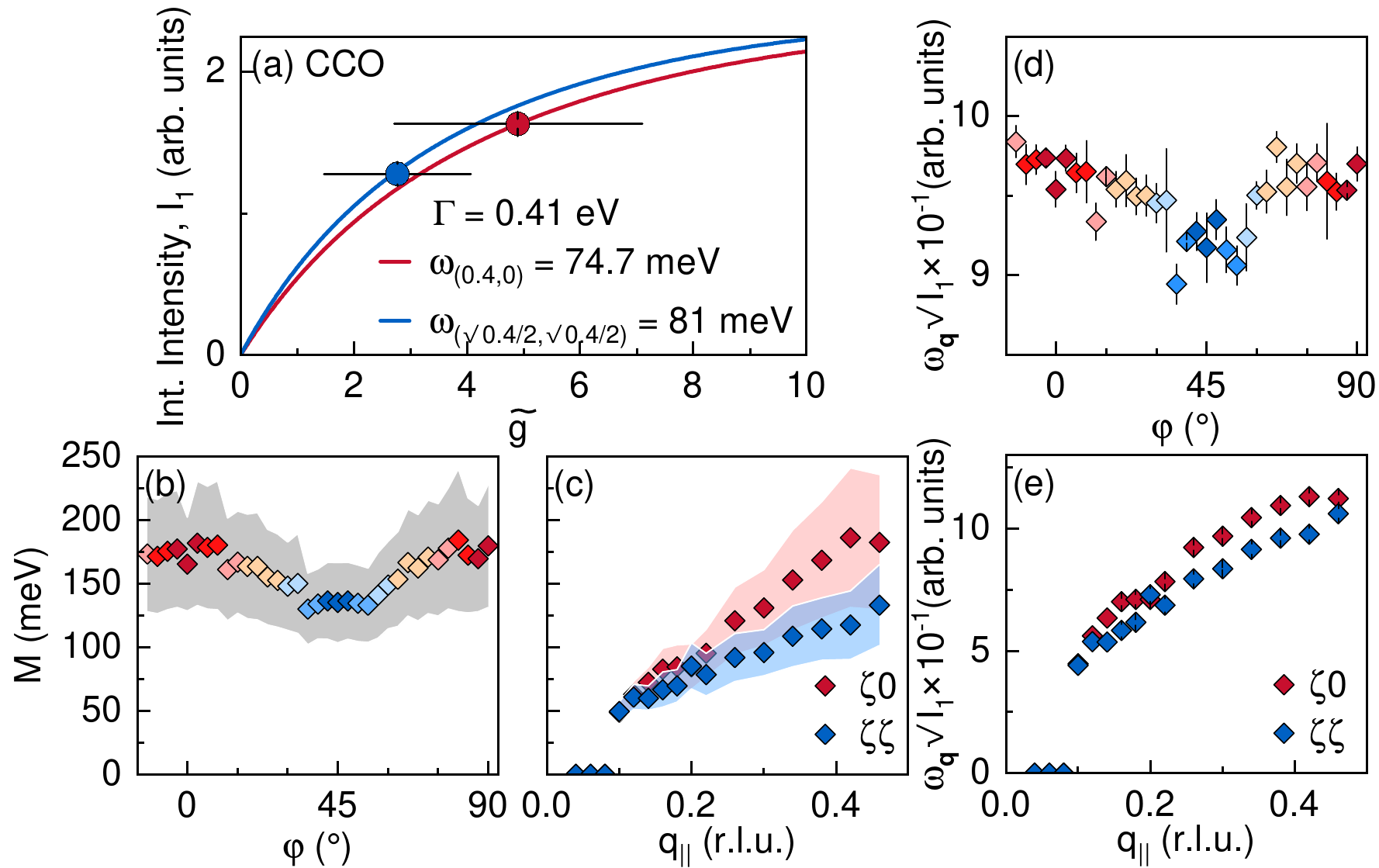}
    \caption{(a) Curves of the Ament model for CCO calculated for two different $\omega_{\textbf{q}}$. The circles indicate the EPC $\tilde{g}$ with error bars for two $\textbf{q}_{\parallel}$. (b), (c) Evolution of the breathing phonon EPC strngth $M$ as a function of $\mathbf{q}_{\parallel}$ along the azimuth and ($\zeta$,0), ($\zeta$,$\zeta$) directions. (d), (e) Evolution with $\varphi$ and $q_{\parallel}$ of the product $\omega_{\textbf{q}}\sqrt{I_1}$ which is defined as a simplified version of the Ament model in the main text.}
      \label{fig:RIXS_CCO2}
\end{figure}
 
Figure\,\ref{fig:Detuning1}\,(d),\,(f),\,(g) displays RIXS spectra for the antiferromagnetic cuprates as a function of detuning energy $\Delta$. The strongest inelastic scattering signal is observed between -1 and -3.5\,eV which arises from crystal field $dd$-excitation. Similarly, the spectral weight below -3.5\,eV (not shown here) is attributed to charge-transfer (CT) excitations. However, in this work we discuss more in details the energy region above -1\,eV (see insets) which  includes contributions from elastic (0\,eV), phonons (between -0.03 and -0.08\,eV), magnon ($\sim -0.35$\,eV) and bimagnon ($\sim -0.55$\,eV) peaks. \par
In order to extract quantitative information from our experimental results, we modeled the spectra using a standard multi-peak fitting approach (see Fig.\,\ref{fig:Detuning1}\,(e)). In particular, we used resolution-limited Gaussian profiles for the elastic (orange) and a low energy phonon branches (cyan), a broader Gaussian curve for the breathing phonon (red) and its overtone (gray) and damped harmonic oscillators for magnon and bimagnon excitations (pink and blue, respectively). Following the same approach as in our previous works \cite{Braicovich_PRR, Rossi_PRL}, we extract the integrated intensities (i.e., peak areas, $I_{\Delta}$) of the inelastic excitations for each detuning energy $\Delta$. These intensities are normalized to their on resonance value $I_0$, and the resulting ratio $I_{\Delta}/I_0$ is plotted as a function of $\Delta$, compared with the XAS spectrum normalized in the same way (see Fig.\,\ref{fig:Detuning2}\,(a)–(c)). We note that while the detuning dependence of $dd$ and CT excitations coincide with the XAS decrease, in CCO and LCO-AF the magnon decays more slowly upon detuning and it resonates before the maximum of the XAS. 
On the other hand, the breathing phonon and bimagnon decay faster than XAS, highlighting the slow nature of the excitation process. Surprisingly, the maximum of the bimagnon peak seems to be shifted at higher energy with respect to the XAS maximum, an unusual behavior considering that Ref.\cite{Peng_PRB22, Peng_PRL, Feng_PRL} report opposite displacement for phonons.  \par
The detuning dependence of the magnon integrated intensity can be partially explained by Ref.\,\cite{Bisogni_PRL2014}. It suggests that the single spin-flip excitation is mostly unaffected by the lifetime of the intermediate state, as it is primary formed during the decay process. The reason why this feature has not been observed in YBCO-AF and NBCO \cite{Rossi_PRL} could be a less sharp single magnon peak that overlaps with bimagnon. This hallmark becomes particularly relevant when analyzing detuning data in highly doped cuprates, as, for example, in Ref.\,\cite{Peng_PRL}. In such cases, the broad paramagnon significantly overlaps with the breathing phonon, making it difficult to separate the two in a multi-peak fitting analysis. By accident, a part of the magnetic spectral weight may be captured within the phonon Gaussian leading to slower decay of the phonon area with detuning and, consequently, to the higher EPC. For these reasons, we selected an experimental geometry where magnons are at high energies and the phonons cross-section is maximized.\par
\par

With these measurements, we can estimate the EPC in CCO, LCO-AF and YBCO-AF using Eq.\,\ref{eq:RIXS_Int_1} (see Fig.\,\ref{fig:Detuning2}\,(d)-(f)). As discussed above, the value of $\tilde{g}$ is sensitive to the choice of $\Gamma$. Here, we use the fitting procedure carried out with $\Gamma=0.41$\,eV, which was determined experimentally. Considering that in literature smaller $\Gamma$ are often used, we also re-evaluate the strength of the EPC using $\Gamma = 0.28$\,eV and $\Gamma = 0.34$\,eV (see Table\,\ref{tab:Gamma}). We note that reducing $\Gamma$ leads to an increase in the estimated value of $M$, although the overall variation is moderate. \par
\begin{table}[h]
\begin{center}
\tabcolsep 1.5 mm
\caption{Extracted EPC strength from detuning analysis, assuming different values of the HWHM $\Gamma$.}

\label{tab:Gamma}
\begin{tabular} {c c c c c}
\hline \hline
  &  $\Gamma$\,(eV)   & $\tilde{g}$ & $M$\,(eV)    \\
\hline
\vspace{-8pt} \\
 CCO   & 0.28 & 7.48 $(\pm1.97)$ & 0.20 $(\pm0.03)$ \\
       & 0.34 & 6.34 $(\pm2.03)$ & 0.19 $(\pm0.03)$ \\
       & 0.41 & 4.9 $(\pm2.19)$ & 0.17 $(\pm0.04)$ \\
LCO-AF & 0.28 & 5.6 $(\pm1.48)$ & 0.19 $(\pm0.02)$ \\
       & 0.34 & 4.7 $(\pm1.44)$ & 0.17 $(\pm0.03)$ \\
       & 0.41 & 3.51 $(\pm1.46)$ & 0.15 $(\pm0.03)$ \\
YBCO-AF& 0.28 & 8.93 $(\pm1.06)$ & 0.19 $(\pm0.01)$ \\
      & 0.34 & 7.35 $(\pm1.06)$ & 0.18 $(\pm0.01)$ \\
      & 0.41 & 5.38 $(\pm1.30)$ & 0.15 $(\pm0.02)$ \\
\vspace{-8pt} \\
\hline \hline
\end{tabular}
\vspace{-0.6cm}
\end{center}
\end{table}

\subsection{\label{subsec:MomentumDependence}Momentum-dependence of the EPC}
\begin{figure}
    %\centering
    \textbf{EPC in La$_2$CuO$_4$ and YBa$_2$Cu$_3$O$_6$. }\\[0.5em]
    \includegraphics[width=\columnwidth]{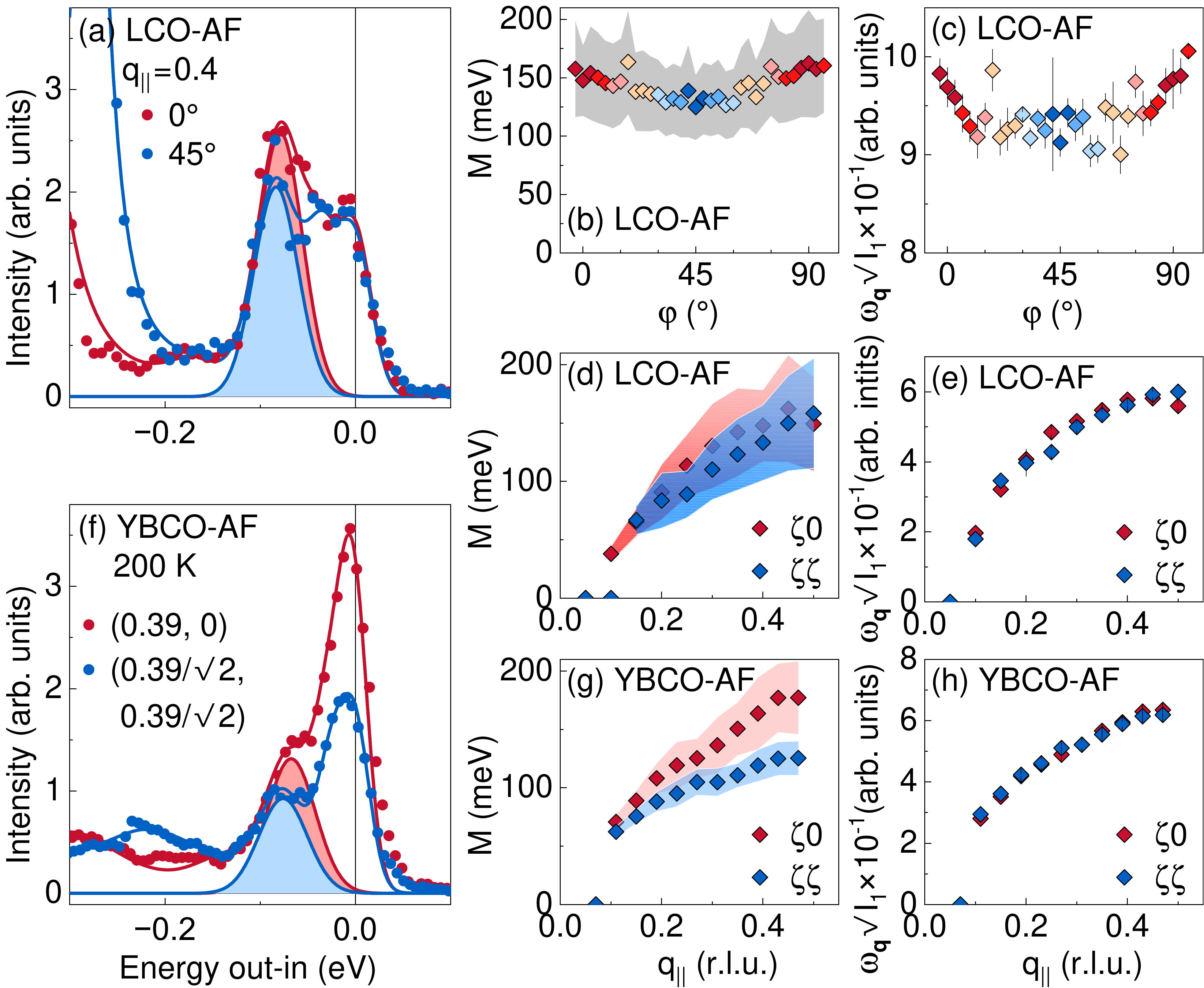}
    \caption{(a) Fit (solid lines) of RIXS data (markers) of LCO-AF corresponding to $\varphi=0^{\circ}$ (red) and $\varphi=45^{\circ}$ (blue). The phonon Gaussians are highlighted with filled area, while other fitting features are omitted for clarity.(b) Evolution of the breathing phonon EPC strngth $M$ along the azimuth path. (c) $\varphi$ evolution of the product $\omega_{\mathbf{q}}\sqrt{I_1}$. (d), (f) Same as panels (b) and (c), but for ($\zeta$,0) and ($\zeta$,$\zeta$) directions. All for LCO-AF are taken at 20\,K. $\varphi$ scan was performed at constant $2\theta=149.5^{\circ}$ and $q_{\parallel}=-0.4$\,r.l.u. ($\zeta$,0) and ($\zeta$,$\zeta$) scans were performed at constant $L = 1.3$\,r.l.u. ($\mathbf{q}$-component along $c$) and negative $\zeta$. (f), (g), (h) Same as panels (a), (d) and (e), respectively, but for YBCO-AF.  ($\zeta$,0) and ($\zeta$,$\zeta$) scans were performed  at constant $2\theta=149.5^{\circ}$ and positive $\zeta$. The data for YBCO-AF are taken at 200\,K.}
    \label{fig:LCO_YBCO}
\end{figure}
\begin{figure}
    %\centering
  \textbf{Qualitative evolution of the EPC in La$_{2-x}$Sr$_x$CuO$_{4+\delta}$. }\\[0.5em]   
\includegraphics[width=\columnwidth]{ 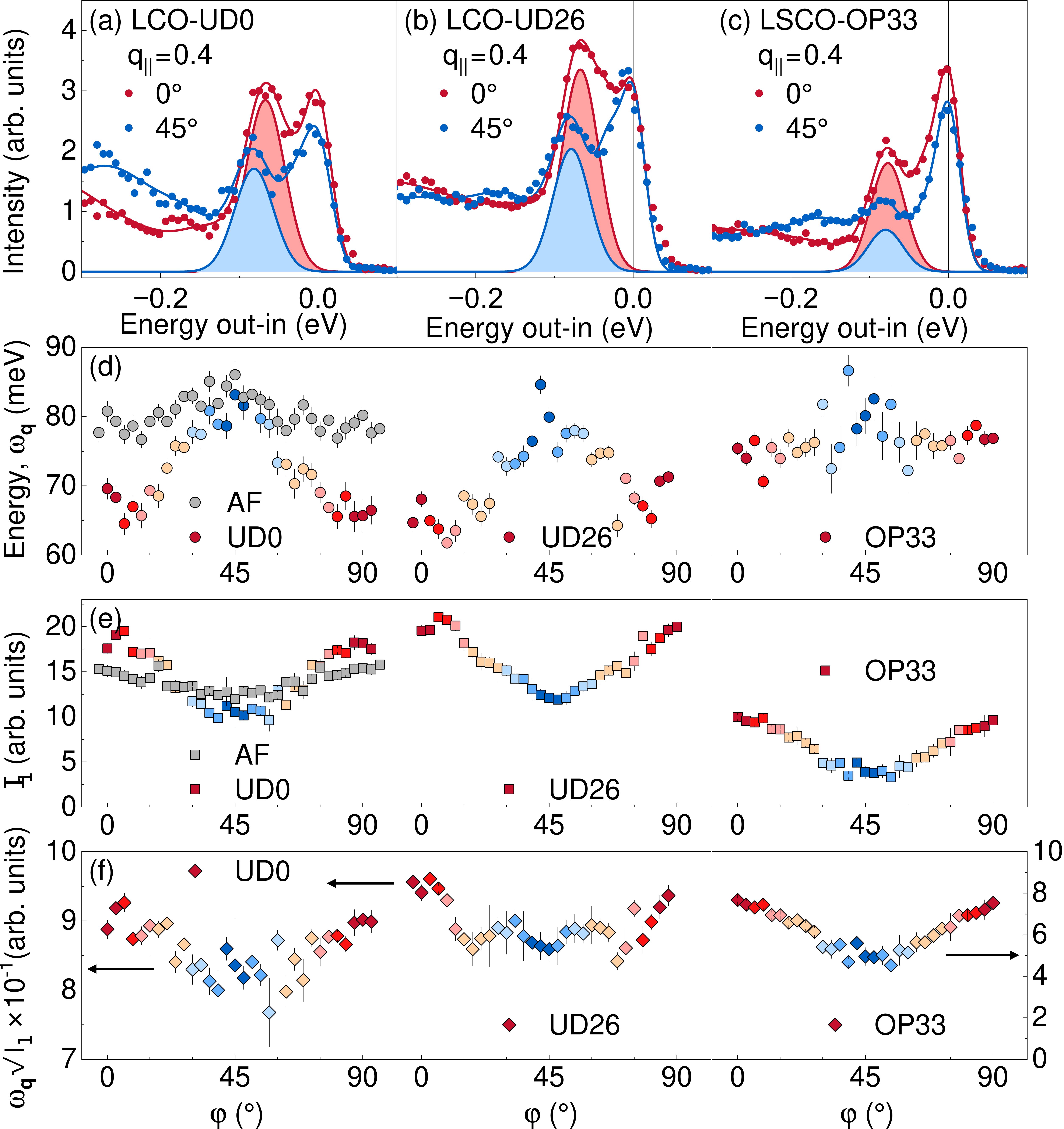}
    \caption{(a)-(c) Collection of the experimental data for LCO-UD0, LCO-UD26 and LSCO-OP33.  (d) The extracted one-phonon energies $\omega_{\textbf{q}}$ for three doped samples (red-orange-blue circles) and for LCO-AF (gray circles). (e) Extracted integrated intensities $I_1$ for three doped samples (red-orange-blue squares) and for LCO-AF (gray squares). (f) The product $\omega_{\textbf{q}}\sqrt{I_1}$ for the three doped samples (red-orange-blue diamonds). All data are taken at 20\,K and 2$\theta=149.5^{\circ}$.}
    \label{fig:LCO_doped}
\end{figure}

Figure\,\ref{fig:RIXS_CCO1}\,(a)-(c) displays the RIXS intensity maps and (d) a stack of RIXS spectra collected on CCO as a function of energy loss and momentum along high-symmetry directions (a) ($\zeta$, 0), (b) ($\zeta$,$\zeta$) and (c), (d) along the azimuthal direction with fixed $q_{\parallel} = -0.4$\,r.l.u.. To evaluate the momentum-dependence of the EPC we extract from the fit of the data the breathing phonon energies and integrated intensities, see Fig.\,\ref{fig:RIXS_CCO1}\,(e)-(h). Since we know the EPC $\tilde{g}$ for \textbf{q$_{\parallel}$}\,=\,(-0.4,0)\,r.l.u. we calculate the proportionality factor between theoretical and experimental integrated intensity and then use the Ament model to evaluate $\tilde{g}(\mathbf{q_{\parallel}})$ and $M(\mathbf{q_{\parallel}})$, $M(\mathbf{q_{\parallel}})=\omega_{\mathbf{q}}\sqrt{\tilde{g}(\mathbf{q_{\parallel}})}$, for all \textbf{q$_{\parallel}$}, see Fig.\ref{fig:RIXS_CCO2}\,(a). Figure\,\ref{fig:RIXS_CCO2}\,(b), (c) shows the evolution of $M$ as a function of $\varphi$ and $q_{\parallel}$ where the colored surrounding indicates the error bars. Despite the large uncertainty that propagates from the error in the detuning result, we see a minimum in the $\varphi$ dependence of $M$ at $\varphi =45^{\circ}$. In the same way, in $q_{\parallel}$ scans the extracted $M$ along the ($\zeta$, 0) direction is larger comparing to the ($\zeta$,$\zeta$) direction. We highlight that all data are plotted for a constant $q_{\parallel}$, such that along the ($\zeta$,0) direction the coordinates are ($q_x=\zeta$, $q_y=0$), while along the ($\zeta$,$\zeta$) direction we have ($q_x=\zeta/\sqrt{2}$, $q_y=\zeta/\sqrt{2}$).\par
In addition to the quantitative analysys discussed above, we look at the data in a simplified way assuming $M(\mathbf{q_{\parallel}})$ to be proportional to $\omega_{\textbf{q}}\sqrt{I_1(\mathbf{q_{\parallel}})}$. As shown in Fig.\,\ref{fig:RIXS_CCO2}\,(d), (e) the simplification leads to a smaller difference between ($\zeta$,0) and  ($\zeta$,$\zeta$) directions and the $\varphi$ scans become flatter, though the minimum at $\varphi=45^{\circ}$ can still be observed. As mentioned above, the deviation between the two approaches increases for higher value of the experimental EPC $\tilde{g}$.\par 
The $\textbf{q}$-dependence of the EPC in LCO-AF and YBCO-AF is shown in Fig.\,\ref{fig:LCO_YBCO}. For LCO-AF (see panels (a)-(e)) the EPC is quasi-isotropic in $\varphi$ for each $q_{\parallel}$, while for YBCO-AF (see panels (f)-(h)) with the Ament model we clearly observe a stronger EPC along the ($\zeta$, 0) with respect to ($\zeta$,$\zeta$). Due to the large value of $\tilde{g}$ obtained with the detuning approach, the results from the simplified model deviate from those of the full Ament model.\par
To investigate the effect of doping on the $\textbf{q}$-dependence of the EPC, we performed additional measurements on three doped L(S)CO samples. Since we do not have detuning data for these samples, our discussion is limited to qualitative observations. Figure\,\ref{fig:LCO_doped} shows the collection of the experimental data (a)-(c) and the extracted one-phonon energies $\omega_{\textbf{q}}$ (d), integrated intensities $I_1$ (e) and their product $\omega_{\textbf{q}}\sqrt{I_1}$ (f) for doped and antiferromagnetic L(S)CO. In LCO-UD0 and LCO-UD26 we detect a large softening of $\omega_{\textbf{q}}$ and an enhancement of $I_1$ compared to LCO-AF. However, in two cases, the product $\omega_{\textbf{q}}\sqrt{I_1}$ remains almost constant in $\varphi$, suggesting that in this approximation oxygen-doping does not change the $\textbf{q}$-dependence of the EPC.\par
In LSCO-OP33 the overall RIXS intensity, including $dd$ region, normalized to the incident photon flux  is significantly weaker. In principle, RIXS spectral weight might be strongly sample dependent, for instance significantly changing with the thickness of the film and the level of doping. Here, even though $\omega_{\textbf{q}}$ remains nearly unchanged from LCO-AF, we find that the anisotropy in $I_1$ is more pronounced when comparing $\varphi = 0^{\circ}$ and $45^{\circ}$. As a result, $\omega_{\textbf{q}}\sqrt{I_1}$ shows a clear minimum at $\varphi = 45^{\circ}$.\par

\subsection{\label{subsec:Theory}Theoretical predictions}
\begin{figure}
    %\centering
     \textbf{EPC with TB model. }\\[0.5em]   
    \includegraphics[width=\columnwidth]{ 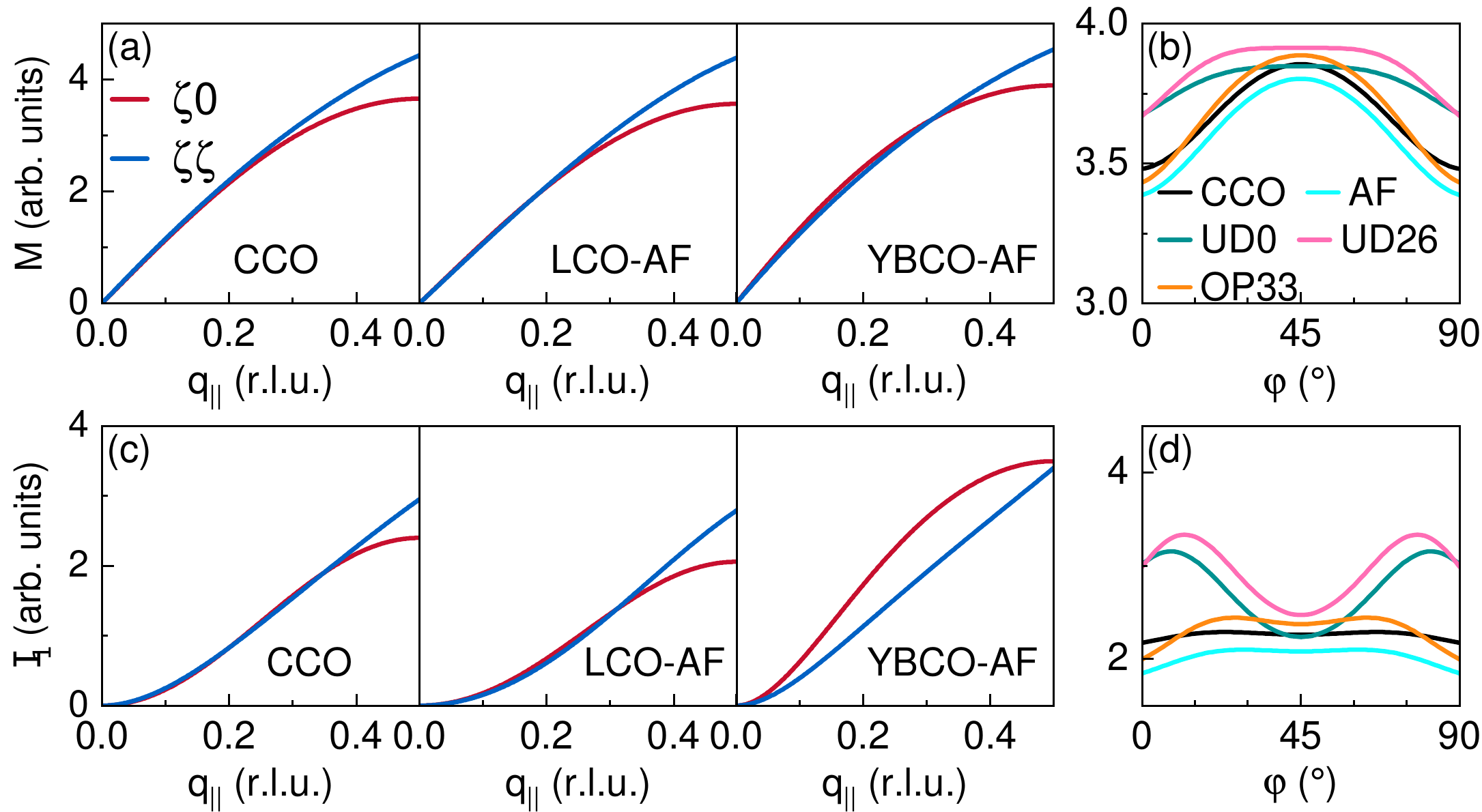}
    \caption{(a), (b) Theoretical evolution of the EPC strength $M$ along the ($\zeta$, 0), ($\zeta$,$\zeta$), and azimuthal directions for all investigated cuprates from the TB model.  The material dependence is taken into account via their phonon dispersions. (c), (d) The corresponding one-phonon integrated intensity for all investigated cuprates. Due to non negligible dispersion its evolution with $q_{\parallel}$ and $\varphi$ can be opposite to the evolution of $M$.}
    \label{fig:EPC_theory}
\end{figure}
Fig.\,\ref{fig:EPC_theory} shows the evolution of the EPC strength $M$ according to the TB model (see Eq.\,\ref{eq:M_br_fermi}) and the integrated one-phonon intensity $I_1$ ($I_1$ is almost proportional to $\tilde{g}=M^2/\omega_{\textbf{q}}^2\propto \mathrm{sin}^2 (q\pi)/\omega_{\textbf{q}}^3$) along the ($\zeta$,0), ($\zeta$,$\zeta$), and azimuthal directions, using the phonon dispersion determined from the fitting of our spectra. It is important to note that since the breathing mode is characterized by a relatively large dispersion, the $\textbf{q}$-dependence of $I_1$ and $M$ \emph{do not coincide} due to the $\omega_{\textbf{q}}^3$ term in the denominator.\par
One prominent feature of our experimental data is that the estimated EPC $M$ of the breathing phonon at $\varphi=45^{\circ}$ is \emph{smaller} than at $\varphi=0^{\circ}$ which is \emph{opposite} to the theoretical prediction based on the TB model (see Fig.\,\ref{fig:EPC_theory}\,(b)). In particular,the experimental difference $M(45^{\circ})<M(0^{\circ})$ at $q_{\parallel}\approx 0.4$\,r.l.u. is $\sim 20$\,\% in CCO, $\sim 15$\,\% in LCO-AF and $\sim 27$\,\% in YBCO-AF with respect to theoretical curves where $M(45^{\circ})>M(0^{\circ})$ by $\sim 12$\,\%. The same effect is observed for large $q_{\parallel}$ along the ($\zeta$,0) and ($\zeta$,$\zeta$) directions (see Fig.\,\ref{fig:EPC_theory}\,(a)). However, if we consider separately the evolution of the phonon peaks along the ($\zeta$,0) and ($\zeta$,$\zeta$) the shape is fitted well by the theory. This explains why this discrepancy was not detected in previous works, which only reported measurements along a single high-symmetry direction \cite{Braicovich_PRR,Rossi_PRL}.\par
The dependence on $\textbf{q}_{\parallel}$ is qualitatively the same in all cuprates, strongly suggesting that there is a general discrepancy between the one-phonon experimental RIXS EPC and the theoretical one, calculated by combining Eq.\,\ref{eq:RIXS_Int_1} from Ref.\,\cite{Ament_EPL} and  Eq.\,\ref{eq:M_br_fermi} from Ref.\,\cite{Johnston_PRB,Rosch_PRB,Ishihara_PRB,Devereaux_PRL_2004}. We note also that hints of this discrepancy can be found in supplementary figures of some recent articles on charge order in cuprates \cite{arpaia2023signature,Scott_SciAdv2023}.\par

\section{\label{sec:Methods}Theoretical developments beyond tight-binding}

\subsection{Density functional theory}
\label{subsec:DFT}

\begin{figure*}
    %\centering
         \textbf{EPC with DFT. }\\[0.5em]   
\includegraphics[width=0.75\textwidth]{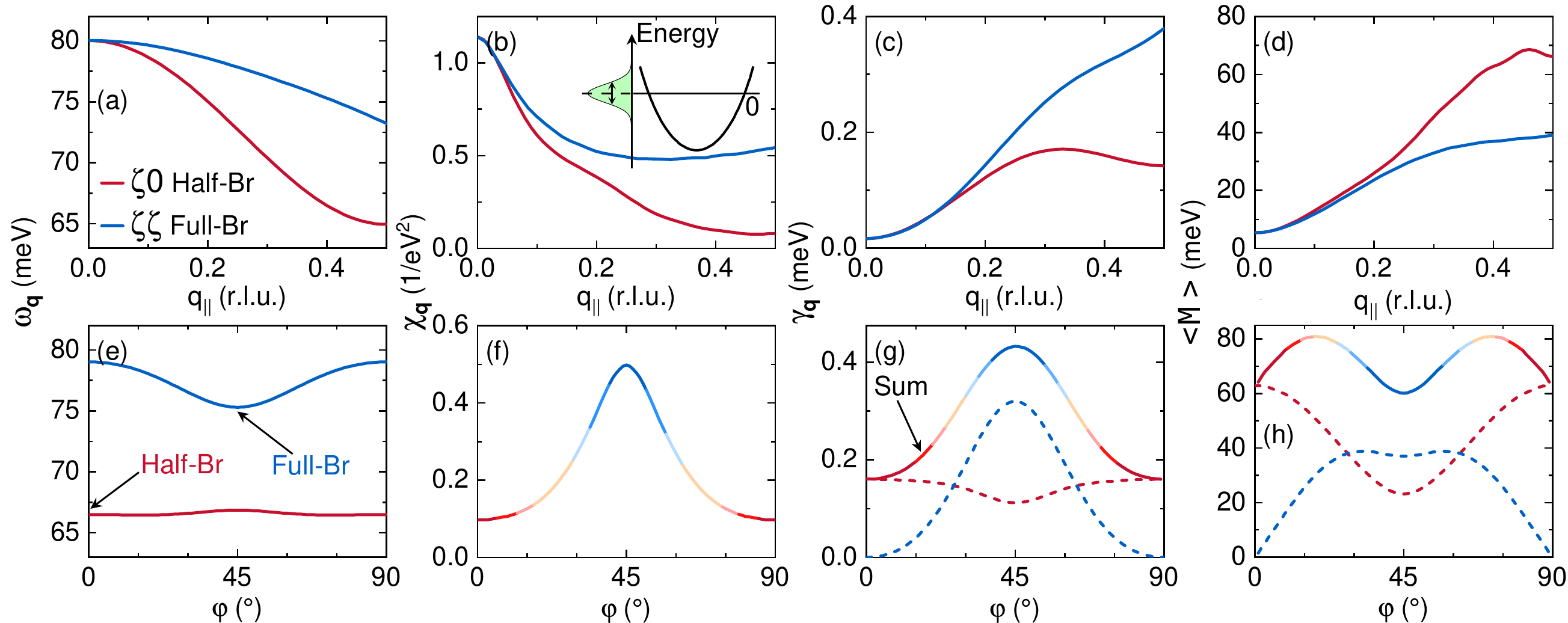}
    \caption{(a)-(h) The EPC quantities calculated with DFT along (a)-(d) the ($\zeta$,0), ($\zeta$,$\zeta$)  and (e)-(h) $\varphi$ ($q_{\parallel} = 0.4)$ lines. The \textbf{q}-dependence of the energy of the bond-stretching mode is in a agreement with previous INS measurements \cite{REICHARDT_PhC,PINTSCHOVIUS_PhC,Pintschovius_PRB,Braden_PRB}. The inset in (b) shows Gaussian region around the Fermi level considered in our DFT calculations. For the ($\zeta$,0) and ($\zeta$,$\zeta$) scans, the red and blue lines correspond to the half-breathing and full-breathing modes, respectively, while BS$_{01}$ and BS$_{11}$ modes are omitted. The dashed lines in (g) and (h) indicate $\gamma_{\boldsymbol{\mathrm{q}}}$ and $\left<M\right>$ for two close in energy phonon branches at intermediate $\varphi$ between 0 and 45$^\circ$. The solid lines are the sum of two branches, separately unresolved in our experiments. See main text for details.}
    \label{fig:RIXS_DFT}
\end{figure*}
To move beyond TB, the RIXS phonon intensity can be calculated using Ament's formula with the EPC obtained from DFT \cite{Heid2008,Bohnen2003,Giustino2008,Sterling_PRB}. Compared to TB, DFT provides a more complete description of the electronic structure, although the effects of correlation are largely missing in both models.%  An important \textit{caveat} is that, in standard DFT, electron correlation is not fully accounted for, so that, for example, undoped cuprates are found in calculations to be metallic rather than insulating as in real materials.\par

DFT calculations are routinely performed to estimate the phonon linewidth in the INS and IXS experiments, and it is worth emphasizing the difference between these calculations and those required for the RIXS amplitude.  The linewidth $\gamma_{\textbf{q}}$ is related to the EPC constant as follows

\begin{align}
\begin{split}
    \gamma_{\boldsymbol{\mathrm{q}}j} = 2\pi\omega_{\boldsymbol{\mathrm{q}}j}\frac{1}{N_k}\sum_{\boldsymbol{\mathrm{k}}}\sum_{\mu\nu}\left| g_{\boldsymbol{\mathrm{k}}\mu,\boldsymbol{\mathrm{k}}+\boldsymbol{\mathrm{q}}\nu}^{\boldsymbol{\mathrm{q}}j}\right|^2\times\\ \times  \delta(\epsilon_{\boldsymbol{\mathrm{k}}\mu}-\epsilon_{F})\delta(\epsilon_{\boldsymbol{\mathrm{k}}+\boldsymbol{\mathrm{q}}\nu}-\epsilon_{F}),
     \end{split}\label{eq:gamma_INS}
    \end{align}
where the indices $\mu$ and $\nu$ label  electronic bands occupied by the electron before and after the scattering with a phonon, respectively. Eq.\,\ref{eq:gamma_INS}~ contains a sum over scattering processes between electronic states \emph{very close to the Fermi energy} $\epsilon_{F}$, weighted by the squared EPC matrix elements $g (\boldsymbol{\mathrm{k}},\boldsymbol{\mathrm{q}})$ for phonon branch $j$. In practice, in the calculations the delta functions are replaced by Gaussians confined to a rather small energy range around the Fermi level (here with a width of 0.2\,eV, see inset in Fig.\,\ref{fig:RIXS_DFT}\,(d)). In the following we neglect the index $j$ and refer exclusively to the bond-stretching mode.\par 
For our purposes, we can define a joint density of states (JDOS) $\chi_{\boldsymbol{\mathrm{q}}}$:
\begin{align}
\begin{split}
    \chi_{\boldsymbol{\mathrm{q}}} = \frac{1}{N_k}\sum_{\boldsymbol{\mathrm{k}}}\sum_{\mu\nu}\delta(\epsilon_{\boldsymbol{\mathrm{k}}\mu}-\epsilon_{F})\delta(\epsilon_{\boldsymbol{\mathrm{k}}+\boldsymbol{\mathrm{q}}\nu}-\epsilon_{F})
     \end{split}\label{eq:chi_INS}
    \end{align}
Eq.\,\ref{eq:chi_INS} looks similar to the formula for $\gamma_{\textbf{q}}$, but without the EPC matrix elements. $\chi_{\boldsymbol{\mathrm{q}}}$ is a purely electronic property that depends only on $\boldsymbol{\mathrm{q}}$, not on the specific phonon branch.
$\chi_{\textbf{q}}$ can be used to define an average of the EPC matrix elements:
\begin{align}
\begin{split}
    \left<M\right>=\sqrt{\left< g_{\boldsymbol{\mathrm{q}}}^2 \right >} = \sqrt{\frac{\gamma_{\boldsymbol{\mathrm{q}}}}{2\pi\chi_{\boldsymbol{\mathrm{q}}}\omega_{\boldsymbol{\mathrm{q}}}}}.
     \end{split}\label{eq:g2_DFT}
    \end{align}

We performed DFT calculations for \ch{La_2CuO_4} in the high-temperature tetragonal (HTT) phase. DFT calculations of phonon properties and EPC matrix elements were based on density functional perturbation theory \cite{Baroni2001} as implemented for a mixed-basis pseudopotential method \cite{Meyer,Heid1999}, following previous work on phonons of cuprates \cite{Bohnen2003,LeTacon_Nature,Miao_PRB}. The exchange-correlation functional is treated in the local-density approximation \cite{Perdew1992}. We focused on the ($\zeta$,0), ($\zeta$,$\zeta$) and $\varphi$ lines for $q_{\parallel} = 0.4$\,r.l.u. and $L=0.76$ r.l.u. to follow closely the experimental paths, though the $L$-value is not very important in our calculations. To take into account the energy spread of the RIXS process, which is not limited to states very close to the Fermi energy, we set the width of the Gaussian to 0.2\,eV.  We found that only minor details in magnitude and momentum dependence are sensitive to the details of the mesh and Gaussian energy width (e.g. setting it to 0.1\, eV), indicating that the overall trends are robust.  For each $\boldsymbol{\mathrm{q}}$ point, we calculated 
 $\omega_{\boldsymbol{\mathrm{q}}}$ (directly from DFT), $\gamma_{\boldsymbol{\mathrm{q}}}$ (from Eq.\,\ref{eq:gamma_INS}), $\chi_{\boldsymbol{\mathrm{q}}}$ (from Eq.\,\ref{eq:chi_INS}), and the EPC constant $\left<M\right>$ (from Eq.\,\ref{eq:g2_DFT}). \par
 
 The results of our DFT calculations, which we denote as $M$-DFT, are displayed in Fig.\,\ref{fig:RIXS_DFT}\,(a)-(h). For the ($\zeta$,0) and ($\zeta$,$\zeta$) scans, the dark red and blue lines correspond to the pure half-breathing and full-breathing modes, respectively, while the BS$_{10}$ and BS$_{11}$ modes  are not shown. Since the BS$_{10}$ and BS$_{11}$ modes are assumed to be silent in RIXS (see Methods\,\ref{sec:Methods}), we focus our analysis only on the breathing modes. However, in the $\varphi$ scan, these four modes overlap in our calculations, resulting in a small overestimate of $M$ at $\varphi=45^{\circ}$.\par 
 It is worth emphasising that i) $\gamma_{\boldsymbol{\mathrm{q}}}$ is larger along ($\zeta$,$\zeta$) as compared to ($\zeta$,0), so that a $\varphi$ scan would have a maximum at 45$^\circ$; 
 ii) the JDOS $\chi_{\boldsymbol{\mathrm{q}}}$ also has  a strong $\boldsymbol{\mathrm{q}}$-dependence, showing a marked drop along ($\zeta$,0) towards $\zeta=0.5$\,r.l.u., while being larger and flatter along ($\zeta$,$\zeta$); 
 iii) $\left<M\right>$ shows a reversed behavior as compared to $\gamma_{\boldsymbol{\mathrm{q}}}$, being larger at ($\zeta$,0) than at ($\zeta$,$\zeta$), while the sum over both branches has a minimum for $\varphi=45^\circ$. This looks qualitatively different from the TB model (Eq.\,\ref{eq:M_br_fermi}) and similar to the experimental data.\par
 Notably, the $M$-DFT results also show good \emph{quantitative} agreement with the tight-binding calculations, even though it slightly underestimates the experimental value of the EPC. At $\mathbf{q_{\parallel}}=(0.4,0)$ r.l.u. both approaches find $M$ of the order of 80\,meV while from detuning we extract $\sim 150$\,meV. The possible reasons for this result are explained in the earlier works \cite{Reznik2008,Johnston_PRB} and are related to the fact that DFT always treats cuprates as metals. In particular, the high mobility of the electron induces a strong Coulomb screening that may consequently \emph{reduce} the coupling of electrons to phonons. \par
We note that our DFT calculations were performed on the ideal HTT structure with formally no doping. Since doping  affects the shape of the Fermi surface, the $\boldsymbol{\mathrm{q}}$-dependence of the EPC quantities can in principle also be affected, and can also be slightly different for different cuprate families. However, based on our experimental data, there are no major differences in the general shape of the EPC vs $\varphi$ for all samples.\par

\subsection{\label{subsec:RGZ model} Resonant form factor modulation}

Having analyzed the merits and limits of Ament's model as combined with TB or DFT, and having found that it works quite well despite the evident contradiction of adopting a local description of the scattering process and of the vibrational excitation (Einstein phonon) to study a non-local property such as the \textbf{q}-dependence of the EPC, we can try to further simplify the approach by getting rid of the EPC, while still capturing the collective properties of phonons. The idea is to bluntly ignore the electronic structure and focus totally on the description of the RIXS process at local level, but to take into account the \textbf{q}-dependent symmetry properties of the phonon mode. Therefore, we propose a method to calculate the RIXS one-phonon intensity $I_1(\boldsymbol{\mathrm{q}})$ that does not include the EPC explicitly and does not use Ament's formula (Eq.\,\ref{eq:RIXS_Int_1}). This model, which we name resonant form factor modulation (RFFM) for reasons that will soon become clear, is directly inspired by the INS one-phonon structure factor, which, apart from a pre-factor $k_{\mathrm{p}}'/k_{\mathrm{p}}$, is identical to the structure factor of a \emph{static} modulation with the same wavevector and displacement pattern of the phonon.  When calculating the resonant \emph{elastic} x-ray scattering (REXS) form factor for the same static modulation, the neutron diffraction (ND) expression requires a modification, because, unlike the Fermi length, the resonant atomic scattering factor (or resonant form factor, for short) depends on the local environment. Note that the resonant form factor depends upon the incident and scattered x-ray polarization, thus encoding information about the different polarization channels (not discussed here). Importantly, we assume the modes to be \emph{dispersionless} with the constant frequency $\omega_{\mathbf{q}}$.\par

In general, the linearized expression of the REXS structure factor for a generic scattering vector $\boldsymbol{\mathrm{Q}}$ (capital letter to indicate the total wave vector, not reduced to the first Brillouin zone as indicated elsewhere by $\mathbf{q}$) can be written as the sum of two terms (see Methods\,\ref{sec:Methods} for further details):
\begin{align}
\begin{split}
F_{\mathrm{tot}}^\pm(\boldsymbol{\mathrm{Q}}) &= F_{I}(\boldsymbol{\mathrm{Q}})+F_{II}^\pm(\boldsymbol{\mathrm{Q}})
     \end{split}\label{eq:I_PR}
    \end{align}
where 
\begin{align}
\begin{split}
F_{I}(\boldsymbol{\mathrm{Q}}) &=\sum_i \left(\frac{\partial f_i}{\partial \Qc}\right)e^{i\boldsymbol{\mathrm{Q}}\cdot\boldsymbol{\mathrm{r}}_i} \Qc^0\\
\end{split}\label{eq:I_PR2a}
\end{align}
\begin{align}
\begin{split}
F_{II}^\pm(\boldsymbol{\mathrm{Q}}) &= \sum_j (\boldsymbol{\mathbf{\varepsilon}}_j\cdot\boldsymbol{\mathrm{Q}}) f_j^0 
 e^{\mp i \varphi_j} e^{i\boldsymbol{\mathrm{Q}}\cdot\boldsymbol{\mathrm{r}}_j} \Qc ^0
\end{split}\label{eq:I_PR2b}
\end{align}
and
 \begin{align}
\begin{split}
 I_1(\boldsymbol{\mathrm{q}}) &\propto  \frac{k'_{\mathrm{p}}}{k_{\mathrm{p}}}\left|F_{\mathrm{tot}} (\boldsymbol{\mathrm{Q}})\right|^2 \delta(E_f-E_i-\hbar\omega_{\boldsymbol{\mathrm{q}}})
  \end{split}\label{eq:I_PR_int}
    \end{align} 

Here, $f^0$  is the total form factor (non-resonant plus resonant) in the ground-state (undistorted) structure, $\partial f/\partial \Qc$ is the derivative of the resonant form factor with respect to the phonon coordinate $\Qc$, and $\Qc^0\propto \sqrt{1/\omega\,m_{\mathrm{eff}}}$ is the one-phonon amplitude, where $m_{\mathrm{eff}}$ is the mode effective mass. The indices $i$ and $j$ run over the resonant sites and over all sites, respectively, $\boldsymbol{\mathrm{r}}_j$ are the relative position vectors of the sites in the unit cell, and $\varphi_j$ are the phases of the phonon for the different sites.  The two signs indicated by $\pm$ are for the two directions of the propagation vector.  As for the case of INS, the phonon is modeled as a rigid shift of the atoms from their equilibrium position with a pattern defined by the mode branch and the wavevector $\boldsymbol{\mathrm{Q}}$ (frozen phonon). The $F_{II}^\pm(\boldsymbol{\mathrm{Q}})$ defined in Eq.\,\ref{eq:I_PR2b} contains both resonant and non-resonant contributions but is formally identical to the neutron scattering term (with the Fermi length replacing the form factor) and is entirely due to the ionic displacements. By contrast, $F_{I}(\boldsymbol{\mathrm{Q}})$ can be neglected for INS and IXS, because it emerges only at resonance as due to the modulation of the resonant form factor arising from the structural distortion around each resonantly scattering site. Therefore, for IXS and INS the one-phonon intensity is given by Eqs.\,\ref{eq:I_PR} and \ref{eq:I_PR_int} 
with $F_{I}(\boldsymbol{\mathrm{q}})=0$; the term $F_{II}(\boldsymbol{\mathrm{q}})$ is known as \emph{dynamical structure factor}. By contrast, for RIXS both $F_{I} (\boldsymbol{\mathrm{q}})$ and $F_{II} (\boldsymbol{\mathrm{q}})$ in Eq.\,\ref{eq:I_PR} need to be retained, and $F_{I} (\boldsymbol{\mathrm{q}})$ is totally dominant in certain conditions. 

%\begin{align}
%\begin{split}
% I(\boldsymbol{\mathrm{q}}) &\propto  \frac{k'}{k}\left|F_{II} (\boldsymbol{\mathrm{q}})\right|^2 \delta(E_f-E_i-\hbar\omega_{\boldsymbol{\mathrm{q}}})
%  \end{split}\label{eq:I_PR_int}
%    \end{align} \par

For RIXS measurements on cuprates at the Cu L$_3$ edge, the variation of the RIXS resonant form factor $f_{\mathrm{Cu}}$ is almost entirely due to the breathing displacement of the nearby oxygen atoms. Moreover, since the high-energy breathing modes involve mainly the motion of the oxygens, the displacement of copper is $\boldsymbol{\varepsilon_{\mathrm{q}}} \simeq 0$, the contribution of the second term in Eq.\,\ref{eq:I_PR} can be neglected. Therefore, in our RFFM model, considering that there is only one resonant Cu atom per unit cell, the RIXS intensity is taken to be given by Eqs.\,\ref{eq:I_PR} and \ref{eq:I_PR_int} with $F_{I}(\boldsymbol{\mathrm{q}})=\left( \partial f_{\mathrm{Cu}}/{\partial \Qc}\right) \Qc^0$.

\par

The RFFM model has the merit of being extremely simple, especially with the further simplification we make here below.  Before proceeding further, it is worth considering to what extent this model can also be realistic.  Since the dynamical structure factor is obtained from a frozen phonon model, it is in principle exact for zero energy transfer, and should be a valid approximation provided that the inverse of the phonon frequency is larger than the core hole lifetime, since in this scenario the scattering process could be thought as `seeing' an effectively frozen phonon (this is exactly the approximation that is made for INS and IXS).  %However, for a given phonon displacement pattern, it should be $\partial f_{\mathrm{Cu}}/\partial \Qc \propto 1/\gamma_{\mathrm{ch}}$, where $\gamma_{\mathrm{ch}}$ is the resonant linewidth, largely determined by the inverse of the core hole lifetime.  Therefore, the core hole lifetime cannot be too short, lest the phonon be essentially invisible.  In conclusion, we cannot state with full confidence that RFFM model is realistic in the regime we consider, and core-hole effect may need to be considered from the outset. 
Even though the microscopic basis of the RFFM model is yet to be determined, as we shall see, it is in remarkable agreement with TB, and very close to DFT (both models being based on the Ament's formula), indicating that the $\boldsymbol{\mathrm{q}}$-dependence of the phonon intensities is captured by the same underlying features in all models.

\emph{A priori}, $\partial f_{\mathrm{Cu}}/\partial \Qc$ is strongly energy-dependent, and calculating its $\boldsymbol{\mathrm{q}}$-dependence for a given energy is far from straightforward. We therefore make a further, bold simplification by re-expressing the oxygen displacements in terms of the phonon eigenvectors and assuming that $\partial f_{Cu}/\partial \Qc$  is proportional to the mean squared breathing displacement of oxygen around Cu. 

In an extremely simplified version (s-RFFM), we construct a `toy model' of the phonon (see Methods\,\ref{sec:Methods}), which, for the half- and full-breathing modes, yields 

\begin{align}
    \begin{split}
I_{\mathrm{HB}}&\propto\sin^2 (\pi \zeta)=\frac{1}{2}\sin^2 (\pi q_\parallel)\\
I_{\mathrm{FB}}&\propto 2\sin^2 (\pi \zeta)=\sin^2 (\pi q_\parallel/\sqrt{2})\\
     \end{split}\label{eq: approx_RFFM_text}
\end{align}  
which produces an identical result to Eq.\,\ref{eq:M_br_fermi} assuming $\omega_{\mathbf{q}}$ as a constant.\par

For the s-RFFM mode, we can also verify straightforwardly that the RIXS phonon intensity is expected to be zero for the other two bond-stretching branches (shown in Fig.\,\ref{fig:nomenclature} with dashed lines).

One can make further progress by inserting the more realistic phonon displacement parameters from DFT into the RFFM model (which we denote DFT-RFFM).  In this case, we employed DFT to calculate the $\boldsymbol{\mathrm{q}}$-dependence of the eigenvectors, including the term $\Qc^0$ in Eq.\,\ref{eq:I_PR2a}, for the two highest phonon branches along ($\zeta$,0), ($\zeta$,$\zeta$), as well as the dependence of the RFFM intensity on the azimuth $\varphi$ at fixed $q_\parallel$. Note that, at intermediate values of $\varphi$, the contribution of the second bond-stretching branch to the intensity is not exactly zero. Therefore, we have calculated the intensity as the superposition of the two phonons, since they are too close in energy to be resolved.

\subsection{Comparison between different theoretical models and with experiments}

\begin{figure}
    %\centering
    \textbf{Comparison between the TB, s-RFFM, and DFT-RFFM models.}\\[0.5em] 
    \includegraphics[width=\columnwidth]{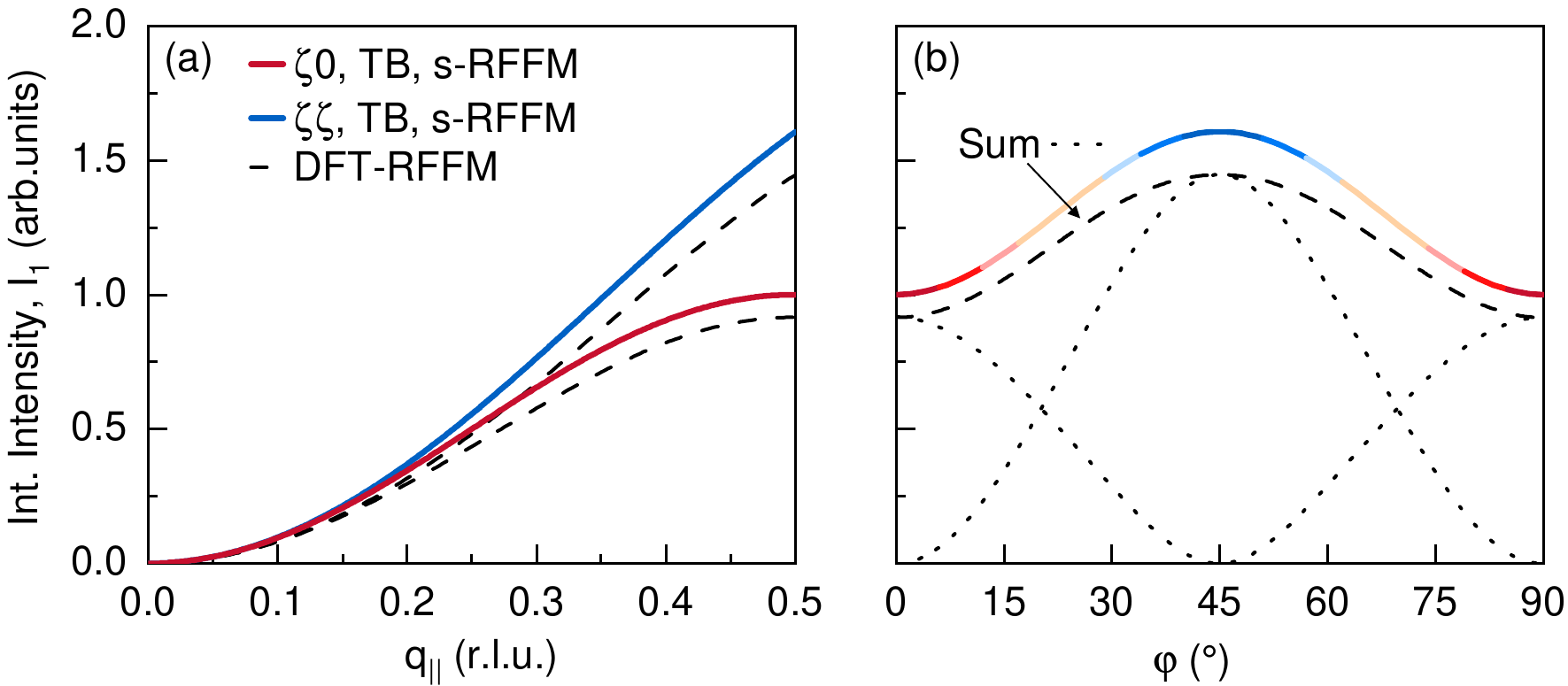}
    \caption{(a) The evolution of the RIXS one-phonon integrated intensity along the high-symmetry directions in TB, s-RFFM (red and blue solid lines) and DFT-RFFM (black dashed lines) models. TB and s-RFFM results perfectly overlap for an appropriate choice of the RFFM parameters. (b) The same as (a) for $\varphi$ scans at $q_{\parallel}=0.5$\,r.l.u.. The dotted curves calculated with DFT indicate two close in energy phonon branches, unresolved in our experiment. We note that assuming $\omega_{\textbf{q}}=\mathrm{const}$ the discrepancy between theory and experiments increases (compare with Fig.\,\ref{fig:EPC_theory}\,(c),\,(d)). }
    \label{fig:RIXS_CrystalField2}
\end{figure}

Figure\,\ref{fig:RIXS_CrystalField2}\,(a)-(b) shows a comparison between the RIXS phonon $\mathbf{q}$-dependent intensities calculated using the TB, s-RFFM, and DFT-RFFM models. Remarkably, the TB and s-RFFM models yield identical results (for $\omega_{\textbf{q}}=\mathrm{const}$, while DFT-RFFM is only slightly different, indicating that the displacement patterns of our `toy model' is a good approximation of the breathing phonons displacements. This suggests that the difference in intensities between the two branches arises primarily from the different projection of the phonon displacements onto the active $x^2-y^2$ orbitals, which is captured by the models in a very similar way. However, the azimuthal dependence in TB, s-RFFM, and DFT-RFFM models is \emph{opposite} to the experimental data, which are rather better described by $M$-DFT approach.\par

As pointed out above, the main difference between the TB approach and $M$-DFT calculations is how the electronic band structure of cuprates is taken into account. Indeed, one of the approximations in the derivation of Eq.\,\ref{eq:M_br_fermi} is the averaging over the Fermi surface of a more general expression of the EPC of the breathing modes \cite{Johnston_PRB}. As a consequence, the dependence on the electron wavevector \textbf{k} completely disappears for this specific case. $M$-DFT does not go through such a simplification, since $g (\boldsymbol{\mathrm{k}},\boldsymbol{\mathrm{q}})$ for each $\boldsymbol{\mathrm{k}}$ is calculated numerically within a certain bandwidth around the Fermi level, and the $\boldsymbol{\mathrm{k}}$ averaging is performed afterwards. Likewise, the RFFM model in either the simplified version (s-RFFM) does not consider the electronic structure at all, while  DFT-RFFM includes it only indirectly. However, it takes into account the phonon symmetries and displacement patterns correctly.  This explains the similarities of the RFFM with TB Eq.\,\ref{eq:M_br_fermi} and the differences with $M$-DFT.\par 

\par

\section{\label{sec:summary and conclusions} Discussion and conclusions}

This work had several objectives: i) to quantitatively determine EPC in widely studied antiferromagnetic cuprates; ii) to measure its momentum dependence throughout the entire BZ; iii) to assess the reliability of the Ament formula \cite{Ament_EPL} which allows to extract EPC from one-phonon RIXS intensities. We specifically investigated 
the breathing phonon mode in CCO, LCO-AF and YBCO-AF with the detuning method at the Cu L$_3$ edge. The EPC values in all three cases are nearly identical, ranging from 0.15 to 0.17\,eV at $\mathbf{q}_{\parallel} = (0.4, 0)$\,r.l.u.. Next, we focused on the evolution of the EPC and phonon intensity with $\boldsymbol{\mathrm{q_{\parallel}}}$ comparing them with different theoretical models, including two (based on TB and DFT) that calculate intensities using Ament formula and one that does not. Experimentally, we observed that EPC $M$ is a monotonic increasing function of $\mathbf{q}_{\parallel}$ and it is consistently $\sim$20\,\% higher at $\mathbf{q}_{\parallel} = (-0.4, 0)$ than at $\mathbf{q}_{\parallel} = (-0.4/\sqrt{2}, 0.4/\sqrt{2})$\,r.l.u. across all samples. We further investigated the $\mathbf{q}$-dependence in differently doped L(S)CO. Although a quantitative estimation of the EPC is not available for these samples, the overall trend of $M(\mathbf{q}_{\parallel})$ appears to be very similar.\par
We found that the theories we tested lead to rather consistent results but are not fully satisfactory. Indeed we found that the gross features of the relative $\boldsymbol{\mathrm{q}}$-dependent RIXS intensities can be captured equally well by all the models. Remarkably our new model of the RIXS dynamical structure factor that \emph{is not based explicitly on the EPC and is not based on Ament's formula} gives results identical to the prediction of Ament's formula combined with tight-binding calculations of the EPC \cite{LastComment}. Moreover, the fixed $q_{\parallel}$ $\varphi$-scans of our experiments are not reproduced by the TB model \cite{Song_PRB1995, Johnston_PRB,Rosch_PRB,Ishihara_PRB,Devereaux_PRL_2004,horsch2005}, while DFT calculations reproduce the $\boldsymbol{\mathrm{q}}$-dependent intensities somewhat more accurately, most likely due to a better modeling of the electronic band structure of cuprates.\par 

We note that the influence of the electronic band structure on  the phonon peak intensity in RIXS is even more complex if several resonances are present in the absorption spectrum. For example, in C K edge RIXS of graphite the multi-phonon features are different at the $\pi^*$ and $\sigma^ *$ resonances, which select different bands for the intermediate state \cite{DashwoodPRX2024}. Similarly, at the O K edge in the 1D compound \ch{Ca_{2+5x}Y_{2-5x}Cu_5O_{10}} the EPC was found different at the upper and lower Hubbard band \cite{WSLee_PRL13}. Therefore, in such cases more sophisticated calculations like Green’s function or exact diagonalization methods have to be employed to reliably extract the EPC from the experimental data. \par

Our results raise questions about the interplay between the ubiquitous charge order (CO) in cuprates and lattice excitations. In doped cuprates, the breathing phonon typically exhibits a pronounced softening in energy and a concurrent enhancement in intensity near the CO wave vector. However, under the simplifying assumption that $M \propto \omega_{\mathbf{q}} \sqrt{I_1}$, these two effects tend to compensate each other, suggesting that the EPC $M$ does not necessarily increase.\par
 
We conclude that, pending further theoretical developments, although care should be taken in interpreting one-phonon RIXS intensities as a direct measure of the EPC, the $\mathbf{q}$-dependence encodes the main symmetry properties of the phonon modes and is also sensitive to the details of the electronic structure. In particular for the in-plane bond-stretching mode in cuprates, different families are characterized by a similar $\varphi$-dependence of the EPC irrespective of the doping. It suggests that the main features that influence the EPC are of the Cu-O plane, common among all high-T$_c$ superconductors and their parent compounds. Finally, we note that at the O K edge the extraction of the EPC can be challenging because of the multiple absorption resonances in the XAS, associated to the hybridization of oxygen with different nearby ions \cite{Peng_PRB22}. In those circumstances, resonances are partly overlapping and tuning to the maximum of one absorption peak inevitably means detuning from others. Disentangling which phonon modes are excited at each resonance and how the incident energy affects the various phonon intensities can be a difficult task. Therefore, results at O K and Cu L$_3$ edges have to be compared with care to derive consistent and reliable estimations of the EPC in cuprates. \par 

\section*{Author contributions}
L.B. and G.G. conceptualized the study. R.A., N.A., M.B., D.D.C., A.K. and F.L. synthesized and characterized the samples. M.Z., G.M. R.A., N.B.B., K.K., L.M. F.R., M.R, F.Y.H., M.M.S. and G.G. performed the experiments. M.Z. analyzed the experimental data with inputs from G.M., R.A., L.M., L.B. and G.G.. R.H. performed the DFT calculations. P.G.R. and G.G. conceived the simplified theoretical model for RIXS scattering from phonons. G.G. supervised the project. M.Z., R.H., G.M., R.A., L.M., L.B., M.M.S., P.G.R. and G.G. contributed to the data interpretation. M.Z., P.G.R. and G.G. wrote the manuscript with input from all authors.

\begin{acknowledgments}
We are deeply grateful to Tom Devereaux for the enlightening and stimulating exchanges of ideas on RIXS, cuprates and phonons over the years. We thank Sofia Michaela Souliou, Alexei Bosak and and Jeroen van den Brink for useful discussions. The experimental data were collected at the beam line ID32 of the ESRF during experiments HC4000, HC4149, HC4412\,\cite{HC4412}, HC4607\,\cite{HC4607}, HC5029\,\cite{HC5029}, HC5221\,\cite{HC5221} and HC5627\,\cite{HC5627}; we thank the whole technical and scientific staff of ID32 for their excellent support. Myfab is acknowledged for support and for access to the nanofabrication laboratories at Chalmers, where the LSCO and the YBCO films have been grown. 
\end{acknowledgments}

\section*{Funding Declaration}
D.D.C., L.M., M.M.S. and G.G. acknowledge support by the project PRIN2020 ``QT-FLUO`` ID 20207ZXT4Z of the Ministry for University and Research (MUR) of Italy. R.A. and F.L. acknowledge support by the Swedish Research Council (VR) under the Project 2020-04945. R.H. acknowledges support by the state of Baden-Württemberg through bwHPC. The work by G.M. was jointly supported by Politecnico di Milano and European X-ray Free Electron Laser Facility GmbH. The funders played no role in study design, data collection, analysis and interpretation of data, or the writing of this manuscript.

\section*{Competing interests}
All authors declare no financial or non-financial competing interests. 

\section*{Data availability}
 RIXS data of some experiments are available at ESRF portal HC4412\,\cite{HC4412}, HC4607\,\cite{HC4607}, HC5029\,\cite{HC5029}, HC5221\,\cite{HC5221} and HC5627\,\cite{HC5627}. The data from experiments HC4000 and HC4149 are available from corresponding authors upon reasonable request.

\bibliography{PHNbiblio}% Produces the bibliography via BibTeX.

\begin{thebibliography}{116}%
\makeatletter
\providecommand \@ifxundefined [1]{%
 \@ifx{#1\undefined}
}%
\providecommand \@ifnum [1]{%
 \ifnum #1\expandafter \@firstoftwo
 \else \expandafter \@secondoftwo
 \fi
}%
\providecommand \@ifx [1]{%
 \ifx #1\expandafter \@firstoftwo
 \else \expandafter \@secondoftwo
 \fi
}%
\providecommand \natexlab [1]{#1}%
\providecommand \enquote  [1]{``#1''}%
\providecommand \bibnamefont  [1]{#1}%
\providecommand \bibfnamefont [1]{#1}%
\providecommand \citenamefont [1]{#1}%
\providecommand \href@noop [0]{\@secondoftwo}%
\providecommand \href [0]{\begingroup \@sanitize@url \@href}%
\providecommand \@href[1]{\@@startlink{#1}\@@href}%
\providecommand \@@href[1]{\endgroup#1\@@endlink}%
\providecommand \@sanitize@url [0]{\catcode `\\12\catcode `\$12\catcode `\&12\catcode `\#12\catcode `\^12\catcode `\_12\catcode `\%12\relax}%
\providecommand \@@startlink[1]{}%
\providecommand \@@endlink[0]{}%
\providecommand \url  [0]{\begingroup\@sanitize@url \@url }%
\providecommand \@url [1]{\endgroup\@href {#1}{\urlprefix }}%
\providecommand \urlprefix  [0]{URL }%
\providecommand \Eprint [0]{\href }%
\providecommand \doibase [0]{https://doi.org/}%
\providecommand \selectlanguage [0]{\@gobble}%
\providecommand \bibinfo  [0]{\@secondoftwo}%
\providecommand \bibfield  [0]{\@secondoftwo}%
\providecommand \translation [1]{[#1]}%
\providecommand \BibitemOpen [0]{}%
\providecommand \bibitemStop [0]{}%
\providecommand \bibitemNoStop [0]{.\EOS\space}%
\providecommand \EOS [0]{\spacefactor3000\relax}%
\providecommand \BibitemShut  [1]{\csname bibitem#1\endcsname}%
\let\auto@bib@innerbib\@empty
%</preamble>
\bibitem [{\citenamefont {Bednorz}\ and\ \citenamefont {M{\"u}ller}(1986)}]{Bednorz1986}%
  \BibitemOpen
  \bibfield  {author} {\bibinfo {author} {\bibfnamefont {J.~G.}\ \bibnamefont {Bednorz}}\ and\ \bibinfo {author} {\bibfnamefont {K.~A.}\ \bibnamefont {M{\"u}ller}},\ }\bibfield  {title} {\bibinfo {title} {{Possible high \ch{Tc} superconductivity in the \ch{Ba}- \ch{La}- \ch{Cu}- \ch{O} system}},\ }\href {https://doi.org/10.1007/BF01303701} {\bibfield  {journal} {\bibinfo  {journal} {Zeitschrift f{\"u}r Physik B Condensed Matter}\ }\textbf {\bibinfo {volume} {64}},\ \bibinfo {pages} {189} (\bibinfo {year} {1986})}\BibitemShut {NoStop}%
\bibitem [{\citenamefont {Savrasov}\ and\ \citenamefont {Andersen}(1996)}]{Savrasov_PRL}%
  \BibitemOpen
  \bibfield  {author} {\bibinfo {author} {\bibfnamefont {S.~Y.}\ \bibnamefont {Savrasov}}\ and\ \bibinfo {author} {\bibfnamefont {O.~K.}\ \bibnamefont {Andersen}},\ }\bibfield  {title} {\bibinfo {title} {{Linear-Response Calculation of the Electron-Phonon Coupling in Doped \ch{CaCuO_2}}},\ }\href {https://doi.org/10.1103/PhysRevLett.77.4430} {\bibfield  {journal} {\bibinfo  {journal} {Phys. Rev. Lett.}\ }\textbf {\bibinfo {volume} {77}},\ \bibinfo {pages} {4430} (\bibinfo {year} {1996})}\BibitemShut {NoStop}%
\bibitem [{\citenamefont {Sakai}\ \emph {et~al.}(1997)\citenamefont {Sakai}, \citenamefont {Poilblanc},\ and\ \citenamefont {Scalapino}}]{Sakai_PRB}%
  \BibitemOpen
  \bibfield  {author} {\bibinfo {author} {\bibfnamefont {T.}~\bibnamefont {Sakai}}, \bibinfo {author} {\bibfnamefont {D.}~\bibnamefont {Poilblanc}},\ and\ \bibinfo {author} {\bibfnamefont {D.~J.}\ \bibnamefont {Scalapino}},\ }\bibfield  {title} {\bibinfo {title} {{Hole pairing and phonon dynamics in generalized two-dimensional t-\ch{J} Holstein models}},\ }\href {https://doi.org/10.1103/PhysRevB.55.8445} {\bibfield  {journal} {\bibinfo  {journal} {Phys. Rev. B}\ }\textbf {\bibinfo {volume} {55}},\ \bibinfo {pages} {8445} (\bibinfo {year} {1997})}\BibitemShut {NoStop}%
\bibitem [{\citenamefont {Jepsen}\ \emph {et~al.}(1998)\citenamefont {Jepsen}, \citenamefont {Andersen}, \citenamefont {Dasgupta},\ and\ \citenamefont {Savrasov}}]{Jepsen1998}%
  \BibitemOpen
  \bibfield  {author} {\bibinfo {author} {\bibfnamefont {O.}~\bibnamefont {Jepsen}}, \bibinfo {author} {\bibfnamefont {O.}~\bibnamefont {Andersen}}, \bibinfo {author} {\bibfnamefont {I.}~\bibnamefont {Dasgupta}},\ and\ \bibinfo {author} {\bibfnamefont {S.}~\bibnamefont {Savrasov}},\ }\bibfield  {title} {\bibinfo {title} {{Buckling and d-wave pairing in \ch{HiTc} superconductors}},\ }\href {https://doi.org/10.1016/S0022-3697%2898%2900089-4} {\bibfield  {journal} {\bibinfo  {journal} {Journal of Physics and Chemistry of Solids}\ }\textbf {\bibinfo {volume} {59}},\ \bibinfo {pages} {1718} (\bibinfo {year} {1998})}\BibitemShut {NoStop}%
\bibitem [{\citenamefont {Ishihara}\ and\ \citenamefont {Nagaosa}(2004)}]{Ishihara_PRB}%
  \BibitemOpen
  \bibfield  {author} {\bibinfo {author} {\bibfnamefont {S.}~\bibnamefont {Ishihara}}\ and\ \bibinfo {author} {\bibfnamefont {N.}~\bibnamefont {Nagaosa}},\ }\bibfield  {title} {\bibinfo {title} {{Interplay of electron-phonon interaction and electron correlation in high-temperature superconductivity}},\ }\href {https://doi.org/10.1103/PhysRevB.69.144520} {\bibfield  {journal} {\bibinfo  {journal} {Phys. Rev. B}\ }\textbf {\bibinfo {volume} {69}},\ \bibinfo {pages} {144520} (\bibinfo {year} {2004})}\BibitemShut {NoStop}%
\bibitem [{\citenamefont {Johnston}\ \emph {et~al.}(2010)\citenamefont {Johnston}, \citenamefont {Vernay}, \citenamefont {Moritz}, \citenamefont {Shen}, \citenamefont {Nagaosa}, \citenamefont {Zaanen},\ and\ \citenamefont {Devereaux}}]{Johnston_PRB}%
  \BibitemOpen
  \bibfield  {author} {\bibinfo {author} {\bibfnamefont {S.}~\bibnamefont {Johnston}}, \bibinfo {author} {\bibfnamefont {F.}~\bibnamefont {Vernay}}, \bibinfo {author} {\bibfnamefont {B.}~\bibnamefont {Moritz}}, \bibinfo {author} {\bibfnamefont {Z.-X.}\ \bibnamefont {Shen}}, \bibinfo {author} {\bibfnamefont {N.}~\bibnamefont {Nagaosa}}, \bibinfo {author} {\bibfnamefont {J.}~\bibnamefont {Zaanen}},\ and\ \bibinfo {author} {\bibfnamefont {T.~P.}\ \bibnamefont {Devereaux}},\ }\bibfield  {title} {\bibinfo {title} {{Systematic study of electron-phonon coupling to oxygen modes across the cuprates}},\ }\href {https://doi.org/10.1103/PhysRevB.82.064513} {\bibfield  {journal} {\bibinfo  {journal} {Phys. Rev. B}\ }\textbf {\bibinfo {volume} {82}},\ \bibinfo {pages} {064513} (\bibinfo {year} {2010})}\BibitemShut {NoStop}%
\bibitem [{\citenamefont {Pashkin}\ \emph {et~al.}(2010)\citenamefont {Pashkin}, \citenamefont {Porer}, \citenamefont {Beyer}, \citenamefont {Kim}, \citenamefont {Dubroka}, \citenamefont {Bernhard}, \citenamefont {Yao}, \citenamefont {Dagan}, \citenamefont {Hackl}, \citenamefont {Erb}, \citenamefont {Demsar}, \citenamefont {Huber},\ and\ \citenamefont {Leitenstorfer}}]{Pashkin_PRL}%
  \BibitemOpen
  \bibfield  {author} {\bibinfo {author} {\bibfnamefont {A.}~\bibnamefont {Pashkin}}, \bibinfo {author} {\bibfnamefont {M.}~\bibnamefont {Porer}}, \bibinfo {author} {\bibfnamefont {M.}~\bibnamefont {Beyer}}, \bibinfo {author} {\bibfnamefont {K.~W.}\ \bibnamefont {Kim}}, \bibinfo {author} {\bibfnamefont {A.}~\bibnamefont {Dubroka}}, \bibinfo {author} {\bibfnamefont {C.}~\bibnamefont {Bernhard}}, \bibinfo {author} {\bibfnamefont {X.}~\bibnamefont {Yao}}, \bibinfo {author} {\bibfnamefont {Y.}~\bibnamefont {Dagan}}, \bibinfo {author} {\bibfnamefont {R.}~\bibnamefont {Hackl}}, \bibinfo {author} {\bibfnamefont {A.}~\bibnamefont {Erb}}, \bibinfo {author} {\bibfnamefont {J.}~\bibnamefont {Demsar}}, \bibinfo {author} {\bibfnamefont {R.}~\bibnamefont {Huber}},\ and\ \bibinfo {author} {\bibfnamefont {A.}~\bibnamefont {Leitenstorfer}},\ }\bibfield  {title} {\bibinfo {title} {{Femtosecond Response of Quasiparticles and Phonons in Superconducting \ch{YBa_2Cu_3O_{7-$\delta$}} Studied by Wideband Terahertz Spectroscopy}},\
  }\href {https://doi.org/10.1103/PhysRevLett.105.067001} {\bibfield  {journal} {\bibinfo  {journal} {Phys. Rev. Lett.}\ }\textbf {\bibinfo {volume} {105}},\ \bibinfo {pages} {067001} (\bibinfo {year} {2010})}\BibitemShut {NoStop}%
\bibitem [{\citenamefont {Fausti}\ \emph {et~al.}(2011)\citenamefont {Fausti}, \citenamefont {Tobey}, \citenamefont {Dean}, \citenamefont {Kaiser}, \citenamefont {Dienst}, \citenamefont {Hoffmann}, \citenamefont {Pyon}, \citenamefont {Takayama}, \citenamefont {Takagi},\ and\ \citenamefont {Cavalleri}}]{Fausti_Science}%
  \BibitemOpen
  \bibfield  {author} {\bibinfo {author} {\bibfnamefont {D.}~\bibnamefont {Fausti}}, \bibinfo {author} {\bibfnamefont {R.}~\bibnamefont {Tobey}}, \bibinfo {author} {\bibfnamefont {N.}~\bibnamefont {Dean}}, \bibinfo {author} {\bibfnamefont {S.}~\bibnamefont {Kaiser}}, \bibinfo {author} {\bibfnamefont {A.}~\bibnamefont {Dienst}}, \bibinfo {author} {\bibfnamefont {M.~C.}\ \bibnamefont {Hoffmann}}, \bibinfo {author} {\bibfnamefont {S.}~\bibnamefont {Pyon}}, \bibinfo {author} {\bibfnamefont {T.}~\bibnamefont {Takayama}}, \bibinfo {author} {\bibfnamefont {H.}~\bibnamefont {Takagi}},\ and\ \bibinfo {author} {\bibfnamefont {A.}~\bibnamefont {Cavalleri}},\ }\bibfield  {title} {\bibinfo {title} {{Light-induced superconductivity in a stripe-ordered cuprate}},\ }\href {https://www.science.org/doi/10.1126/science.1197294} {\bibfield  {journal} {\bibinfo  {journal} {Science}\ }\textbf {\bibinfo {volume} {331}},\ \bibinfo {pages} {189} (\bibinfo {year} {2011})}\BibitemShut {NoStop}%
\bibitem [{\citenamefont {Kaiser}\ \emph {et~al.}(2014)\citenamefont {Kaiser}, \citenamefont {Hunt}, \citenamefont {Nicoletti}, \citenamefont {Hu}, \citenamefont {Gierz}, \citenamefont {Liu}, \citenamefont {Le~Tacon}, \citenamefont {Loew}, \citenamefont {Haug}, \citenamefont {Keimer},\ and\ \citenamefont {Cavalleri}}]{Kaiser_PRB}%
  \BibitemOpen
  \bibfield  {author} {\bibinfo {author} {\bibfnamefont {S.}~\bibnamefont {Kaiser}}, \bibinfo {author} {\bibfnamefont {C.~R.}\ \bibnamefont {Hunt}}, \bibinfo {author} {\bibfnamefont {D.}~\bibnamefont {Nicoletti}}, \bibinfo {author} {\bibfnamefont {W.}~\bibnamefont {Hu}}, \bibinfo {author} {\bibfnamefont {I.}~\bibnamefont {Gierz}}, \bibinfo {author} {\bibfnamefont {H.~Y.}\ \bibnamefont {Liu}}, \bibinfo {author} {\bibfnamefont {M.}~\bibnamefont {Le~Tacon}}, \bibinfo {author} {\bibfnamefont {T.}~\bibnamefont {Loew}}, \bibinfo {author} {\bibfnamefont {D.}~\bibnamefont {Haug}}, \bibinfo {author} {\bibfnamefont {B.}~\bibnamefont {Keimer}},\ and\ \bibinfo {author} {\bibfnamefont {A.}~\bibnamefont {Cavalleri}},\ }\bibfield  {title} {\bibinfo {title} {{Optically induced coherent transport far above ${T}_{c}$ in underdoped \ch{YBa_{2}Cu_{3}O_{6+$\delta$}}}},\ }\href {https://doi.org/10.1103/PhysRevB.89.184516} {\bibfield  {journal} {\bibinfo  {journal} {Phys. Rev. B}\ }\textbf {\bibinfo {volume} {89}},\ \bibinfo
  {pages} {184516} (\bibinfo {year} {2014})}\BibitemShut {NoStop}%
\bibitem [{\citenamefont {Hu}\ \emph {et~al.}(2014)\citenamefont {Hu}, \citenamefont {Kaiser}, \citenamefont {Nicoletti}, \citenamefont {Hunt}, \citenamefont {Gierz}, \citenamefont {Hoffmann}, \citenamefont {Le~Tacon}, \citenamefont {Loew}, \citenamefont {Keimer},\ and\ \citenamefont {Cavalleri}}]{Hu_Nature}%
  \BibitemOpen
  \bibfield  {author} {\bibinfo {author} {\bibfnamefont {W.}~\bibnamefont {Hu}}, \bibinfo {author} {\bibfnamefont {S.}~\bibnamefont {Kaiser}}, \bibinfo {author} {\bibfnamefont {D.}~\bibnamefont {Nicoletti}}, \bibinfo {author} {\bibfnamefont {C.~R.}\ \bibnamefont {Hunt}}, \bibinfo {author} {\bibfnamefont {I.}~\bibnamefont {Gierz}}, \bibinfo {author} {\bibfnamefont {M.~C.}\ \bibnamefont {Hoffmann}}, \bibinfo {author} {\bibfnamefont {M.}~\bibnamefont {Le~Tacon}}, \bibinfo {author} {\bibfnamefont {T.}~\bibnamefont {Loew}}, \bibinfo {author} {\bibfnamefont {B.}~\bibnamefont {Keimer}},\ and\ \bibinfo {author} {\bibfnamefont {A.}~\bibnamefont {Cavalleri}},\ }\bibfield  {title} {\bibinfo {title} {{Optically enhanced coherent transport in \ch{YBa_2Cu_3O_{6.5}} by ultrafast redistribution of interlayer coupling}},\ }\href {https://doi.org/10.1038/NMAT3963} {\bibfield  {journal} {\bibinfo  {journal} {Nature Materials}\ }\textbf {\bibinfo {volume} {13}},\ \bibinfo {pages} {705} (\bibinfo {year} {2014})}\BibitemShut
  {NoStop}%
\bibitem [{\citenamefont {Liu}\ \emph {et~al.}(2020)\citenamefont {Liu}, \citenamefont {F\"orst}, \citenamefont {Fechner}, \citenamefont {Nicoletti}, \citenamefont {Porras}, \citenamefont {Loew}, \citenamefont {Keimer},\ and\ \citenamefont {Cavalleri}}]{Liu_PRX}%
  \BibitemOpen
  \bibfield  {author} {\bibinfo {author} {\bibfnamefont {B.}~\bibnamefont {Liu}}, \bibinfo {author} {\bibfnamefont {M.}~\bibnamefont {F\"orst}}, \bibinfo {author} {\bibfnamefont {M.}~\bibnamefont {Fechner}}, \bibinfo {author} {\bibfnamefont {D.}~\bibnamefont {Nicoletti}}, \bibinfo {author} {\bibfnamefont {J.}~\bibnamefont {Porras}}, \bibinfo {author} {\bibfnamefont {T.}~\bibnamefont {Loew}}, \bibinfo {author} {\bibfnamefont {B.}~\bibnamefont {Keimer}},\ and\ \bibinfo {author} {\bibfnamefont {A.}~\bibnamefont {Cavalleri}},\ }\bibfield  {title} {\bibinfo {title} {{Pump Frequency Resonances for Light-Induced Incipient Superconductivity in \ch{YBa_2Cu_3O_{6.5}}}},\ }\href {https://doi.org/10.1103/PhysRevX.10.011053} {\bibfield  {journal} {\bibinfo  {journal} {Phys. Rev. X}\ }\textbf {\bibinfo {volume} {10}},\ \bibinfo {pages} {011053} (\bibinfo {year} {2020})}\BibitemShut {NoStop}%
\bibitem [{\citenamefont {Fava}\ \emph {et~al.}(2024)\citenamefont {Fava}, \citenamefont {De~Vecchi}, \citenamefont {Jotzu}, \citenamefont {Buzzi}, \citenamefont {Gebert}, \citenamefont {Liu}, \citenamefont {Keimer},\ and\ \citenamefont {Cavalleri}}]{Fava_Nature}%
  \BibitemOpen
  \bibfield  {author} {\bibinfo {author} {\bibfnamefont {S.}~\bibnamefont {Fava}}, \bibinfo {author} {\bibfnamefont {G.}~\bibnamefont {De~Vecchi}}, \bibinfo {author} {\bibfnamefont {G.}~\bibnamefont {Jotzu}}, \bibinfo {author} {\bibfnamefont {M.}~\bibnamefont {Buzzi}}, \bibinfo {author} {\bibfnamefont {T.}~\bibnamefont {Gebert}}, \bibinfo {author} {\bibfnamefont {Y.}~\bibnamefont {Liu}}, \bibinfo {author} {\bibfnamefont {B.}~\bibnamefont {Keimer}},\ and\ \bibinfo {author} {\bibfnamefont {A.}~\bibnamefont {Cavalleri}},\ }\bibfield  {title} {\bibinfo {title} {{Magnetic field expulsion in optically driven \ch{YBa_2Cu_3O_{6.48}}}},\ }\href {https://doi.org/10.48550/arXiv.2405.00848} {\bibfield  {journal} {\bibinfo  {journal} {Nature}\ ,\ \bibinfo {pages} {1}} (\bibinfo {year} {2024})}\BibitemShut {NoStop}%
\bibitem [{\citenamefont {Huang}\ \emph {et~al.}(2003)\citenamefont {Huang}, \citenamefont {Hanke}, \citenamefont {Arrigoni},\ and\ \citenamefont {Scalapino}}]{Huang_PRB}%
  \BibitemOpen
  \bibfield  {author} {\bibinfo {author} {\bibfnamefont {Z.~B.}\ \bibnamefont {Huang}}, \bibinfo {author} {\bibfnamefont {W.}~\bibnamefont {Hanke}}, \bibinfo {author} {\bibfnamefont {E.}~\bibnamefont {Arrigoni}},\ and\ \bibinfo {author} {\bibfnamefont {D.~J.}\ \bibnamefont {Scalapino}},\ }\bibfield  {title} {\bibinfo {title} {{Electron-phonon vertex in the two-dimensional one-band Hubbard model}},\ }\href {https://doi.org/10.1103/PhysRevB.68.220507} {\bibfield  {journal} {\bibinfo  {journal} {Phys. Rev. B}\ }\textbf {\bibinfo {volume} {68}},\ \bibinfo {pages} {220507} (\bibinfo {year} {2003})}\BibitemShut {NoStop}%
\bibitem [{\citenamefont {Marsiglio}(1990)}]{Marsiglio_PRB}%
  \BibitemOpen
  \bibfield  {author} {\bibinfo {author} {\bibfnamefont {F.}~\bibnamefont {Marsiglio}},\ }\bibfield  {title} {\bibinfo {title} {{Pairing and charge-density-wave correlations in the Holstein model at half-filling}},\ }\href {https://doi.org/10.1103/PhysRevB.42.2416} {\bibfield  {journal} {\bibinfo  {journal} {Phys. Rev. B}\ }\textbf {\bibinfo {volume} {42}},\ \bibinfo {pages} {2416} (\bibinfo {year} {1990})}\BibitemShut {NoStop}%
\bibitem [{\citenamefont {Esterlis}\ \emph {et~al.}(2018)\citenamefont {Esterlis}, \citenamefont {Nosarzewski}, \citenamefont {Huang}, \citenamefont {Moritz}, \citenamefont {Devereaux}, \citenamefont {Scalapino},\ and\ \citenamefont {Kivelson}}]{Esterlis_PRB}%
  \BibitemOpen
  \bibfield  {author} {\bibinfo {author} {\bibfnamefont {I.}~\bibnamefont {Esterlis}}, \bibinfo {author} {\bibfnamefont {B.}~\bibnamefont {Nosarzewski}}, \bibinfo {author} {\bibfnamefont {E.~W.}\ \bibnamefont {Huang}}, \bibinfo {author} {\bibfnamefont {B.}~\bibnamefont {Moritz}}, \bibinfo {author} {\bibfnamefont {T.~P.}\ \bibnamefont {Devereaux}}, \bibinfo {author} {\bibfnamefont {D.~J.}\ \bibnamefont {Scalapino}},\ and\ \bibinfo {author} {\bibfnamefont {S.~A.}\ \bibnamefont {Kivelson}},\ }\bibfield  {title} {\bibinfo {title} {{Breakdown of the Migdal-Eliashberg theory: A determinant quantum Monte Carlo study}},\ }\href {https://doi.org/10.1103/PhysRevB.97.140501} {\bibfield  {journal} {\bibinfo  {journal} {Phys. Rev. B}\ }\textbf {\bibinfo {volume} {97}},\ \bibinfo {pages} {140501} (\bibinfo {year} {2018})}\BibitemShut {NoStop}%
\bibitem [{\citenamefont {Nosarzewski}\ \emph {et~al.}(2021)\citenamefont {Nosarzewski}, \citenamefont {Sch\"uler},\ and\ \citenamefont {Devereaux}}]{Nosarzewski_PRB}%
  \BibitemOpen
  \bibfield  {author} {\bibinfo {author} {\bibfnamefont {B.}~\bibnamefont {Nosarzewski}}, \bibinfo {author} {\bibfnamefont {M.}~\bibnamefont {Sch\"uler}},\ and\ \bibinfo {author} {\bibfnamefont {T.~P.}\ \bibnamefont {Devereaux}},\ }\bibfield  {title} {\bibinfo {title} {{Spectral properties and enhanced superconductivity in renormalized Migdal-Eliashberg theory}},\ }\href {https://doi.org/10.1103/PhysRevB.103.024520} {\bibfield  {journal} {\bibinfo  {journal} {Phys. Rev. B}\ }\textbf {\bibinfo {volume} {103}},\ \bibinfo {pages} {024520} (\bibinfo {year} {2021})}\BibitemShut {NoStop}%
\bibitem [{\citenamefont {Capone}\ \emph {et~al.}(2004)\citenamefont {Capone}, \citenamefont {Sangiovanni}, \citenamefont {Castellani}, \citenamefont {Di~Castro},\ and\ \citenamefont {Grilli}}]{Capone_PRL}%
  \BibitemOpen
  \bibfield  {author} {\bibinfo {author} {\bibfnamefont {M.}~\bibnamefont {Capone}}, \bibinfo {author} {\bibfnamefont {G.}~\bibnamefont {Sangiovanni}}, \bibinfo {author} {\bibfnamefont {C.}~\bibnamefont {Castellani}}, \bibinfo {author} {\bibfnamefont {C.}~\bibnamefont {Di~Castro}},\ and\ \bibinfo {author} {\bibfnamefont {M.}~\bibnamefont {Grilli}},\ }\bibfield  {title} {\bibinfo {title} {{Phase Separation Close to the Density-Driven Mott Transition in the Hubbard-Holstein Model}},\ }\href {https://doi.org/10.1103/PhysRevLett.92.106401} {\bibfield  {journal} {\bibinfo  {journal} {Phys. Rev. Lett.}\ }\textbf {\bibinfo {volume} {92}},\ \bibinfo {pages} {106401} (\bibinfo {year} {2004})}\BibitemShut {NoStop}%
\bibitem [{\citenamefont {Sangiovanni}\ \emph {et~al.}(2005)\citenamefont {Sangiovanni}, \citenamefont {Capone}, \citenamefont {Castellani},\ and\ \citenamefont {Grilli}}]{Sangiovanni_PRL}%
  \BibitemOpen
  \bibfield  {author} {\bibinfo {author} {\bibfnamefont {G.}~\bibnamefont {Sangiovanni}}, \bibinfo {author} {\bibfnamefont {M.}~\bibnamefont {Capone}}, \bibinfo {author} {\bibfnamefont {C.}~\bibnamefont {Castellani}},\ and\ \bibinfo {author} {\bibfnamefont {M.}~\bibnamefont {Grilli}},\ }\bibfield  {title} {\bibinfo {title} {{Electron-Phonon Interaction Close to a Mott Transition}},\ }\href {https://doi.org/10.1103/PhysRevLett.94.026401} {\bibfield  {journal} {\bibinfo  {journal} {Phys. Rev. Lett.}\ }\textbf {\bibinfo {volume} {94}},\ \bibinfo {pages} {026401} (\bibinfo {year} {2005})}\BibitemShut {NoStop}%
\bibitem [{\citenamefont {Werner}\ and\ \citenamefont {Millis}(2007)}]{Werner_PRL}%
  \BibitemOpen
  \bibfield  {author} {\bibinfo {author} {\bibfnamefont {P.}~\bibnamefont {Werner}}\ and\ \bibinfo {author} {\bibfnamefont {A.~J.}\ \bibnamefont {Millis}},\ }\bibfield  {title} {\bibinfo {title} {{Efficient Dynamical Mean Field Simulation of the Holstein-Hubbard Model}},\ }\href {https://doi.org/10.1103/PhysRevLett.99.146404} {\bibfield  {journal} {\bibinfo  {journal} {Phys. Rev. Lett.}\ }\textbf {\bibinfo {volume} {99}},\ \bibinfo {pages} {146404} (\bibinfo {year} {2007})}\BibitemShut {NoStop}%
\bibitem [{\citenamefont {Macridin}\ \emph {et~al.}(2006)\citenamefont {Macridin}, \citenamefont {Moritz}, \citenamefont {Jarrell},\ and\ \citenamefont {Maier}}]{Macridin_PRL}%
  \BibitemOpen
  \bibfield  {author} {\bibinfo {author} {\bibfnamefont {A.}~\bibnamefont {Macridin}}, \bibinfo {author} {\bibfnamefont {B.}~\bibnamefont {Moritz}}, \bibinfo {author} {\bibfnamefont {M.}~\bibnamefont {Jarrell}},\ and\ \bibinfo {author} {\bibfnamefont {T.}~\bibnamefont {Maier}},\ }\bibfield  {title} {\bibinfo {title} {{Synergistic Polaron Formation in the Hubbard-Holstein Model at Small Doping}},\ }\href {https://doi.org/10.1103/PhysRevLett.97.056402} {\bibfield  {journal} {\bibinfo  {journal} {Phys. Rev. Lett.}\ }\textbf {\bibinfo {volume} {97}},\ \bibinfo {pages} {056402} (\bibinfo {year} {2006})}\BibitemShut {NoStop}%
\bibitem [{\citenamefont {Macridin}\ \emph {et~al.}(2012)\citenamefont {Macridin}, \citenamefont {Moritz}, \citenamefont {Jarrell},\ and\ \citenamefont {Maier}}]{Macridin2012}%
  \BibitemOpen
  \bibfield  {author} {\bibinfo {author} {\bibfnamefont {A.}~\bibnamefont {Macridin}}, \bibinfo {author} {\bibfnamefont {B.}~\bibnamefont {Moritz}}, \bibinfo {author} {\bibfnamefont {M.}~\bibnamefont {Jarrell}},\ and\ \bibinfo {author} {\bibfnamefont {T.}~\bibnamefont {Maier}},\ }\bibfield  {title} {\bibinfo {title} {{Suppression of superconductivity in the Hubbard model by buckling and breathing phonons}},\ }\href {https://doi.org/10.1088/0953-8984/24/47/475603} {\bibfield  {journal} {\bibinfo  {journal} {Journal of Physics: Condensed Matter}\ }\textbf {\bibinfo {volume} {24}},\ \bibinfo {pages} {475603} (\bibinfo {year} {2012})}\BibitemShut {NoStop}%
\bibitem [{\citenamefont {Heid}\ \emph {et~al.}(2008)\citenamefont {Heid}, \citenamefont {Bohnen}, \citenamefont {Zeyher},\ and\ \citenamefont {Manske}}]{Heid2008}%
  \BibitemOpen
  \bibfield  {author} {\bibinfo {author} {\bibfnamefont {R.}~\bibnamefont {Heid}}, \bibinfo {author} {\bibfnamefont {K.-P.}\ \bibnamefont {Bohnen}}, \bibinfo {author} {\bibfnamefont {R.}~\bibnamefont {Zeyher}},\ and\ \bibinfo {author} {\bibfnamefont {D.}~\bibnamefont {Manske}},\ }\bibfield  {title} {\bibinfo {title} {{Momentum Dependence of the Electron-Phonon Coupling and Self-Energy Effects in Superconducting \ch{YBa_2Cu_3O_7} within the Local Density Approximation}},\ }\href {https://doi.org/10.1103/PhysRevLett.100.137001} {\bibfield  {journal} {\bibinfo  {journal} {Physical Review Letters}\ }\textbf {\bibinfo {volume} {100}},\ \bibinfo {pages} {137001} (\bibinfo {year} {2008})}\BibitemShut {NoStop}%
\bibitem [{\citenamefont {Bohnen}\ \emph {et~al.}(2003)\citenamefont {Bohnen}, \citenamefont {Heid},\ and\ \citenamefont {Krauss}}]{Bohnen2003}%
  \BibitemOpen
  \bibfield  {author} {\bibinfo {author} {\bibfnamefont {K.-P.}\ \bibnamefont {Bohnen}}, \bibinfo {author} {\bibfnamefont {R.}~\bibnamefont {Heid}},\ and\ \bibinfo {author} {\bibfnamefont {M.}~\bibnamefont {Krauss}},\ }\bibfield  {title} {\bibinfo {title} {{Phonon dispersion and electron-phonon interaction for \ch{YBa_2Cu_3O_7} from first-principles calculations}},\ }\href {https://doi.org/10.1209/epl/i2003-00143-x} {\bibfield  {journal} {\bibinfo  {journal} {Europhysics Letters}\ }\textbf {\bibinfo {volume} {64}},\ \bibinfo {pages} {104} (\bibinfo {year} {2003})}\BibitemShut {NoStop}%
\bibitem [{\citenamefont {Giustino}\ \emph {et~al.}(2008)\citenamefont {Giustino}, \citenamefont {Cohen},\ and\ \citenamefont {Louie}}]{Giustino2008}%
  \BibitemOpen
  \bibfield  {author} {\bibinfo {author} {\bibfnamefont {F.}~\bibnamefont {Giustino}}, \bibinfo {author} {\bibfnamefont {M.~L.}\ \bibnamefont {Cohen}},\ and\ \bibinfo {author} {\bibfnamefont {S.~G.}\ \bibnamefont {Louie}},\ }\bibfield  {title} {\bibinfo {title} {{Small phonon contribution to the photoemission kink in the copper oxide superconductors}},\ }\href {https://doi.org/10.1038/nature06874} {\bibfield  {journal} {\bibinfo  {journal} {Nature}\ }\textbf {\bibinfo {volume} {452}},\ \bibinfo {pages} {975} (\bibinfo {year} {2008})}\BibitemShut {NoStop}%
\bibitem [{\citenamefont {Sterling}\ and\ \citenamefont {Reznik}(2021{\natexlab{a}})}]{Sterling_PRB}%
  \BibitemOpen
  \bibfield  {author} {\bibinfo {author} {\bibfnamefont {T.~C.}\ \bibnamefont {Sterling}}\ and\ \bibinfo {author} {\bibfnamefont {D.}~\bibnamefont {Reznik}},\ }\bibfield  {title} {\bibinfo {title} {{Effect of the electronic charge gap on LO bond-stretching phonons in undoped \ch{La_2CuO_4} calculated using \ch{LDA}+\ch{U}}},\ }\href {https://doi.org/10.1103/PhysRevB.104.134311} {\bibfield  {journal} {\bibinfo  {journal} {Phys. Rev. B}\ }\textbf {\bibinfo {volume} {104}},\ \bibinfo {pages} {134311} (\bibinfo {year} {2021}{\natexlab{a}})}\BibitemShut {NoStop}%
\bibitem [{\citenamefont {Bohnen}\ \emph {et~al.}(2001)\citenamefont {Bohnen}, \citenamefont {Heid},\ and\ \citenamefont {Renker}}]{Bohnen_PRB_2001}%
  \BibitemOpen
  \bibfield  {author} {\bibinfo {author} {\bibfnamefont {K.-P.}\ \bibnamefont {Bohnen}}, \bibinfo {author} {\bibfnamefont {R.}~\bibnamefont {Heid}},\ and\ \bibinfo {author} {\bibfnamefont {B.}~\bibnamefont {Renker}},\ }\bibfield  {title} {\bibinfo {title} {{Phonon Dispersion and Electron-Phonon Coupling in \ch{MgB_2} and \ch{AlB_2}}},\ }\href {https://doi.org/10.1103/PhysRevLett.86.5771} {\bibfield  {journal} {\bibinfo  {journal} {Phys. Rev. Lett.}\ }\textbf {\bibinfo {volume} {86}},\ \bibinfo {pages} {5771} (\bibinfo {year} {2001})}\BibitemShut {NoStop}%
\bibitem [{\citenamefont {Reznik}\ \emph {et~al.}(2008)\citenamefont {Reznik}, \citenamefont {Sangiovanni}, \citenamefont {Gunnarsson},\ and\ \citenamefont {Devereaux}}]{Reznik2008}%
  \BibitemOpen
  \bibfield  {author} {\bibinfo {author} {\bibfnamefont {D.}~\bibnamefont {Reznik}}, \bibinfo {author} {\bibfnamefont {G.}~\bibnamefont {Sangiovanni}}, \bibinfo {author} {\bibfnamefont {O.}~\bibnamefont {Gunnarsson}},\ and\ \bibinfo {author} {\bibfnamefont {T.}~\bibnamefont {Devereaux}},\ }\bibfield  {title} {\bibinfo {title} {{Photoemission kinks and phonons in cuprates}},\ }\href {https://doi.org/10.1038/nature07364} {\bibfield  {journal} {\bibinfo  {journal} {Nature}\ }\textbf {\bibinfo {volume} {455}},\ \bibinfo {pages} {E6} (\bibinfo {year} {2008})}\BibitemShut {NoStop}%
\bibitem [{\citenamefont {Pintschovius}(2005)}]{Pintschovius2005}%
  \BibitemOpen
  \bibfield  {author} {\bibinfo {author} {\bibfnamefont {L.}~\bibnamefont {Pintschovius}},\ }\bibfield  {title} {\bibinfo {title} {{Electron--phonon coupling effects explored by inelastic neutron scattering}},\ }\href {https://doi.org/10.1002/pssb.200404951} {\bibfield  {journal} {\bibinfo  {journal} {physica status solidi (b)}\ }\textbf {\bibinfo {volume} {242}},\ \bibinfo {pages} {30} (\bibinfo {year} {2005})}\BibitemShut {NoStop}%
\bibitem [{\citenamefont {Reznik}(2010)}]{Reznik2010}%
  \BibitemOpen
  \bibfield  {author} {\bibinfo {author} {\bibfnamefont {D.}~\bibnamefont {Reznik}},\ }\bibfield  {title} {\bibinfo {title} {{Giant Electron-Phonon Anomaly in Doped \ch{La_2CuO_4} and Other Cuprates}},\ }\href {https://doi.org/10.1155/2010/523549} {\bibfield  {journal} {\bibinfo  {journal} {Advances in Condensed Matter Physics}\ }\textbf {\bibinfo {volume} {2010}},\ \bibinfo {pages} {523549} (\bibinfo {year} {2010})}\BibitemShut {NoStop}%
\bibitem [{\citenamefont {Ahmadova}\ \emph {et~al.}(2020)\citenamefont {Ahmadova}, \citenamefont {Sterling}, \citenamefont {Sokolik}, \citenamefont {Abernathy}, \citenamefont {Greven},\ and\ \citenamefont {Reznik}}]{Ahmadova2020}%
  \BibitemOpen
  \bibfield  {author} {\bibinfo {author} {\bibfnamefont {I.}~\bibnamefont {Ahmadova}}, \bibinfo {author} {\bibfnamefont {T.}~\bibnamefont {Sterling}}, \bibinfo {author} {\bibfnamefont {A.}~\bibnamefont {Sokolik}}, \bibinfo {author} {\bibfnamefont {D.}~\bibnamefont {Abernathy}}, \bibinfo {author} {\bibfnamefont {M.}~\bibnamefont {Greven}},\ and\ \bibinfo {author} {\bibfnamefont {D.}~\bibnamefont {Reznik}},\ }\bibfield  {title} {\bibinfo {title} {{Phonon spectrum of underdoped \ch{HgBa_2CuO_{4+$\delta$}} investigated by neutron scattering}},\ }\href {https://doi.org/10.1103/PhysRevB.101.184508} {\bibfield  {journal} {\bibinfo  {journal} {Physical Review B}\ }\textbf {\bibinfo {volume} {101}},\ \bibinfo {pages} {184508} (\bibinfo {year} {2020})}\BibitemShut {NoStop}%
\bibitem [{\citenamefont {Le~Tacon}\ \emph {et~al.}(2014)\citenamefont {Le~Tacon}, \citenamefont {Bosak}, \citenamefont {Souliou}, \citenamefont {Dellea}, \citenamefont {Loew}, \citenamefont {Heid}, \citenamefont {Bohnen}, \citenamefont {Ghiringhelli}, \citenamefont {Krisch},\ and\ \citenamefont {Keimer}}]{LeTacon_Nature}%
  \BibitemOpen
  \bibfield  {author} {\bibinfo {author} {\bibfnamefont {M.}~\bibnamefont {Le~Tacon}}, \bibinfo {author} {\bibfnamefont {A.}~\bibnamefont {Bosak}}, \bibinfo {author} {\bibfnamefont {S.}~\bibnamefont {Souliou}}, \bibinfo {author} {\bibfnamefont {G.}~\bibnamefont {Dellea}}, \bibinfo {author} {\bibfnamefont {T.}~\bibnamefont {Loew}}, \bibinfo {author} {\bibfnamefont {R.}~\bibnamefont {Heid}}, \bibinfo {author} {\bibfnamefont {K.}~\bibnamefont {Bohnen}}, \bibinfo {author} {\bibfnamefont {G.}~\bibnamefont {Ghiringhelli}}, \bibinfo {author} {\bibfnamefont {M.}~\bibnamefont {Krisch}},\ and\ \bibinfo {author} {\bibfnamefont {B.}~\bibnamefont {Keimer}},\ }\bibfield  {title} {\bibinfo {title} {{Inelastic X-ray scattering in \ch{YBa_2Cu_3O_{6.6}} reveals giant phonon anomalies and elastic central peak due to charge-density-wave formation}},\ }\href {https://doi.org/10.1038/nphys2805} {\bibfield  {journal} {\bibinfo  {journal} {Nature Physics}\ }\textbf {\bibinfo {volume} {10}},\ \bibinfo {pages} {52} (\bibinfo {year}
  {2014})}\BibitemShut {NoStop}%
\bibitem [{\citenamefont {Miao}\ \emph {et~al.}(2018)\citenamefont {Miao}, \citenamefont {Ishikawa}, \citenamefont {Heid}, \citenamefont {Le~Tacon}, \citenamefont {Fabbris}, \citenamefont {Meyers}, \citenamefont {Gu}, \citenamefont {Baron},\ and\ \citenamefont {Dean}}]{Miao_PRB}%
  \BibitemOpen
  \bibfield  {author} {\bibinfo {author} {\bibfnamefont {H.}~\bibnamefont {Miao}}, \bibinfo {author} {\bibfnamefont {D.}~\bibnamefont {Ishikawa}}, \bibinfo {author} {\bibfnamefont {R.}~\bibnamefont {Heid}}, \bibinfo {author} {\bibfnamefont {M.}~\bibnamefont {Le~Tacon}}, \bibinfo {author} {\bibfnamefont {G.}~\bibnamefont {Fabbris}}, \bibinfo {author} {\bibfnamefont {D.}~\bibnamefont {Meyers}}, \bibinfo {author} {\bibfnamefont {G.~D.}\ \bibnamefont {Gu}}, \bibinfo {author} {\bibfnamefont {A.~Q.~R.}\ \bibnamefont {Baron}},\ and\ \bibinfo {author} {\bibfnamefont {M.~P.~M.}\ \bibnamefont {Dean}},\ }\bibfield  {title} {\bibinfo {title} {{Incommensurate Phonon Anomaly and the Nature of Charge Density Waves in Cuprates}},\ }\href {https://doi.org/10.1103/PhysRevX.8.011008} {\bibfield  {journal} {\bibinfo  {journal} {Phys. Rev. X}\ }\textbf {\bibinfo {volume} {8}},\ \bibinfo {pages} {011008} (\bibinfo {year} {2018})}\BibitemShut {NoStop}%
\bibitem [{\citenamefont {Souliou}\ \emph {et~al.}(2020)\citenamefont {Souliou}, \citenamefont {Bosak}, \citenamefont {Garbarino},\ and\ \citenamefont {Le~Tacon}}]{Souliou2020}%
  \BibitemOpen
  \bibfield  {author} {\bibinfo {author} {\bibfnamefont {S.}~\bibnamefont {Souliou}}, \bibinfo {author} {\bibfnamefont {A.}~\bibnamefont {Bosak}}, \bibinfo {author} {\bibfnamefont {G.}~\bibnamefont {Garbarino}},\ and\ \bibinfo {author} {\bibfnamefont {M.}~\bibnamefont {Le~Tacon}},\ }\bibfield  {title} {\bibinfo {title} {{Inelastic x-ray scattering studies of phonon dispersions in superconductors at high pressures}},\ }\href {https://doi.org/10.1088/1361-6668/abbdc3} {\bibfield  {journal} {\bibinfo  {journal} {Superconductor Science and Technology}\ }\textbf {\bibinfo {volume} {33}},\ \bibinfo {pages} {124004} (\bibinfo {year} {2020})}\BibitemShut {NoStop}%
\bibitem [{\citenamefont {Souliou}\ \emph {et~al.}(2021)\citenamefont {Souliou}, \citenamefont {Sen}, \citenamefont {Heid}, \citenamefont {Nakata}, \citenamefont {Wang}, \citenamefont {Kim}, \citenamefont {Uchiyama}, \citenamefont {Merz}, \citenamefont {Minola}, \citenamefont {Keimer} \emph {et~al.}}]{Souliou2021}%
  \BibitemOpen
  \bibfield  {author} {\bibinfo {author} {\bibfnamefont {S.-M.}\ \bibnamefont {Souliou}}, \bibinfo {author} {\bibfnamefont {K.}~\bibnamefont {Sen}}, \bibinfo {author} {\bibfnamefont {R.}~\bibnamefont {Heid}}, \bibinfo {author} {\bibfnamefont {S.}~\bibnamefont {Nakata}}, \bibinfo {author} {\bibfnamefont {L.}~\bibnamefont {Wang}}, \bibinfo {author} {\bibfnamefont {H.-h.}\ \bibnamefont {Kim}}, \bibinfo {author} {\bibfnamefont {H.}~\bibnamefont {Uchiyama}}, \bibinfo {author} {\bibfnamefont {M.}~\bibnamefont {Merz}}, \bibinfo {author} {\bibfnamefont {M.}~\bibnamefont {Minola}}, \bibinfo {author} {\bibfnamefont {B.}~\bibnamefont {Keimer}}, \emph {et~al.},\ }\bibfield  {title} {\bibinfo {title} {{In-plane isotropy of the low energy phonon anomalies in \ch{YBa_2Cu_3O_{6+x}}}},\ }\href {https://doi.org/10.7566/JPSJ.90.111006} {\bibfield  {journal} {\bibinfo  {journal} {Journal of the Physical Society of Japan}\ }\textbf {\bibinfo {volume} {90}},\ \bibinfo {pages} {111006} (\bibinfo {year} {2021})}\BibitemShut
  {NoStop}%
\bibitem [{\citenamefont {Peng}\ \emph {et~al.}(2024)\citenamefont {Peng}, \citenamefont {Boukahil}, \citenamefont {Krongchon}, \citenamefont {Xiao}, \citenamefont {Husain}, \citenamefont {Lee}, \citenamefont {Li}, \citenamefont {Alatas}, \citenamefont {Said}, \citenamefont {Yan}, \citenamefont {Ding}, \citenamefont {Zhao}, \citenamefont {Zhou}, \citenamefont {Devereaux}, \citenamefont {Wagner}, \citenamefont {Pemmaraju},\ and\ \citenamefont {Abbamonte}}]{Peng_PRM}%
  \BibitemOpen
  \bibfield  {author} {\bibinfo {author} {\bibfnamefont {Y.~Y.}\ \bibnamefont {Peng}}, \bibinfo {author} {\bibfnamefont {I.}~\bibnamefont {Boukahil}}, \bibinfo {author} {\bibfnamefont {K.}~\bibnamefont {Krongchon}}, \bibinfo {author} {\bibfnamefont {Q.}~\bibnamefont {Xiao}}, \bibinfo {author} {\bibfnamefont {A.~A.}\ \bibnamefont {Husain}}, \bibinfo {author} {\bibfnamefont {S.}~\bibnamefont {Lee}}, \bibinfo {author} {\bibfnamefont {Q.~Z.}\ \bibnamefont {Li}}, \bibinfo {author} {\bibfnamefont {A.}~\bibnamefont {Alatas}}, \bibinfo {author} {\bibfnamefont {A.~H.}\ \bibnamefont {Said}}, \bibinfo {author} {\bibfnamefont {H.~T.}\ \bibnamefont {Yan}}, \bibinfo {author} {\bibfnamefont {Y.}~\bibnamefont {Ding}}, \bibinfo {author} {\bibfnamefont {L.}~\bibnamefont {Zhao}}, \bibinfo {author} {\bibfnamefont {X.~J.}\ \bibnamefont {Zhou}}, \bibinfo {author} {\bibfnamefont {T.~P.}\ \bibnamefont {Devereaux}}, \bibinfo {author} {\bibfnamefont {L.~K.}\ \bibnamefont {Wagner}}, \bibinfo {author} {\bibfnamefont {C.~D.}\
  \bibnamefont {Pemmaraju}},\ and\ \bibinfo {author} {\bibfnamefont {P.}~\bibnamefont {Abbamonte}},\ }\bibfield  {title} {\bibinfo {title} {{Observation of van der Waals phonons in the single-layer cuprate \ch{(Bi,Pb)_2(Sr,La)_2CuO_{6+$\delta$}}}},\ }\href {https://doi.org/10.1103/PhysRevMaterials.8.024804} {\bibfield  {journal} {\bibinfo  {journal} {Phys. Rev. Mater.}\ }\textbf {\bibinfo {volume} {8}},\ \bibinfo {pages} {024804} (\bibinfo {year} {2024})}\BibitemShut {NoStop}%
\bibitem [{\citenamefont {Cardona}\ \emph {et~al.}(1988)\citenamefont {Cardona}, \citenamefont {Liu}, \citenamefont {Thomsen}, \citenamefont {Bauer}, \citenamefont {Genzel}, \citenamefont {K{\"o}nig}, \citenamefont {Wittlin}, \citenamefont {Amador}, \citenamefont {Barahona}, \citenamefont {Fernandez} \emph {et~al.}}]{Cardona1988}%
  \BibitemOpen
  \bibfield  {author} {\bibinfo {author} {\bibfnamefont {M.}~\bibnamefont {Cardona}}, \bibinfo {author} {\bibfnamefont {R.}~\bibnamefont {Liu}}, \bibinfo {author} {\bibfnamefont {C.}~\bibnamefont {Thomsen}}, \bibinfo {author} {\bibfnamefont {M.}~\bibnamefont {Bauer}}, \bibinfo {author} {\bibfnamefont {L.}~\bibnamefont {Genzel}}, \bibinfo {author} {\bibfnamefont {W.}~\bibnamefont {K{\"o}nig}}, \bibinfo {author} {\bibfnamefont {A.}~\bibnamefont {Wittlin}}, \bibinfo {author} {\bibfnamefont {U.}~\bibnamefont {Amador}}, \bibinfo {author} {\bibfnamefont {M.}~\bibnamefont {Barahona}}, \bibinfo {author} {\bibfnamefont {F.}~\bibnamefont {Fernandez}}, \emph {et~al.},\ }\bibfield  {title} {\bibinfo {title} {{Infrared and Raman spectra of the new superconducting cuprate perovskites \ch{MBa_2Cu_3O_7}}, \ch{M= Nd, Dy, Er, Tm}},\ }\href {https://doi.org/10.1016/0038-1098(88)90591-1} {\bibfield  {journal} {\bibinfo  {journal} {Solid state communications}\ }\textbf {\bibinfo {volume} {65}},\ \bibinfo {pages} {71} (\bibinfo
  {year} {1988})}\BibitemShut {NoStop}%
\bibitem [{\citenamefont {Motida}\ and\ \citenamefont {Suzuki}(1999)}]{Motida1999}%
  \BibitemOpen
  \bibfield  {author} {\bibinfo {author} {\bibfnamefont {K.}~\bibnamefont {Motida}}\ and\ \bibinfo {author} {\bibfnamefont {K.}~\bibnamefont {Suzuki}},\ }\bibfield  {title} {\bibinfo {title} {{Evidence for strong electron--phonon coupling in double-layered cuprate superconductors}},\ }\href {https://doi.org/10.1016/S0921-4526(98)01459-8} {\bibfield  {journal} {\bibinfo  {journal} {Physica B: Condensed Matter}\ }\textbf {\bibinfo {volume} {263}},\ \bibinfo {pages} {772} (\bibinfo {year} {1999})}\BibitemShut {NoStop}%
\bibitem [{\citenamefont {Zhang}\ and\ \citenamefont {Zhang}(2013)}]{Zhang2013}%
  \BibitemOpen
  \bibfield  {author} {\bibinfo {author} {\bibfnamefont {A.-M.}\ \bibnamefont {Zhang}}\ and\ \bibinfo {author} {\bibfnamefont {Q.-M.}\ \bibnamefont {Zhang}},\ }\bibfield  {title} {\bibinfo {title} {{Electron—phonon coupling in cuprate and iron-based superconductors revealed by Raman scattering}},\ }\href {https://iopscience.iop.org/article/10.1088/1674-1056/22/8/087103} {\bibfield  {journal} {\bibinfo  {journal} {Chinese Physics B}\ }\textbf {\bibinfo {volume} {22}},\ \bibinfo {pages} {087103} (\bibinfo {year} {2013})}\BibitemShut {NoStop}%
\bibitem [{\citenamefont {Farina}\ \emph {et~al.}(2018)\citenamefont {Farina}, \citenamefont {De~Filippis}, \citenamefont {Mishchenko}, \citenamefont {Nagaosa}, \citenamefont {Yang}, \citenamefont {Reznik}, \citenamefont {Wolf},\ and\ \citenamefont {Cataudella}}]{Farina_PRB}%
  \BibitemOpen
  \bibfield  {author} {\bibinfo {author} {\bibfnamefont {D.}~\bibnamefont {Farina}}, \bibinfo {author} {\bibfnamefont {G.}~\bibnamefont {De~Filippis}}, \bibinfo {author} {\bibfnamefont {A.~S.}\ \bibnamefont {Mishchenko}}, \bibinfo {author} {\bibfnamefont {N.}~\bibnamefont {Nagaosa}}, \bibinfo {author} {\bibfnamefont {J.-A.}\ \bibnamefont {Yang}}, \bibinfo {author} {\bibfnamefont {D.}~\bibnamefont {Reznik}}, \bibinfo {author} {\bibfnamefont {T.}~\bibnamefont {Wolf}},\ and\ \bibinfo {author} {\bibfnamefont {V.}~\bibnamefont {Cataudella}},\ }\bibfield  {title} {\bibinfo {title} {{Electron-phonon coupling in the undoped cuprate \ch{YBa_2Cu_3O_6} estimated from Raman and optical conductivity spectra}},\ }\href {https://doi.org/10.1103/PhysRevB.98.121104} {\bibfield  {journal} {\bibinfo  {journal} {Phys. Rev. B}\ }\textbf {\bibinfo {volume} {98}},\ \bibinfo {pages} {121104} (\bibinfo {year} {2018})}\BibitemShut {NoStop}%
\bibitem [{\citenamefont {Damascelli}\ \emph {et~al.}(2003)\citenamefont {Damascelli}, \citenamefont {Hussain},\ and\ \citenamefont {Shen}}]{Damascelli_RMP}%
  \BibitemOpen
  \bibfield  {author} {\bibinfo {author} {\bibfnamefont {A.}~\bibnamefont {Damascelli}}, \bibinfo {author} {\bibfnamefont {Z.}~\bibnamefont {Hussain}},\ and\ \bibinfo {author} {\bibfnamefont {Z.-X.}\ \bibnamefont {Shen}},\ }\bibfield  {title} {\bibinfo {title} {{Angle-resolved photoemission studies of the cuprate superconductors}},\ }\href {https://doi.org/10.1103/RevModPhys.75.473} {\bibfield  {journal} {\bibinfo  {journal} {Rev. Mod. Phys.}\ }\textbf {\bibinfo {volume} {75}},\ \bibinfo {pages} {473} (\bibinfo {year} {2003})}\BibitemShut {NoStop}%
\bibitem [{\citenamefont {Cuk}\ \emph {et~al.}(2005)\citenamefont {Cuk}, \citenamefont {Lu}, \citenamefont {Zhou}, \citenamefont {Shen}, \citenamefont {Devereaux},\ and\ \citenamefont {Nagaosa}}]{Cuk2005}%
  \BibitemOpen
  \bibfield  {author} {\bibinfo {author} {\bibfnamefont {T.}~\bibnamefont {Cuk}}, \bibinfo {author} {\bibfnamefont {D.}~\bibnamefont {Lu}}, \bibinfo {author} {\bibfnamefont {X.}~\bibnamefont {Zhou}}, \bibinfo {author} {\bibfnamefont {Z.-X.}\ \bibnamefont {Shen}}, \bibinfo {author} {\bibfnamefont {T.}~\bibnamefont {Devereaux}},\ and\ \bibinfo {author} {\bibfnamefont {N.}~\bibnamefont {Nagaosa}},\ }\bibfield  {title} {\bibinfo {title} {{A review of electron--phonon coupling seen in the high-Tc superconductors by angle-resolved photoemission studies (ARPES)}},\ }\href {https://doi.org/10.1002/pssb.200404959} {\bibfield  {journal} {\bibinfo  {journal} {physica status solidi (b)}\ }\textbf {\bibinfo {volume} {242}},\ \bibinfo {pages} {11} (\bibinfo {year} {2005})}\BibitemShut {NoStop}%
\bibitem [{\citenamefont {Lee}\ \emph {et~al.}(2006)\citenamefont {Lee}, \citenamefont {Fujita}, \citenamefont {McElroy}, \citenamefont {Slezak}, \citenamefont {Wang}, \citenamefont {Aiura}, \citenamefont {Bando}, \citenamefont {Ishikado}, \citenamefont {Masui}, \citenamefont {Zhu} \emph {et~al.}}]{JLee_Nature}%
  \BibitemOpen
  \bibfield  {author} {\bibinfo {author} {\bibfnamefont {J.}~\bibnamefont {Lee}}, \bibinfo {author} {\bibfnamefont {K.}~\bibnamefont {Fujita}}, \bibinfo {author} {\bibfnamefont {K.}~\bibnamefont {McElroy}}, \bibinfo {author} {\bibfnamefont {J.}~\bibnamefont {Slezak}}, \bibinfo {author} {\bibfnamefont {M.}~\bibnamefont {Wang}}, \bibinfo {author} {\bibfnamefont {Y.}~\bibnamefont {Aiura}}, \bibinfo {author} {\bibfnamefont {H.}~\bibnamefont {Bando}}, \bibinfo {author} {\bibfnamefont {M.}~\bibnamefont {Ishikado}}, \bibinfo {author} {\bibfnamefont {T.}~\bibnamefont {Masui}}, \bibinfo {author} {\bibfnamefont {J.-X.}\ \bibnamefont {Zhu}}, \emph {et~al.},\ }\bibfield  {title} {\bibinfo {title} {{Interplay of electron--lattice interactions and superconductivity in \ch{Bi_2Sr_2CaCu_2O_{8+ $\delta$}}}},\ }\href {https://doi.org/10.1038/nature04973} {\bibfield  {journal} {\bibinfo  {journal} {Nature}\ }\textbf {\bibinfo {volume} {442}},\ \bibinfo {pages} {546} (\bibinfo {year} {2006})}\BibitemShut {NoStop}%
\bibitem [{\citenamefont {Eickhoff}\ \emph {et~al.}(2020)\citenamefont {Eickhoff}, \citenamefont {Kolodzeiski}, \citenamefont {Esat}, \citenamefont {Fournier}, \citenamefont {Wagner}, \citenamefont {Deilmann}, \citenamefont {Temirov}, \citenamefont {Rohlfing}, \citenamefont {Tautz},\ and\ \citenamefont {Anders}}]{Eickhoff_PRB}%
  \BibitemOpen
  \bibfield  {author} {\bibinfo {author} {\bibfnamefont {F.}~\bibnamefont {Eickhoff}}, \bibinfo {author} {\bibfnamefont {E.}~\bibnamefont {Kolodzeiski}}, \bibinfo {author} {\bibfnamefont {T.}~\bibnamefont {Esat}}, \bibinfo {author} {\bibfnamefont {N.}~\bibnamefont {Fournier}}, \bibinfo {author} {\bibfnamefont {C.}~\bibnamefont {Wagner}}, \bibinfo {author} {\bibfnamefont {T.}~\bibnamefont {Deilmann}}, \bibinfo {author} {\bibfnamefont {R.}~\bibnamefont {Temirov}}, \bibinfo {author} {\bibfnamefont {M.}~\bibnamefont {Rohlfing}}, \bibinfo {author} {\bibfnamefont {F.~S.}\ \bibnamefont {Tautz}},\ and\ \bibinfo {author} {\bibfnamefont {F.~B.}\ \bibnamefont {Anders}},\ }\bibfield  {title} {\bibinfo {title} {{Inelastic electron tunneling spectroscopy for probing strongly correlated many-body systems by scanning tunneling microscopy}},\ }\href {https://doi.org/10.1103/PhysRevB.101.125405} {\bibfield  {journal} {\bibinfo  {journal} {Phys. Rev. B}\ }\textbf {\bibinfo {volume} {101}},\ \bibinfo {pages} {125405} (\bibinfo
  {year} {2020})}\BibitemShut {NoStop}%
\bibitem [{\citenamefont {Ament}\ \emph {et~al.}(2011)\citenamefont {Ament}, \citenamefont {van Veenendaal},\ and\ \citenamefont {Brink}}]{Ament_EPL}%
  \BibitemOpen
  \bibfield  {author} {\bibinfo {author} {\bibfnamefont {L.}~\bibnamefont {Ament}}, \bibinfo {author} {\bibfnamefont {M.}~\bibnamefont {van Veenendaal}},\ and\ \bibinfo {author} {\bibfnamefont {J.}~\bibnamefont {Brink}},\ }\bibfield  {title} {\bibinfo {title} {{Determining the electron-phonon coupling strength from Resonant Inelastic X-ray Scattering at transition metal L-edges}},\ }\href {https://doi.org/10.1209/0295-5075/95/27008} {\bibfield  {journal} {\bibinfo  {journal} {EPL (Europhysics Letters)}\ }\textbf {\bibinfo {volume} {95}},\ \bibinfo {pages} {27008} (\bibinfo {year} {2011})}\BibitemShut {NoStop}%
\bibitem [{\citenamefont {Devereaux}\ \emph {et~al.}(2016)\citenamefont {Devereaux}, \citenamefont {Shvaika}, \citenamefont {Wu}, \citenamefont {Wohlfeld}, \citenamefont {Jia}, \citenamefont {Wang}, \citenamefont {Moritz}, \citenamefont {Chaix}, \citenamefont {Lee}, \citenamefont {Shen} \emph {et~al.}}]{Devereaux2016_PRX}%
  \BibitemOpen
  \bibfield  {author} {\bibinfo {author} {\bibfnamefont {T.}~\bibnamefont {Devereaux}}, \bibinfo {author} {\bibfnamefont {A.}~\bibnamefont {Shvaika}}, \bibinfo {author} {\bibfnamefont {K.}~\bibnamefont {Wu}}, \bibinfo {author} {\bibfnamefont {K.}~\bibnamefont {Wohlfeld}}, \bibinfo {author} {\bibfnamefont {C.}~\bibnamefont {Jia}}, \bibinfo {author} {\bibfnamefont {Y.}~\bibnamefont {Wang}}, \bibinfo {author} {\bibfnamefont {B.}~\bibnamefont {Moritz}}, \bibinfo {author} {\bibfnamefont {L.}~\bibnamefont {Chaix}}, \bibinfo {author} {\bibfnamefont {W.-S.}\ \bibnamefont {Lee}}, \bibinfo {author} {\bibfnamefont {Z.-X.}\ \bibnamefont {Shen}}, \emph {et~al.},\ }\bibfield  {title} {\bibinfo {title} {{Directly characterizing the relative strength and momentum dependence of electron-phonon coupling using resonant inelastic x-ray scattering}},\ }\href {https://doi.org/10.1103/PhysRevX.6.041019} {\bibfield  {journal} {\bibinfo  {journal} {Physical Review X}\ }\textbf {\bibinfo {volume} {6}},\ \bibinfo {pages} {041019}
  (\bibinfo {year} {2016})}\BibitemShut {NoStop}%
\bibitem [{\citenamefont {Geondzhian}\ and\ \citenamefont {Gilmore}(2020)}]{Geondzhian_PRB}%
  \BibitemOpen
  \bibfield  {author} {\bibinfo {author} {\bibfnamefont {A.}~\bibnamefont {Geondzhian}}\ and\ \bibinfo {author} {\bibfnamefont {K.}~\bibnamefont {Gilmore}},\ }\bibfield  {title} {\bibinfo {title} {{Generalization of the Franck-Condon model for phonon excitations by resonant inelastic x-ray scattering}},\ }\href {https://doi.org/10.1103/PhysRevB.101.214307} {\bibfield  {journal} {\bibinfo  {journal} {Phys. Rev. B}\ }\textbf {\bibinfo {volume} {101}},\ \bibinfo {pages} {214307} (\bibinfo {year} {2020})}\BibitemShut {NoStop}%
\bibitem [{\citenamefont {Gilmore}(2023)}]{Gilmore2023}%
  \BibitemOpen
  \bibfield  {author} {\bibinfo {author} {\bibfnamefont {K.}~\bibnamefont {Gilmore}},\ }\bibfield  {title} {\bibinfo {title} {{Quantifying vibronic coupling with resonant inelastic X-ray scattering}},\ }\href {https://doi.org/10.1039/D2CP00968D} {\bibfield  {journal} {\bibinfo  {journal} {Physical Chemistry Chemical Physics}\ }\textbf {\bibinfo {volume} {25}},\ \bibinfo {pages} {217} (\bibinfo {year} {2023})}\BibitemShut {NoStop}%
\bibitem [{\citenamefont {Braicovich}\ \emph {et~al.}(2020)\citenamefont {Braicovich}, \citenamefont {Rossi}, \citenamefont {Fumagalli}, \citenamefont {Peng}, \citenamefont {Wang}, \citenamefont {Arpaia}, \citenamefont {Betto}, \citenamefont {De~Luca}, \citenamefont {Di~Castro}, \citenamefont {Kummer}, \citenamefont {Moretti~Sala}, \citenamefont {Pagetti}, \citenamefont {Balestrino}, \citenamefont {Brookes}, \citenamefont {Salluzzo}, \citenamefont {Johnston}, \citenamefont {van~den Brink},\ and\ \citenamefont {Ghiringhelli}}]{Braicovich_PRR}%
  \BibitemOpen
  \bibfield  {author} {\bibinfo {author} {\bibfnamefont {L.}~\bibnamefont {Braicovich}}, \bibinfo {author} {\bibfnamefont {M.}~\bibnamefont {Rossi}}, \bibinfo {author} {\bibfnamefont {R.}~\bibnamefont {Fumagalli}}, \bibinfo {author} {\bibfnamefont {Y.}~\bibnamefont {Peng}}, \bibinfo {author} {\bibfnamefont {Y.}~\bibnamefont {Wang}}, \bibinfo {author} {\bibfnamefont {R.}~\bibnamefont {Arpaia}}, \bibinfo {author} {\bibfnamefont {D.}~\bibnamefont {Betto}}, \bibinfo {author} {\bibfnamefont {G.~M.}\ \bibnamefont {De~Luca}}, \bibinfo {author} {\bibfnamefont {D.}~\bibnamefont {Di~Castro}}, \bibinfo {author} {\bibfnamefont {K.}~\bibnamefont {Kummer}}, \bibinfo {author} {\bibfnamefont {M.}~\bibnamefont {Moretti~Sala}}, \bibinfo {author} {\bibfnamefont {M.}~\bibnamefont {Pagetti}}, \bibinfo {author} {\bibfnamefont {G.}~\bibnamefont {Balestrino}}, \bibinfo {author} {\bibfnamefont {N.~B.}\ \bibnamefont {Brookes}}, \bibinfo {author} {\bibfnamefont {M.}~\bibnamefont {Salluzzo}}, \bibinfo {author} {\bibfnamefont
  {S.}~\bibnamefont {Johnston}}, \bibinfo {author} {\bibfnamefont {J.}~\bibnamefont {van~den Brink}},\ and\ \bibinfo {author} {\bibfnamefont {G.}~\bibnamefont {Ghiringhelli}},\ }\bibfield  {title} {\bibinfo {title} {{Determining the electron-phonon coupling in superconducting cuprates by resonant inelastic x-ray scattering: Methods and results on \ch{Nd_{1+x}Ba_{2-x}Cu_{3}O_{7-$\delta$}}}},\ }\href {https://doi.org/10.1103/PhysRevResearch.2.023231} {\bibfield  {journal} {\bibinfo  {journal} {Phys. Rev. Res.}\ }\textbf {\bibinfo {volume} {2}},\ \bibinfo {pages} {023231} (\bibinfo {year} {2020})}\BibitemShut {NoStop}%
\bibitem [{\citenamefont {Tanaka}\ and\ \citenamefont {Kayanuma}(2005)}]{Tanaka_PRB}%
  \BibitemOpen
  \bibfield  {author} {\bibinfo {author} {\bibfnamefont {S.}~\bibnamefont {Tanaka}}\ and\ \bibinfo {author} {\bibfnamefont {Y.}~\bibnamefont {Kayanuma}},\ }\bibfield  {title} {\bibinfo {title} {{Dynamics in resonant x-ray emission of the core exciton state: Competition between electron itinerancy and lattice relaxation}},\ }\href {https://doi.org/10.1103/PhysRevB.71.024302} {\bibfield  {journal} {\bibinfo  {journal} {Phys. Rev. B}\ }\textbf {\bibinfo {volume} {71}},\ \bibinfo {pages} {024302} (\bibinfo {year} {2005})}\BibitemShut {NoStop}%
\bibitem [{\citenamefont {Lee}\ \emph {et~al.}(2013)\citenamefont {Lee}, \citenamefont {Johnston}, \citenamefont {Moritz}, \citenamefont {Lee}, \citenamefont {Yi}, \citenamefont {Zhou}, \citenamefont {Schmitt}, \citenamefont {Patthey}, \citenamefont {Strocov}, \citenamefont {Kudo}, \citenamefont {Koike}, \citenamefont {van~den Brink}, \citenamefont {Devereaux},\ and\ \citenamefont {Shen}}]{WSLee_PRL13}%
  \BibitemOpen
  \bibfield  {author} {\bibinfo {author} {\bibfnamefont {W.~S.}\ \bibnamefont {Lee}}, \bibinfo {author} {\bibfnamefont {S.}~\bibnamefont {Johnston}}, \bibinfo {author} {\bibfnamefont {B.}~\bibnamefont {Moritz}}, \bibinfo {author} {\bibfnamefont {J.}~\bibnamefont {Lee}}, \bibinfo {author} {\bibfnamefont {M.}~\bibnamefont {Yi}}, \bibinfo {author} {\bibfnamefont {K.~J.}\ \bibnamefont {Zhou}}, \bibinfo {author} {\bibfnamefont {T.}~\bibnamefont {Schmitt}}, \bibinfo {author} {\bibfnamefont {L.}~\bibnamefont {Patthey}}, \bibinfo {author} {\bibfnamefont {V.}~\bibnamefont {Strocov}}, \bibinfo {author} {\bibfnamefont {K.}~\bibnamefont {Kudo}}, \bibinfo {author} {\bibfnamefont {Y.}~\bibnamefont {Koike}}, \bibinfo {author} {\bibfnamefont {J.}~\bibnamefont {van~den Brink}}, \bibinfo {author} {\bibfnamefont {T.~P.}\ \bibnamefont {Devereaux}},\ and\ \bibinfo {author} {\bibfnamefont {Z.~X.}\ \bibnamefont {Shen}},\ }\bibfield  {title} {\bibinfo {title} {{Role of Lattice Coupling in Establishing Electronic and Magnetic
  Properties in Quasi-One-Dimensional Cuprates}},\ }\href {https://doi.org/10.1103/PhysRevLett.110.265502} {\bibfield  {journal} {\bibinfo  {journal} {Phys. Rev. Lett.}\ }\textbf {\bibinfo {volume} {110}},\ \bibinfo {pages} {265502} (\bibinfo {year} {2013})}\BibitemShut {NoStop}%
\bibitem [{\citenamefont {Johnston}\ \emph {et~al.}(2016)\citenamefont {Johnston}, \citenamefont {Monney}, \citenamefont {Bisogni}, \citenamefont {Zhou}, \citenamefont {Kraus}, \citenamefont {Behr}, \citenamefont {Strocov}, \citenamefont {M{\'a}lek}, \citenamefont {Drechsler}, \citenamefont {Geck} \emph {et~al.}}]{Johnston_Nature2016}%
  \BibitemOpen
  \bibfield  {author} {\bibinfo {author} {\bibfnamefont {S.}~\bibnamefont {Johnston}}, \bibinfo {author} {\bibfnamefont {C.}~\bibnamefont {Monney}}, \bibinfo {author} {\bibfnamefont {V.}~\bibnamefont {Bisogni}}, \bibinfo {author} {\bibfnamefont {K.-J.}\ \bibnamefont {Zhou}}, \bibinfo {author} {\bibfnamefont {R.}~\bibnamefont {Kraus}}, \bibinfo {author} {\bibfnamefont {G.}~\bibnamefont {Behr}}, \bibinfo {author} {\bibfnamefont {V.~N.}\ \bibnamefont {Strocov}}, \bibinfo {author} {\bibfnamefont {J.}~\bibnamefont {M{\'a}lek}}, \bibinfo {author} {\bibfnamefont {S.-L.}\ \bibnamefont {Drechsler}}, \bibinfo {author} {\bibfnamefont {J.}~\bibnamefont {Geck}}, \emph {et~al.},\ }\bibfield  {title} {\bibinfo {title} {{Electron-lattice interactions strongly renormalize the charge-transfer energy in the spin-chain cuprate \ch{Li_2CuO_2}}},\ }\href {https://doi.org/https://doi.org/10.1038/ncomms10563} {\bibfield  {journal} {\bibinfo  {journal} {Nature Communications}\ }\textbf {\bibinfo {volume} {7}},\ \bibinfo {pages}
  {10563} (\bibinfo {year} {2016})}\BibitemShut {NoStop}%
\bibitem [{\citenamefont {Scott}\ \emph {et~al.}(2024)\citenamefont {Scott}, \citenamefont {Kisiel}, \citenamefont {Yakhou}, \citenamefont {Agrestini}, \citenamefont {Garcia-Fernandez}, \citenamefont {Kummer}, \citenamefont {Choi}, \citenamefont {Zhong}, \citenamefont {Schneeloch}, \citenamefont {Gu}, \citenamefont {Zhou}, \citenamefont {Brookes}, \citenamefont {Kemper}, \citenamefont {Minola}, \citenamefont {Boschini}, \citenamefont {Frano}, \citenamefont {Gozar},\ and\ \citenamefont {da~Silva~Neto}}]{Scott_PRB2024}%
  \BibitemOpen
  \bibfield  {author} {\bibinfo {author} {\bibfnamefont {K.}~\bibnamefont {Scott}}, \bibinfo {author} {\bibfnamefont {E.}~\bibnamefont {Kisiel}}, \bibinfo {author} {\bibfnamefont {F.}~\bibnamefont {Yakhou}}, \bibinfo {author} {\bibfnamefont {S.}~\bibnamefont {Agrestini}}, \bibinfo {author} {\bibfnamefont {M.}~\bibnamefont {Garcia-Fernandez}}, \bibinfo {author} {\bibfnamefont {K.}~\bibnamefont {Kummer}}, \bibinfo {author} {\bibfnamefont {J.}~\bibnamefont {Choi}}, \bibinfo {author} {\bibfnamefont {R.~D.}\ \bibnamefont {Zhong}}, \bibinfo {author} {\bibfnamefont {J.~A.}\ \bibnamefont {Schneeloch}}, \bibinfo {author} {\bibfnamefont {G.~D.}\ \bibnamefont {Gu}}, \bibinfo {author} {\bibfnamefont {K.-J.}\ \bibnamefont {Zhou}}, \bibinfo {author} {\bibfnamefont {N.~B.}\ \bibnamefont {Brookes}}, \bibinfo {author} {\bibfnamefont {A.~F.}\ \bibnamefont {Kemper}}, \bibinfo {author} {\bibfnamefont {M.}~\bibnamefont {Minola}}, \bibinfo {author} {\bibfnamefont {F.}~\bibnamefont {Boschini}}, \bibinfo {author} {\bibfnamefont
  {A.}~\bibnamefont {Frano}}, \bibinfo {author} {\bibfnamefont {A.}~\bibnamefont {Gozar}},\ and\ \bibinfo {author} {\bibfnamefont {E.~H.}\ \bibnamefont {da~Silva~Neto}},\ }\bibfield  {title} {\bibinfo {title} {{Detection of a two-phonon mode in a cuprate superconductor via polarimetric resonant inelastic x-ray scattering}},\ }\href {https://doi.org/10.1103/PhysRevB.109.125126} {\bibfield  {journal} {\bibinfo  {journal} {Phys. Rev. B}\ }\textbf {\bibinfo {volume} {109}},\ \bibinfo {pages} {125126} (\bibinfo {year} {2024})}\BibitemShut {NoStop}%
\bibitem [{\citenamefont {Dashwood}\ \emph {et~al.}(2021)\citenamefont {Dashwood}, \citenamefont {Geondzhian}, \citenamefont {Vale}, \citenamefont {Pakpour-Tabrizi}, \citenamefont {Howard}, \citenamefont {Faure}, \citenamefont {Veiga}, \citenamefont {Meyers}, \citenamefont {Chiuzb\ifmmode~\u{a}\else \u{a}\fi{}ian}, \citenamefont {Nicolaou}, \citenamefont {Jaouen}, \citenamefont {Jackman}, \citenamefont {Nag}, \citenamefont {Garc\'{\i}a-Fern\'andez}, \citenamefont {Zhou}, \citenamefont {Walters}, \citenamefont {Gilmore}, \citenamefont {McMorrow},\ and\ \citenamefont {Dean}}]{DashwoodPRX2024}%
  \BibitemOpen
  \bibfield  {author} {\bibinfo {author} {\bibfnamefont {C.~D.}\ \bibnamefont {Dashwood}}, \bibinfo {author} {\bibfnamefont {A.}~\bibnamefont {Geondzhian}}, \bibinfo {author} {\bibfnamefont {J.~G.}\ \bibnamefont {Vale}}, \bibinfo {author} {\bibfnamefont {A.~C.}\ \bibnamefont {Pakpour-Tabrizi}}, \bibinfo {author} {\bibfnamefont {C.~A.}\ \bibnamefont {Howard}}, \bibinfo {author} {\bibfnamefont {Q.}~\bibnamefont {Faure}}, \bibinfo {author} {\bibfnamefont {L.~S.~I.}\ \bibnamefont {Veiga}}, \bibinfo {author} {\bibfnamefont {D.}~\bibnamefont {Meyers}}, \bibinfo {author} {\bibfnamefont {S.~G.}\ \bibnamefont {Chiuzb\ifmmode~\u{a}\else \u{a}\fi{}ian}}, \bibinfo {author} {\bibfnamefont {A.}~\bibnamefont {Nicolaou}}, \bibinfo {author} {\bibfnamefont {N.}~\bibnamefont {Jaouen}}, \bibinfo {author} {\bibfnamefont {R.~B.}\ \bibnamefont {Jackman}}, \bibinfo {author} {\bibfnamefont {A.}~\bibnamefont {Nag}}, \bibinfo {author} {\bibfnamefont {M.}~\bibnamefont {Garc\'{\i}a-Fern\'andez}}, \bibinfo {author} {\bibfnamefont {K.-J.}\
  \bibnamefont {Zhou}}, \bibinfo {author} {\bibfnamefont {A.~C.}\ \bibnamefont {Walters}}, \bibinfo {author} {\bibfnamefont {K.}~\bibnamefont {Gilmore}}, \bibinfo {author} {\bibfnamefont {D.~F.}\ \bibnamefont {McMorrow}},\ and\ \bibinfo {author} {\bibfnamefont {M.~P.~M.}\ \bibnamefont {Dean}},\ }\bibfield  {title} {\bibinfo {title} {{Probing Electron-Phonon Interactions Away from the Fermi Level with Resonant Inelastic X-Ray Scattering}},\ }\href {https://doi.org/10.1103/PhysRevX.11.041052} {\bibfield  {journal} {\bibinfo  {journal} {Phys. Rev. X}\ }\textbf {\bibinfo {volume} {11}},\ \bibinfo {pages} {041052} (\bibinfo {year} {2021})}\BibitemShut {NoStop}%
\bibitem [{\citenamefont {Allen}(1972)}]{Allen_PRB}%
  \BibitemOpen
  \bibfield  {author} {\bibinfo {author} {\bibfnamefont {P.~B.}\ \bibnamefont {Allen}},\ }\bibfield  {title} {\bibinfo {title} {{Neutron Spectroscopy of Superconductors}},\ }\href {https://doi.org/10.1103/PhysRevB.6.2577} {\bibfield  {journal} {\bibinfo  {journal} {Phys. Rev. B}\ }\textbf {\bibinfo {volume} {6}},\ \bibinfo {pages} {2577} (\bibinfo {year} {1972})}\BibitemShut {NoStop}%
\bibitem [{\citenamefont {Rossi}\ \emph {et~al.}(2019)\citenamefont {Rossi}, \citenamefont {Arpaia}, \citenamefont {Fumagalli}, \citenamefont {Moretti~Sala}, \citenamefont {Betto}, \citenamefont {Kummer}, \citenamefont {De~Luca}, \citenamefont {van~den Brink}, \citenamefont {Salluzzo}, \citenamefont {Brookes}, \citenamefont {Braicovich},\ and\ \citenamefont {Ghiringhelli}}]{Rossi_PRL}%
  \BibitemOpen
  \bibfield  {author} {\bibinfo {author} {\bibfnamefont {M.}~\bibnamefont {Rossi}}, \bibinfo {author} {\bibfnamefont {R.}~\bibnamefont {Arpaia}}, \bibinfo {author} {\bibfnamefont {R.}~\bibnamefont {Fumagalli}}, \bibinfo {author} {\bibfnamefont {M.}~\bibnamefont {Moretti~Sala}}, \bibinfo {author} {\bibfnamefont {D.}~\bibnamefont {Betto}}, \bibinfo {author} {\bibfnamefont {K.}~\bibnamefont {Kummer}}, \bibinfo {author} {\bibfnamefont {G.~M.}\ \bibnamefont {De~Luca}}, \bibinfo {author} {\bibfnamefont {J.}~\bibnamefont {van~den Brink}}, \bibinfo {author} {\bibfnamefont {M.}~\bibnamefont {Salluzzo}}, \bibinfo {author} {\bibfnamefont {N.~B.}\ \bibnamefont {Brookes}}, \bibinfo {author} {\bibfnamefont {L.}~\bibnamefont {Braicovich}},\ and\ \bibinfo {author} {\bibfnamefont {G.}~\bibnamefont {Ghiringhelli}},\ }\bibfield  {title} {\bibinfo {title} {{Experimental Determination of Momentum-Resolved Electron-Phonon Coupling}},\ }\href {https://doi.org/10.1103/PhysRevLett.123.027001} {\bibfield  {journal} {\bibinfo
  {journal} {Phys. Rev. Lett.}\ }\textbf {\bibinfo {volume} {123}},\ \bibinfo {pages} {027001} (\bibinfo {year} {2019})}\BibitemShut {NoStop}%
\bibitem [{\citenamefont {De~Filippis}\ \emph {et~al.}(2012)\citenamefont {De~Filippis}, \citenamefont {Cataudella}, \citenamefont {Nowadnick}, \citenamefont {Devereaux}, \citenamefont {Mishchenko},\ and\ \citenamefont {Nagaosa}}]{De_Filippis_PRL}%
  \BibitemOpen
  \bibfield  {author} {\bibinfo {author} {\bibfnamefont {G.}~\bibnamefont {De~Filippis}}, \bibinfo {author} {\bibfnamefont {V.}~\bibnamefont {Cataudella}}, \bibinfo {author} {\bibfnamefont {E.~A.}\ \bibnamefont {Nowadnick}}, \bibinfo {author} {\bibfnamefont {T.~P.}\ \bibnamefont {Devereaux}}, \bibinfo {author} {\bibfnamefont {A.~S.}\ \bibnamefont {Mishchenko}},\ and\ \bibinfo {author} {\bibfnamefont {N.}~\bibnamefont {Nagaosa}},\ }\bibfield  {title} {\bibinfo {title} {{Quantum Dynamics of the Hubbard-Holstein Model in Equilibrium and Nonequilibrium: Application to Pump-Probe Phenomena}},\ }\href {https://doi.org/10.1103/PhysRevLett.109.176402} {\bibfield  {journal} {\bibinfo  {journal} {Phys. Rev. Lett.}\ }\textbf {\bibinfo {volume} {109}},\ \bibinfo {pages} {176402} (\bibinfo {year} {2012})}\BibitemShut {NoStop}%
\bibitem [{\citenamefont {R\"osch}\ and\ \citenamefont {Gunnarsson}(2004)}]{Rosch_PRB}%
  \BibitemOpen
  \bibfield  {author} {\bibinfo {author} {\bibfnamefont {O.}~\bibnamefont {R\"osch}}\ and\ \bibinfo {author} {\bibfnamefont {O.}~\bibnamefont {Gunnarsson}},\ }\bibfield  {title} {\bibinfo {title} {{Electron-phonon interaction in the three-band model}},\ }\href {https://doi.org/10.1103/PhysRevB.70.224518} {\bibfield  {journal} {\bibinfo  {journal} {Phys. Rev. B}\ }\textbf {\bibinfo {volume} {70}},\ \bibinfo {pages} {224518} (\bibinfo {year} {2004})}\BibitemShut {NoStop}%
\bibitem [{\citenamefont {Devereaux}\ \emph {et~al.}(2004)\citenamefont {Devereaux}, \citenamefont {Cuk}, \citenamefont {Shen},\ and\ \citenamefont {Nagaosa}}]{Devereaux_PRL_2004}%
  \BibitemOpen
  \bibfield  {author} {\bibinfo {author} {\bibfnamefont {T.~P.}\ \bibnamefont {Devereaux}}, \bibinfo {author} {\bibfnamefont {T.}~\bibnamefont {Cuk}}, \bibinfo {author} {\bibfnamefont {Z.-X.}\ \bibnamefont {Shen}},\ and\ \bibinfo {author} {\bibfnamefont {N.}~\bibnamefont {Nagaosa}},\ }\bibfield  {title} {\bibinfo {title} {{Anisotropic Electron-Phonon Interaction in the Cuprates}},\ }\href {https://doi.org/10.1103/PhysRevLett.93.117004} {\bibfield  {journal} {\bibinfo  {journal} {Phys. Rev. Lett.}\ }\textbf {\bibinfo {volume} {93}},\ \bibinfo {pages} {117004} (\bibinfo {year} {2004})}\BibitemShut {NoStop}%
\bibitem [{\citenamefont {Arpaia}\ \emph {et~al.}(2019)\citenamefont {Arpaia}, \citenamefont {Caprara}, \citenamefont {Fumagalli}, \citenamefont {De~Vecchi}, \citenamefont {Peng}, \citenamefont {Andersson}, \citenamefont {Betto}, \citenamefont {De~Luca}, \citenamefont {Brookes}, \citenamefont {Lombardi} \emph {et~al.}}]{ArpaiaScience}%
  \BibitemOpen
  \bibfield  {author} {\bibinfo {author} {\bibfnamefont {R.}~\bibnamefont {Arpaia}}, \bibinfo {author} {\bibfnamefont {S.}~\bibnamefont {Caprara}}, \bibinfo {author} {\bibfnamefont {R.}~\bibnamefont {Fumagalli}}, \bibinfo {author} {\bibfnamefont {G.}~\bibnamefont {De~Vecchi}}, \bibinfo {author} {\bibfnamefont {Y.}~\bibnamefont {Peng}}, \bibinfo {author} {\bibfnamefont {E.}~\bibnamefont {Andersson}}, \bibinfo {author} {\bibfnamefont {D.}~\bibnamefont {Betto}}, \bibinfo {author} {\bibfnamefont {G.}~\bibnamefont {De~Luca}}, \bibinfo {author} {\bibfnamefont {N.}~\bibnamefont {Brookes}}, \bibinfo {author} {\bibfnamefont {F.}~\bibnamefont {Lombardi}}, \emph {et~al.},\ }\bibfield  {title} {\bibinfo {title} {{Dynamical charge density fluctuations pervading the phase diagram of a Cu-based high-T c superconductor}},\ }\href {https://www.science.org/doi/10.1126/science.aav1315} {\bibfield  {journal} {\bibinfo  {journal} {Science}\ }\textbf {\bibinfo {volume} {365}},\ \bibinfo {pages} {906} (\bibinfo {year}
  {2019})}\BibitemShut {NoStop}%
\bibitem [{\citenamefont {Merzoni}\ \emph {et~al.}(2024)\citenamefont {Merzoni}, \citenamefont {Martinelli}, \citenamefont {Braicovich}, \citenamefont {Brookes}, \citenamefont {Lombardi}, \citenamefont {Rosa}, \citenamefont {Arpaia}, \citenamefont {Moretti~Sala},\ and\ \citenamefont {Ghiringhelli}}]{Merzoni_PRB}%
  \BibitemOpen
  \bibfield  {author} {\bibinfo {author} {\bibfnamefont {G.}~\bibnamefont {Merzoni}}, \bibinfo {author} {\bibfnamefont {L.}~\bibnamefont {Martinelli}}, \bibinfo {author} {\bibfnamefont {L.}~\bibnamefont {Braicovich}}, \bibinfo {author} {\bibfnamefont {N.~B.}\ \bibnamefont {Brookes}}, \bibinfo {author} {\bibfnamefont {F.}~\bibnamefont {Lombardi}}, \bibinfo {author} {\bibfnamefont {F.}~\bibnamefont {Rosa}}, \bibinfo {author} {\bibfnamefont {R.}~\bibnamefont {Arpaia}}, \bibinfo {author} {\bibfnamefont {M.}~\bibnamefont {Moretti~Sala}},\ and\ \bibinfo {author} {\bibfnamefont {G.}~\bibnamefont {Ghiringhelli}},\ }\bibfield  {title} {\bibinfo {title} {{Charge response function probed by resonant inelastic x-ray scattering: Signature of electronic gaps of \ch{YBa_2Cu_3O_{7-$\delta$}}}},\ }\href {https://doi.org/10.1103/PhysRevB.109.184506} {\bibfield  {journal} {\bibinfo  {journal} {Physical Review B}\ }\textbf {\bibinfo {volume} {109}},\ \bibinfo {pages} {184506} (\bibinfo {year} {2024})}\BibitemShut {NoStop}%
\bibitem [{\citenamefont {Hepting}\ \emph {et~al.}(2023)\citenamefont {Hepting}, \citenamefont {Boyko}, \citenamefont {Zimmermann}, \citenamefont {Bejas}, \citenamefont {Suyolcu}, \citenamefont {Puphal}, \citenamefont {Green}, \citenamefont {Zinni}, \citenamefont {Kim}, \citenamefont {Casa} \emph {et~al.}}]{Hepting_PRB}%
  \BibitemOpen
  \bibfield  {author} {\bibinfo {author} {\bibfnamefont {M.}~\bibnamefont {Hepting}}, \bibinfo {author} {\bibfnamefont {T.}~\bibnamefont {Boyko}}, \bibinfo {author} {\bibfnamefont {V.}~\bibnamefont {Zimmermann}}, \bibinfo {author} {\bibfnamefont {M.}~\bibnamefont {Bejas}}, \bibinfo {author} {\bibfnamefont {Y.}~\bibnamefont {Suyolcu}}, \bibinfo {author} {\bibfnamefont {P.}~\bibnamefont {Puphal}}, \bibinfo {author} {\bibfnamefont {R.}~\bibnamefont {Green}}, \bibinfo {author} {\bibfnamefont {L.}~\bibnamefont {Zinni}}, \bibinfo {author} {\bibfnamefont {J.}~\bibnamefont {Kim}}, \bibinfo {author} {\bibfnamefont {D.}~\bibnamefont {Casa}}, \emph {et~al.},\ }\bibfield  {title} {\bibinfo {title} {{Evolution of plasmon excitations across the phase diagram of the cuprate superconductor \ch{La_{2- x}Sr_xCuO_4}}},\ }\href {https://doi.org/10.1103/PhysRevB.107.214516} {\bibfield  {journal} {\bibinfo  {journal} {Physical Review B}\ }\textbf {\bibinfo {volume} {107}},\ \bibinfo {pages} {214516} (\bibinfo {year}
  {2023})}\BibitemShut {NoStop}%
\bibitem [{\citenamefont {Sala}\ \emph {et~al.}(2011)\citenamefont {Sala}, \citenamefont {Bisogni}, \citenamefont {Aruta}, \citenamefont {Balestrino}, \citenamefont {Berger}, \citenamefont {Brookes}, \citenamefont {De~Luca}, \citenamefont {Di~Castro}, \citenamefont {Grioni}, \citenamefont {Guarise} \emph {et~al.}}]{Moretti_NJP}%
  \BibitemOpen
  \bibfield  {author} {\bibinfo {author} {\bibfnamefont {M.~M.}\ \bibnamefont {Sala}}, \bibinfo {author} {\bibfnamefont {V.}~\bibnamefont {Bisogni}}, \bibinfo {author} {\bibfnamefont {C.}~\bibnamefont {Aruta}}, \bibinfo {author} {\bibfnamefont {G.}~\bibnamefont {Balestrino}}, \bibinfo {author} {\bibfnamefont {H.}~\bibnamefont {Berger}}, \bibinfo {author} {\bibfnamefont {N.}~\bibnamefont {Brookes}}, \bibinfo {author} {\bibfnamefont {G.}~\bibnamefont {De~Luca}}, \bibinfo {author} {\bibfnamefont {D.}~\bibnamefont {Di~Castro}}, \bibinfo {author} {\bibfnamefont {M.}~\bibnamefont {Grioni}}, \bibinfo {author} {\bibfnamefont {M.}~\bibnamefont {Guarise}}, \emph {et~al.},\ }\bibfield  {title} {\bibinfo {title} {{Energy and symmetry of dd excitations in undoped layered cuprates measured by Cu $L_3$ resonant inelastic x-ray scattering}},\ }\href {https://iopscience.iop.org/article/10.1088/1367-2630/13/4/043026} {\bibfield  {journal} {\bibinfo  {journal} {New Journal of Physics}\ }\textbf {\bibinfo {volume} {13}},\
  \bibinfo {pages} {043026} (\bibinfo {year} {2011})}\BibitemShut {NoStop}%
\bibitem [{\citenamefont {Martinelli}\ \emph {et~al.}(2024{\natexlab{a}})\citenamefont {Martinelli}, \citenamefont {Wohlfeld}, \citenamefont {Pelliciari}, \citenamefont {Arpaia}, \citenamefont {Brookes}, \citenamefont {Di~Castro}, \citenamefont {Fernandez}, \citenamefont {Kang}, \citenamefont {Krockenberger}, \citenamefont {Kummer}, \citenamefont {McNally}, \citenamefont {Paris}, \citenamefont {Schmitt}, \citenamefont {Yamamoto}, \citenamefont {Walters}, \citenamefont {Zhou}, \citenamefont {Braicovich}, \citenamefont {Comin}, \citenamefont {Sala}, \citenamefont {Devereaux}, \citenamefont {Daghofer},\ and\ \citenamefont {Ghiringhelli}}]{Martinelli_PRL2024}%
  \BibitemOpen
  \bibfield  {author} {\bibinfo {author} {\bibfnamefont {L.}~\bibnamefont {Martinelli}}, \bibinfo {author} {\bibfnamefont {K.}~\bibnamefont {Wohlfeld}}, \bibinfo {author} {\bibfnamefont {J.}~\bibnamefont {Pelliciari}}, \bibinfo {author} {\bibfnamefont {R.}~\bibnamefont {Arpaia}}, \bibinfo {author} {\bibfnamefont {N.~B.}\ \bibnamefont {Brookes}}, \bibinfo {author} {\bibfnamefont {D.}~\bibnamefont {Di~Castro}}, \bibinfo {author} {\bibfnamefont {M.~G.}\ \bibnamefont {Fernandez}}, \bibinfo {author} {\bibfnamefont {M.}~\bibnamefont {Kang}}, \bibinfo {author} {\bibfnamefont {Y.}~\bibnamefont {Krockenberger}}, \bibinfo {author} {\bibfnamefont {K.}~\bibnamefont {Kummer}}, \bibinfo {author} {\bibfnamefont {D.~E.}\ \bibnamefont {McNally}}, \bibinfo {author} {\bibfnamefont {E.}~\bibnamefont {Paris}}, \bibinfo {author} {\bibfnamefont {T.}~\bibnamefont {Schmitt}}, \bibinfo {author} {\bibfnamefont {H.}~\bibnamefont {Yamamoto}}, \bibinfo {author} {\bibfnamefont {A.}~\bibnamefont {Walters}}, \bibinfo {author} {\bibfnamefont
  {K.-J.}\ \bibnamefont {Zhou}}, \bibinfo {author} {\bibfnamefont {L.}~\bibnamefont {Braicovich}}, \bibinfo {author} {\bibfnamefont {R.}~\bibnamefont {Comin}}, \bibinfo {author} {\bibfnamefont {M.~M.}\ \bibnamefont {Sala}}, \bibinfo {author} {\bibfnamefont {T.~P.}\ \bibnamefont {Devereaux}}, \bibinfo {author} {\bibfnamefont {M.}~\bibnamefont {Daghofer}},\ and\ \bibinfo {author} {\bibfnamefont {G.}~\bibnamefont {Ghiringhelli}},\ }\bibfield  {title} {\bibinfo {title} {{Collective Nature of Orbital Excitations in Layered Cuprates in the Absence of Apical Oxygens}},\ }\href {https://doi.org/10.1103/PhysRevLett.132.066004} {\bibfield  {journal} {\bibinfo  {journal} {Phys. Rev. Lett.}\ }\textbf {\bibinfo {volume} {132}},\ \bibinfo {pages} {066004} (\bibinfo {year} {2024}{\natexlab{a}})}\BibitemShut {NoStop}%
\bibitem [{\citenamefont {Braicovich}\ \emph {et~al.}(2010)\citenamefont {Braicovich}, \citenamefont {Van Den~Brink}, \citenamefont {Bisogni}, \citenamefont {Sala}, \citenamefont {Ament}, \citenamefont {Brookes}, \citenamefont {De~Luca}, \citenamefont {Salluzzo}, \citenamefont {Schmitt}, \citenamefont {Strocov} \emph {et~al.}}]{Braicovich2010_PRL}%
  \BibitemOpen
  \bibfield  {author} {\bibinfo {author} {\bibfnamefont {L.}~\bibnamefont {Braicovich}}, \bibinfo {author} {\bibfnamefont {J.}~\bibnamefont {Van Den~Brink}}, \bibinfo {author} {\bibfnamefont {V.}~\bibnamefont {Bisogni}}, \bibinfo {author} {\bibfnamefont {M.~M.}\ \bibnamefont {Sala}}, \bibinfo {author} {\bibfnamefont {L.}~\bibnamefont {Ament}}, \bibinfo {author} {\bibfnamefont {N.}~\bibnamefont {Brookes}}, \bibinfo {author} {\bibfnamefont {G.}~\bibnamefont {De~Luca}}, \bibinfo {author} {\bibfnamefont {M.}~\bibnamefont {Salluzzo}}, \bibinfo {author} {\bibfnamefont {T.}~\bibnamefont {Schmitt}}, \bibinfo {author} {\bibfnamefont {V.}~\bibnamefont {Strocov}}, \emph {et~al.},\ }\bibfield  {title} {\bibinfo {title} {{Magnetic excitations and phase separation in the underdoped \ch{La_{2-x}Sr_xCuO_4} superconductor measured by resonant inelastic X-ray scattering}},\ }\href {https://doi.org/10.1103/PhysRevLett.104.077002} {\bibfield  {journal} {\bibinfo  {journal} {Physical review letters}\ }\textbf {\bibinfo {volume}
  {104}},\ \bibinfo {pages} {077002} (\bibinfo {year} {2010})}\BibitemShut {NoStop}%
\bibitem [{\citenamefont {Lee}\ \emph {et~al.}(2014)\citenamefont {Lee}, \citenamefont {Moritz}, \citenamefont {Lee}, \citenamefont {Yi}, \citenamefont {Jia}, \citenamefont {Sorini}, \citenamefont {Kudo}, \citenamefont {Koike}, \citenamefont {Zhou}, \citenamefont {Monney}, \citenamefont {Strocov}, \citenamefont {Patthey}, \citenamefont {Schmitt}, \citenamefont {Devereaux},\ and\ \citenamefont {Shen}}]{JJLee_PRB}%
  \BibitemOpen
  \bibfield  {author} {\bibinfo {author} {\bibfnamefont {J.~J.}\ \bibnamefont {Lee}}, \bibinfo {author} {\bibfnamefont {B.}~\bibnamefont {Moritz}}, \bibinfo {author} {\bibfnamefont {W.~S.}\ \bibnamefont {Lee}}, \bibinfo {author} {\bibfnamefont {M.}~\bibnamefont {Yi}}, \bibinfo {author} {\bibfnamefont {C.~J.}\ \bibnamefont {Jia}}, \bibinfo {author} {\bibfnamefont {A.~P.}\ \bibnamefont {Sorini}}, \bibinfo {author} {\bibfnamefont {K.}~\bibnamefont {Kudo}}, \bibinfo {author} {\bibfnamefont {Y.}~\bibnamefont {Koike}}, \bibinfo {author} {\bibfnamefont {K.~J.}\ \bibnamefont {Zhou}}, \bibinfo {author} {\bibfnamefont {C.}~\bibnamefont {Monney}}, \bibinfo {author} {\bibfnamefont {V.}~\bibnamefont {Strocov}}, \bibinfo {author} {\bibfnamefont {L.}~\bibnamefont {Patthey}}, \bibinfo {author} {\bibfnamefont {T.}~\bibnamefont {Schmitt}}, \bibinfo {author} {\bibfnamefont {T.~P.}\ \bibnamefont {Devereaux}},\ and\ \bibinfo {author} {\bibfnamefont {Z.~X.}\ \bibnamefont {Shen}},\ }\bibfield  {title} {\bibinfo {title}
  {{Charge-orbital-lattice coupling effects in the $dd$ excitation profile of one-dimensional cuprates}},\ }\href {https://doi.org/10.1103/PhysRevB.89.041104} {\bibfield  {journal} {\bibinfo  {journal} {Phys. Rev. B}\ }\textbf {\bibinfo {volume} {89}},\ \bibinfo {pages} {041104} (\bibinfo {year} {2014})}\BibitemShut {NoStop}%
\bibitem [{\citenamefont {Chaix}\ \emph {et~al.}(2017)\citenamefont {Chaix}, \citenamefont {Ghiringhelli}, \citenamefont {Peng}, \citenamefont {Hashimoto}, \citenamefont {Moritz}, \citenamefont {Kummer}, \citenamefont {Brookes}, \citenamefont {He}, \citenamefont {Chen}, \citenamefont {Ishida} \emph {et~al.}}]{Chaix2017}%
  \BibitemOpen
  \bibfield  {author} {\bibinfo {author} {\bibfnamefont {L.}~\bibnamefont {Chaix}}, \bibinfo {author} {\bibfnamefont {G.}~\bibnamefont {Ghiringhelli}}, \bibinfo {author} {\bibfnamefont {Y.}~\bibnamefont {Peng}}, \bibinfo {author} {\bibfnamefont {M.}~\bibnamefont {Hashimoto}}, \bibinfo {author} {\bibfnamefont {B.}~\bibnamefont {Moritz}}, \bibinfo {author} {\bibfnamefont {K.}~\bibnamefont {Kummer}}, \bibinfo {author} {\bibfnamefont {N.~B.}\ \bibnamefont {Brookes}}, \bibinfo {author} {\bibfnamefont {Y.}~\bibnamefont {He}}, \bibinfo {author} {\bibfnamefont {S.}~\bibnamefont {Chen}}, \bibinfo {author} {\bibfnamefont {S.}~\bibnamefont {Ishida}}, \emph {et~al.},\ }\bibfield  {title} {\bibinfo {title} {{Dispersive charge density wave excitations in \ch{Bi_2Sr_2CaCu2O_{8+$\delta$}}}},\ }\href {https://doi.org/10.1038/nphys4157} {\bibfield  {journal} {\bibinfo  {journal} {Nature Physics}\ }\textbf {\bibinfo {volume} {13}},\ \bibinfo {pages} {952} (\bibinfo {year} {2017})}\BibitemShut {NoStop}%
\bibitem [{\citenamefont {Ament}(2010)}]{AmentPhD}%
  \BibitemOpen
  \bibfield  {author} {\bibinfo {author} {\bibfnamefont {L.}~\bibnamefont {Ament}},\ }\emph {\bibinfo {title} {{Resonant Inelastic X-ray Scattering Studies of Elementary Excitations}}},\ \href {https://hdl.handle.net/1887/16138} {Ph.D. thesis},\ \bibinfo  {school} {Leiden Institute of Physics (LION), Faculty of Science, Leiden University} (\bibinfo {year} {2010})\BibitemShut {NoStop}%
\bibitem [{\citenamefont {Bieniasz}\ \emph {et~al.}(2022)\citenamefont {Bieniasz}, \citenamefont {Johnston},\ and\ \citenamefont {Berciu}}]{BieniaszPRB2022}%
  \BibitemOpen
  \bibfield  {author} {\bibinfo {author} {\bibfnamefont {K.}~\bibnamefont {Bieniasz}}, \bibinfo {author} {\bibfnamefont {S.}~\bibnamefont {Johnston}},\ and\ \bibinfo {author} {\bibfnamefont {M.}~\bibnamefont {Berciu}},\ }\bibfield  {title} {\bibinfo {title} {{Theory of dispersive optical phonons in resonant inelastic x-ray scattering experiments}},\ }\href {https://doi.org/10.1103/PhysRevB.105.L180302} {\bibfield  {journal} {\bibinfo  {journal} {Phys. Rev. B}\ }\textbf {\bibinfo {volume} {105}},\ \bibinfo {pages} {L180302} (\bibinfo {year} {2022})}\BibitemShut {NoStop}%
\bibitem [{\citenamefont {Peng}\ \emph {et~al.}(2020)\citenamefont {Peng}, \citenamefont {Husain}, \citenamefont {Mitrano}, \citenamefont {Sun}, \citenamefont {Johnson}, \citenamefont {Zakrzewski}, \citenamefont {MacDougall}, \citenamefont {Barbour}, \citenamefont {Jarrige}, \citenamefont {Bisogni},\ and\ \citenamefont {Abbamonte}}]{Peng_PRL}%
  \BibitemOpen
  \bibfield  {author} {\bibinfo {author} {\bibfnamefont {Y.~Y.}\ \bibnamefont {Peng}}, \bibinfo {author} {\bibfnamefont {A.~A.}\ \bibnamefont {Husain}}, \bibinfo {author} {\bibfnamefont {M.}~\bibnamefont {Mitrano}}, \bibinfo {author} {\bibfnamefont {S.~X.-L.}\ \bibnamefont {Sun}}, \bibinfo {author} {\bibfnamefont {T.~A.}\ \bibnamefont {Johnson}}, \bibinfo {author} {\bibfnamefont {A.~V.}\ \bibnamefont {Zakrzewski}}, \bibinfo {author} {\bibfnamefont {G.~J.}\ \bibnamefont {MacDougall}}, \bibinfo {author} {\bibfnamefont {A.}~\bibnamefont {Barbour}}, \bibinfo {author} {\bibfnamefont {I.}~\bibnamefont {Jarrige}}, \bibinfo {author} {\bibfnamefont {V.}~\bibnamefont {Bisogni}},\ and\ \bibinfo {author} {\bibfnamefont {P.}~\bibnamefont {Abbamonte}},\ }\bibfield  {title} {\bibinfo {title} {{Enhanced Electron-Phonon Coupling for Charge-Density-Wave Formation in \ch{La_{1.8-x}Eu_{0.2}Sr_{x}CuO_{4+$\delta$}}}},\ }\href {https://doi.org/10.1103/PhysRevLett.125.097002} {\bibfield  {journal} {\bibinfo  {journal} {Phys. Rev.
  Lett.}\ }\textbf {\bibinfo {volume} {125}},\ \bibinfo {pages} {097002} (\bibinfo {year} {2020})}\BibitemShut {NoStop}%
\bibitem [{\citenamefont {Chaix}\ \emph {et~al.}(2022)\citenamefont {Chaix}, \citenamefont {Lebert}, \citenamefont {Miao}, \citenamefont {Nicolaou}, \citenamefont {Yakhou}, \citenamefont {Cercellier}, \citenamefont {Grenier}, \citenamefont {Brookes}, \citenamefont {Sulpice}, \citenamefont {Tsutsui}, \citenamefont {Bosak}, \citenamefont {Paolasini}, \citenamefont {Santos-Cottin}, \citenamefont {Yamamoto}, \citenamefont {Yamada}, \citenamefont {Azuma}, \citenamefont {Nishikubo}, \citenamefont {Yamamoto}, \citenamefont {Katsumata}, \citenamefont {Dean},\ and\ \citenamefont {d'Astuto}}]{ChaixPRR2022}%
  \BibitemOpen
  \bibfield  {author} {\bibinfo {author} {\bibfnamefont {L.}~\bibnamefont {Chaix}}, \bibinfo {author} {\bibfnamefont {B.}~\bibnamefont {Lebert}}, \bibinfo {author} {\bibfnamefont {H.}~\bibnamefont {Miao}}, \bibinfo {author} {\bibfnamefont {A.}~\bibnamefont {Nicolaou}}, \bibinfo {author} {\bibfnamefont {F.}~\bibnamefont {Yakhou}}, \bibinfo {author} {\bibfnamefont {H.}~\bibnamefont {Cercellier}}, \bibinfo {author} {\bibfnamefont {S.}~\bibnamefont {Grenier}}, \bibinfo {author} {\bibfnamefont {N.~B.}\ \bibnamefont {Brookes}}, \bibinfo {author} {\bibfnamefont {A.}~\bibnamefont {Sulpice}}, \bibinfo {author} {\bibfnamefont {S.}~\bibnamefont {Tsutsui}}, \bibinfo {author} {\bibfnamefont {A.}~\bibnamefont {Bosak}}, \bibinfo {author} {\bibfnamefont {L.}~\bibnamefont {Paolasini}}, \bibinfo {author} {\bibfnamefont {D.}~\bibnamefont {Santos-Cottin}}, \bibinfo {author} {\bibfnamefont {H.}~\bibnamefont {Yamamoto}}, \bibinfo {author} {\bibfnamefont {I.}~\bibnamefont {Yamada}}, \bibinfo {author} {\bibfnamefont
  {M.}~\bibnamefont {Azuma}}, \bibinfo {author} {\bibfnamefont {T.}~\bibnamefont {Nishikubo}}, \bibinfo {author} {\bibfnamefont {T.}~\bibnamefont {Yamamoto}}, \bibinfo {author} {\bibfnamefont {M.}~\bibnamefont {Katsumata}}, \bibinfo {author} {\bibfnamefont {M.~P.~M.}\ \bibnamefont {Dean}},\ and\ \bibinfo {author} {\bibfnamefont {M.}~\bibnamefont {d'Astuto}},\ }\bibfield  {title} {\bibinfo {title} {{Bulk charge density wave and electron-phonon coupling in superconducting copper oxychlorides}},\ }\href {https://doi.org/10.1103/PhysRevResearch.4.033004} {\bibfield  {journal} {\bibinfo  {journal} {Phys. Rev. Res.}\ }\textbf {\bibinfo {volume} {4}},\ \bibinfo {pages} {033004} (\bibinfo {year} {2022})}\BibitemShut {NoStop}%
\bibitem [{\citenamefont {Wang}\ \emph {et~al.}(2021)\citenamefont {Wang}, \citenamefont {von Arx}, \citenamefont {Horio}, \citenamefont {Mukkattukavil}, \citenamefont {K{\"u}spert}, \citenamefont {Sassa}, \citenamefont {Schmitt}, \citenamefont {Nag}, \citenamefont {Pyon}, \citenamefont {Takayama} \emph {et~al.}}]{Wang2021}%
  \BibitemOpen
  \bibfield  {author} {\bibinfo {author} {\bibfnamefont {Q.}~\bibnamefont {Wang}}, \bibinfo {author} {\bibfnamefont {K.}~\bibnamefont {von Arx}}, \bibinfo {author} {\bibfnamefont {M.}~\bibnamefont {Horio}}, \bibinfo {author} {\bibfnamefont {D.~J.}\ \bibnamefont {Mukkattukavil}}, \bibinfo {author} {\bibfnamefont {J.}~\bibnamefont {K{\"u}spert}}, \bibinfo {author} {\bibfnamefont {Y.}~\bibnamefont {Sassa}}, \bibinfo {author} {\bibfnamefont {T.}~\bibnamefont {Schmitt}}, \bibinfo {author} {\bibfnamefont {A.}~\bibnamefont {Nag}}, \bibinfo {author} {\bibfnamefont {S.}~\bibnamefont {Pyon}}, \bibinfo {author} {\bibfnamefont {T.}~\bibnamefont {Takayama}}, \emph {et~al.},\ }\bibfield  {title} {\bibinfo {title} {{Charge order lock-in by electron-phonon coupling in \ch{La_{1.675}Eu_{0.2}Sr_{0.125}CuO_4}}},\ }\href {https://www.science.org/doi/10.1126/sciadv.abg7394} {\bibfield  {journal} {\bibinfo  {journal} {Science Advances}\ }\textbf {\bibinfo {volume} {7}},\ \bibinfo {pages} {eabg7394} (\bibinfo {year}
  {2021})}\BibitemShut {NoStop}%
\bibitem [{\citenamefont {Huang}\ \emph {et~al.}(2021)\citenamefont {Huang}, \citenamefont {Singh}, \citenamefont {Mou}, \citenamefont {Johnston}, \citenamefont {Kemper}, \citenamefont {van~den Brink}, \citenamefont {Chen}, \citenamefont {Lee}, \citenamefont {Okamoto}, \citenamefont {Chu}, \citenamefont {Li}, \citenamefont {Komiya}, \citenamefont {Komarek}, \citenamefont {Fujimori}, \citenamefont {Chen},\ and\ \citenamefont {Huang}}]{Huang_PRX2021}%
  \BibitemOpen
  \bibfield  {author} {\bibinfo {author} {\bibfnamefont {H.~Y.}\ \bibnamefont {Huang}}, \bibinfo {author} {\bibfnamefont {A.}~\bibnamefont {Singh}}, \bibinfo {author} {\bibfnamefont {C.~Y.}\ \bibnamefont {Mou}}, \bibinfo {author} {\bibfnamefont {S.}~\bibnamefont {Johnston}}, \bibinfo {author} {\bibfnamefont {A.~F.}\ \bibnamefont {Kemper}}, \bibinfo {author} {\bibfnamefont {J.}~\bibnamefont {van~den Brink}}, \bibinfo {author} {\bibfnamefont {P.~J.}\ \bibnamefont {Chen}}, \bibinfo {author} {\bibfnamefont {T.~K.}\ \bibnamefont {Lee}}, \bibinfo {author} {\bibfnamefont {J.}~\bibnamefont {Okamoto}}, \bibinfo {author} {\bibfnamefont {Y.~Y.}\ \bibnamefont {Chu}}, \bibinfo {author} {\bibfnamefont {J.~H.}\ \bibnamefont {Li}}, \bibinfo {author} {\bibfnamefont {S.}~\bibnamefont {Komiya}}, \bibinfo {author} {\bibfnamefont {A.~C.}\ \bibnamefont {Komarek}}, \bibinfo {author} {\bibfnamefont {A.}~\bibnamefont {Fujimori}}, \bibinfo {author} {\bibfnamefont {C.~T.}\ \bibnamefont {Chen}},\ and\ \bibinfo {author} {\bibfnamefont
  {D.~J.}\ \bibnamefont {Huang}},\ }\bibfield  {title} {\bibinfo {title} {{Quantum Fluctuations of Charge Order Induce Phonon Softening in a Superconducting Cuprate}},\ }\href {https://doi.org/10.1103/PhysRevX.11.041038} {\bibfield  {journal} {\bibinfo  {journal} {Phys. Rev. X}\ }\textbf {\bibinfo {volume} {11}},\ \bibinfo {pages} {041038} (\bibinfo {year} {2021})}\BibitemShut {NoStop}%
\bibitem [{\citenamefont {Scott}\ \emph {et~al.}(2023)\citenamefont {Scott}, \citenamefont {Kisiel}, \citenamefont {Boyle}, \citenamefont {Basak}, \citenamefont {Jargot}, \citenamefont {Das}, \citenamefont {Agrestini}, \citenamefont {Garcia-Fernandez}, \citenamefont {Choi}, \citenamefont {Pelliciari}, \citenamefont {Li}, \citenamefont {Chuang}, \citenamefont {Zhong}, \citenamefont {Schneeloch}, \citenamefont {Gu}, \citenamefont {Légaré}, \citenamefont {Kemper}, \citenamefont {Zhou}, \citenamefont {Bisogni}, \citenamefont {Blanco-Canosa}, \citenamefont {Frano}, \citenamefont {Boschini},\ and\ \citenamefont {da~Silva~Neto}}]{Scott_SciAdv2023}%
  \BibitemOpen
  \bibfield  {author} {\bibinfo {author} {\bibfnamefont {K.}~\bibnamefont {Scott}}, \bibinfo {author} {\bibfnamefont {E.}~\bibnamefont {Kisiel}}, \bibinfo {author} {\bibfnamefont {T.~J.}\ \bibnamefont {Boyle}}, \bibinfo {author} {\bibfnamefont {R.}~\bibnamefont {Basak}}, \bibinfo {author} {\bibfnamefont {G.}~\bibnamefont {Jargot}}, \bibinfo {author} {\bibfnamefont {S.}~\bibnamefont {Das}}, \bibinfo {author} {\bibfnamefont {S.}~\bibnamefont {Agrestini}}, \bibinfo {author} {\bibfnamefont {M.}~\bibnamefont {Garcia-Fernandez}}, \bibinfo {author} {\bibfnamefont {J.}~\bibnamefont {Choi}}, \bibinfo {author} {\bibfnamefont {J.}~\bibnamefont {Pelliciari}}, \bibinfo {author} {\bibfnamefont {J.}~\bibnamefont {Li}}, \bibinfo {author} {\bibfnamefont {Y.-D.}\ \bibnamefont {Chuang}}, \bibinfo {author} {\bibfnamefont {R.}~\bibnamefont {Zhong}}, \bibinfo {author} {\bibfnamefont {J.~A.}\ \bibnamefont {Schneeloch}}, \bibinfo {author} {\bibfnamefont {G.}~\bibnamefont {Gu}}, \bibinfo {author} {\bibfnamefont {F.}~\bibnamefont
  {Légaré}}, \bibinfo {author} {\bibfnamefont {A.~F.}\ \bibnamefont {Kemper}}, \bibinfo {author} {\bibfnamefont {K.-J.}\ \bibnamefont {Zhou}}, \bibinfo {author} {\bibfnamefont {V.}~\bibnamefont {Bisogni}}, \bibinfo {author} {\bibfnamefont {S.}~\bibnamefont {Blanco-Canosa}}, \bibinfo {author} {\bibfnamefont {A.}~\bibnamefont {Frano}}, \bibinfo {author} {\bibfnamefont {F.}~\bibnamefont {Boschini}},\ and\ \bibinfo {author} {\bibfnamefont {E.~H.}\ \bibnamefont {da~Silva~Neto}},\ }\bibfield  {title} {\bibinfo {title} {{Low-energy quasi-circular electron correlations with charge order wavelength in Bi$_2$Sr$_2$CaCu$_2$O$_{8+\delta}$}},\ }\href {https://doi.org/10.1126/sciadv.adg3710} {\bibfield  {journal} {\bibinfo  {journal} {Science Advances}\ }\textbf {\bibinfo {volume} {9}},\ \bibinfo {pages} {eadg3710} (\bibinfo {year} {2023})},\ \Eprint {https://arxiv.org/abs/https://www.science.org/doi/pdf/10.1126/sciadv.adg3710} {https://www.science.org/doi/pdf/10.1126/sciadv.adg3710} \BibitemShut {NoStop}%
\bibitem [{\citenamefont {Keski-Rahkonen}\ and\ \citenamefont {Krause}(1974)}]{keski1974total}%
  \BibitemOpen
  \bibfield  {author} {\bibinfo {author} {\bibfnamefont {O.}~\bibnamefont {Keski-Rahkonen}}\ and\ \bibinfo {author} {\bibfnamefont {M.~O.}\ \bibnamefont {Krause}},\ }\bibfield  {title} {\bibinfo {title} {{Total and partial atomic-level widths}},\ }\href {https://doi.org/10.1016/0038-1098(82)90154-5} {\bibfield  {journal} {\bibinfo  {journal} {Atomic Data and Nuclear Data Tables}\ }\textbf {\bibinfo {volume} {14}},\ \bibinfo {pages} {139} (\bibinfo {year} {1974})}\BibitemShut {NoStop}%
\bibitem [{\citenamefont {Krause}\ and\ \citenamefont {Oliver}(1979)}]{krause1979natural}%
  \BibitemOpen
  \bibfield  {author} {\bibinfo {author} {\bibfnamefont {M.~O.}\ \bibnamefont {Krause}}\ and\ \bibinfo {author} {\bibfnamefont {J.}~\bibnamefont {Oliver}},\ }\bibfield  {title} {\bibinfo {title} {{Natural widths of atomic K and L levels, K $\alpha$ X-ray lines and several KLL Auger lines}},\ }\href {https://doi.org/10.1063/1.555595} {\bibfield  {journal} {\bibinfo  {journal} {Journal of Physical and Chemical Reference Data}\ }\textbf {\bibinfo {volume} {8}},\ \bibinfo {pages} {329} (\bibinfo {year} {1979})}\BibitemShut {NoStop}%
\bibitem [{\citenamefont {M{\"u}ller}\ \emph {et~al.}(1982)\citenamefont {M{\"u}ller}, \citenamefont {Jepsen},\ and\ \citenamefont {Wilkins}}]{muller1982x}%
  \BibitemOpen
  \bibfield  {author} {\bibinfo {author} {\bibfnamefont {J.}~\bibnamefont {M{\"u}ller}}, \bibinfo {author} {\bibfnamefont {O.}~\bibnamefont {Jepsen}},\ and\ \bibinfo {author} {\bibfnamefont {J.}~\bibnamefont {Wilkins}},\ }\bibfield  {title} {\bibinfo {title} {{X-ray absorption spectra: K-edges of 3d transition metals, L-edges of 3d and 4d metals, and M-edges of palladium}},\ }\href {https://doi.org/10.1016/0038-1098(82)90154-5} {\bibfield  {journal} {\bibinfo  {journal} {Solid State Communications}\ }\textbf {\bibinfo {volume} {42}},\ \bibinfo {pages} {365} (\bibinfo {year} {1982})}\BibitemShut {NoStop}%
\bibitem [{\citenamefont {Fuggle}\ and\ \citenamefont {Alvarado}(1980)}]{Fuggle_PRA}%
  \BibitemOpen
  \bibfield  {author} {\bibinfo {author} {\bibfnamefont {J.~C.}\ \bibnamefont {Fuggle}}\ and\ \bibinfo {author} {\bibfnamefont {S.~F.}\ \bibnamefont {Alvarado}},\ }\bibfield  {title} {\bibinfo {title} {{Core-level lifetimes as determined by x-ray photoelectron spectroscopy measurements}},\ }\href {https://doi.org/10.1103/PhysRevA.22.1615} {\bibfield  {journal} {\bibinfo  {journal} {Phys. Rev. A}\ }\textbf {\bibinfo {volume} {22}},\ \bibinfo {pages} {1615} (\bibinfo {year} {1980})}\BibitemShut {NoStop}%
\bibitem [{\citenamefont {Bisogni}\ \emph {et~al.}(2012)\citenamefont {Bisogni}, \citenamefont {Simonelli}, \citenamefont {Ament}, \citenamefont {Forte}, \citenamefont {Moretti~Sala}, \citenamefont {Minola}, \citenamefont {Huotari}, \citenamefont {van~den Brink}, \citenamefont {Ghiringhelli}, \citenamefont {Brookes},\ and\ \citenamefont {Braicovich}}]{Bisogni_bimagnon}%
  \BibitemOpen
  \bibfield  {author} {\bibinfo {author} {\bibfnamefont {V.}~\bibnamefont {Bisogni}}, \bibinfo {author} {\bibfnamefont {L.}~\bibnamefont {Simonelli}}, \bibinfo {author} {\bibfnamefont {L.~J.~P.}\ \bibnamefont {Ament}}, \bibinfo {author} {\bibfnamefont {F.}~\bibnamefont {Forte}}, \bibinfo {author} {\bibfnamefont {M.}~\bibnamefont {Moretti~Sala}}, \bibinfo {author} {\bibfnamefont {M.}~\bibnamefont {Minola}}, \bibinfo {author} {\bibfnamefont {S.}~\bibnamefont {Huotari}}, \bibinfo {author} {\bibfnamefont {J.}~\bibnamefont {van~den Brink}}, \bibinfo {author} {\bibfnamefont {G.}~\bibnamefont {Ghiringhelli}}, \bibinfo {author} {\bibfnamefont {N.~B.}\ \bibnamefont {Brookes}},\ and\ \bibinfo {author} {\bibfnamefont {L.}~\bibnamefont {Braicovich}},\ }\bibfield  {title} {\bibinfo {title} {{Bimagnon studies in cuprates with resonant inelastic x-ray scattering at the O $K$ edge. I. Assessment on La${}_{2}$CuO${}_{4}$ and comparison with the excitation at Cu ${L}_{3}$ and Cu $K$ edges}},\ }\href
  {https://doi.org/10.1103/PhysRevB.85.214527} {\bibfield  {journal} {\bibinfo  {journal} {Phys. Rev. B}\ }\textbf {\bibinfo {volume} {85}},\ \bibinfo {pages} {214527} (\bibinfo {year} {2012})}\BibitemShut {NoStop}%
\bibitem [{\citenamefont {Thomas}\ \emph {et~al.}(2025)\citenamefont {Thomas}, \citenamefont {Banerjee}, \citenamefont {Nocera},\ and\ \citenamefont {Johnston}}]{Jinu_PRX}%
  \BibitemOpen
  \bibfield  {author} {\bibinfo {author} {\bibfnamefont {J.}~\bibnamefont {Thomas}}, \bibinfo {author} {\bibfnamefont {D.}~\bibnamefont {Banerjee}}, \bibinfo {author} {\bibfnamefont {A.}~\bibnamefont {Nocera}},\ and\ \bibinfo {author} {\bibfnamefont {S.}~\bibnamefont {Johnston}},\ }\bibfield  {title} {\bibinfo {title} {{Theory of Electron-Phonon Interactions in Extended Correlated Systems Probed by Resonant Inelastic X-Ray Scattering}},\ }\href {https://doi.org/10.1103/PhysRevX.15.021030} {\bibfield  {journal} {\bibinfo  {journal} {Phys. Rev. X}\ }\textbf {\bibinfo {volume} {15}},\ \bibinfo {pages} {021030} (\bibinfo {year} {2025})}\BibitemShut {NoStop}%
\bibitem [{\citenamefont {Li}\ \emph {et~al.}(2020)\citenamefont {Li}, \citenamefont {Nag}, \citenamefont {Pelliciari}, \citenamefont {Robarts}, \citenamefont {Walters}, \citenamefont {Garcia-Fernandez}, \citenamefont {Eisaki}, \citenamefont {Song}, \citenamefont {Ding}, \citenamefont {Johnston} \emph {et~al.}}]{Li2020}%
  \BibitemOpen
  \bibfield  {author} {\bibinfo {author} {\bibfnamefont {J.}~\bibnamefont {Li}}, \bibinfo {author} {\bibfnamefont {A.}~\bibnamefont {Nag}}, \bibinfo {author} {\bibfnamefont {J.}~\bibnamefont {Pelliciari}}, \bibinfo {author} {\bibfnamefont {H.}~\bibnamefont {Robarts}}, \bibinfo {author} {\bibfnamefont {A.}~\bibnamefont {Walters}}, \bibinfo {author} {\bibfnamefont {M.}~\bibnamefont {Garcia-Fernandez}}, \bibinfo {author} {\bibfnamefont {H.}~\bibnamefont {Eisaki}}, \bibinfo {author} {\bibfnamefont {D.}~\bibnamefont {Song}}, \bibinfo {author} {\bibfnamefont {H.}~\bibnamefont {Ding}}, \bibinfo {author} {\bibfnamefont {S.}~\bibnamefont {Johnston}}, \emph {et~al.},\ }\bibfield  {title} {\bibinfo {title} {{Multiorbital charge-density wave excitations and concomitant phonon anomalies in \ch{Bi_2Sr_2LaCuO_{6+$\delta$}}}},\ }\href {https://www.pnas.org/doi/full/10.1073/pnas.2001755117} {\bibfield  {journal} {\bibinfo  {journal} {Proceedings of the National Academy of Sciences}\ }\textbf {\bibinfo {volume} {117}},\
  \bibinfo {pages} {16219} (\bibinfo {year} {2020})}\BibitemShut {NoStop}%
\bibitem [{\citenamefont {Song}\ and\ \citenamefont {Annett}(1995)}]{Song_PRB1995}%
  \BibitemOpen
  \bibfield  {author} {\bibinfo {author} {\bibfnamefont {J.}~\bibnamefont {Song}}\ and\ \bibinfo {author} {\bibfnamefont {J.~F.}\ \bibnamefont {Annett}},\ }\bibfield  {title} {\bibinfo {title} {{Electron-phonon coupling and d-wave superconductivity in the cuprates}},\ }\href {https://doi.org/10.1103/PhysRevB.51.3840} {\bibfield  {journal} {\bibinfo  {journal} {Phys. Rev. B}\ }\textbf {\bibinfo {volume} {51}},\ \bibinfo {pages} {3840} (\bibinfo {year} {1995})}\BibitemShut {NoStop}%
\bibitem [{\citenamefont {Horsch}\ and\ \citenamefont {Khaliullin}(2005)}]{horsch2005}%
  \BibitemOpen
  \bibfield  {author} {\bibinfo {author} {\bibfnamefont {P.}~\bibnamefont {Horsch}}\ and\ \bibinfo {author} {\bibfnamefont {G.}~\bibnamefont {Khaliullin}},\ }\bibfield  {title} {\bibinfo {title} {{Doping dependence of density response and bond-stretching phonons in cuprates}},\ }\href {https://doi.org/10.1016/j.physb.2005.01.170} {\bibfield  {journal} {\bibinfo  {journal} {Physica B: Condensed Matter}\ }\textbf {\bibinfo {volume} {359}},\ \bibinfo {pages} {620} (\bibinfo {year} {2005})}\BibitemShut {NoStop}%
\bibitem [{\citenamefont {Peng}\ \emph {et~al.}(2022)\citenamefont {Peng}, \citenamefont {Martinelli}, \citenamefont {Li}, \citenamefont {Rossi}, \citenamefont {Mitrano}, \citenamefont {Arpaia}, \citenamefont {Sala}, \citenamefont {Gao}, \citenamefont {Guo}, \citenamefont {De~Luca}, \citenamefont {Walters}, \citenamefont {Nag}, \citenamefont {Barbour}, \citenamefont {Gu}, \citenamefont {Pelliciari}, \citenamefont {Brookes}, \citenamefont {Abbamonte}, \citenamefont {Salluzzo}, \citenamefont {Zhou}, \citenamefont {Zhou}, \citenamefont {Bisogni}, \citenamefont {Braicovich}, \citenamefont {Johnston},\ and\ \citenamefont {Ghiringhelli}}]{Peng_PRB22}%
  \BibitemOpen
  \bibfield  {author} {\bibinfo {author} {\bibfnamefont {Y.}~\bibnamefont {Peng}}, \bibinfo {author} {\bibfnamefont {L.}~\bibnamefont {Martinelli}}, \bibinfo {author} {\bibfnamefont {Q.}~\bibnamefont {Li}}, \bibinfo {author} {\bibfnamefont {M.}~\bibnamefont {Rossi}}, \bibinfo {author} {\bibfnamefont {M.}~\bibnamefont {Mitrano}}, \bibinfo {author} {\bibfnamefont {R.}~\bibnamefont {Arpaia}}, \bibinfo {author} {\bibfnamefont {M.~M.}\ \bibnamefont {Sala}}, \bibinfo {author} {\bibfnamefont {Q.}~\bibnamefont {Gao}}, \bibinfo {author} {\bibfnamefont {X.}~\bibnamefont {Guo}}, \bibinfo {author} {\bibfnamefont {G.~M.}\ \bibnamefont {De~Luca}}, \bibinfo {author} {\bibfnamefont {A.}~\bibnamefont {Walters}}, \bibinfo {author} {\bibfnamefont {A.}~\bibnamefont {Nag}}, \bibinfo {author} {\bibfnamefont {A.}~\bibnamefont {Barbour}}, \bibinfo {author} {\bibfnamefont {G.}~\bibnamefont {Gu}}, \bibinfo {author} {\bibfnamefont {J.}~\bibnamefont {Pelliciari}}, \bibinfo {author} {\bibfnamefont {N.~B.}\ \bibnamefont {Brookes}},
  \bibinfo {author} {\bibfnamefont {P.}~\bibnamefont {Abbamonte}}, \bibinfo {author} {\bibfnamefont {M.}~\bibnamefont {Salluzzo}}, \bibinfo {author} {\bibfnamefont {X.}~\bibnamefont {Zhou}}, \bibinfo {author} {\bibfnamefont {K.-J.}\ \bibnamefont {Zhou}}, \bibinfo {author} {\bibfnamefont {V.}~\bibnamefont {Bisogni}}, \bibinfo {author} {\bibfnamefont {L.}~\bibnamefont {Braicovich}}, \bibinfo {author} {\bibfnamefont {S.}~\bibnamefont {Johnston}},\ and\ \bibinfo {author} {\bibfnamefont {G.}~\bibnamefont {Ghiringhelli}},\ }\bibfield  {title} {\bibinfo {title} {{Doping dependence of the electron-phonon coupling in two families of bilayer superconducting cuprates}},\ }\href {https://doi.org/10.1103/PhysRevB.105.115105} {\bibfield  {journal} {\bibinfo  {journal} {Phys. Rev. B}\ }\textbf {\bibinfo {volume} {105}},\ \bibinfo {pages} {115105} (\bibinfo {year} {2022})}\BibitemShut {NoStop}%
\bibitem [{\citenamefont {Shen}\ \emph {et~al.}(2002)\citenamefont {Shen}, \citenamefont {Lanzara}, \citenamefont {Ishihara},\ and\ \citenamefont {Nagaosa}}]{Shen2002}%
  \BibitemOpen
  \bibfield  {author} {\bibinfo {author} {\bibfnamefont {Z.-X.}\ \bibnamefont {Shen}}, \bibinfo {author} {\bibfnamefont {A.}~\bibnamefont {Lanzara}}, \bibinfo {author} {\bibfnamefont {S.}~\bibnamefont {Ishihara}},\ and\ \bibinfo {author} {\bibfnamefont {N.}~\bibnamefont {Nagaosa}},\ }\bibfield  {title} {\bibinfo {title} {{Role of the electron-phonon interaction in the strongly correlated cuprate superconductors}},\ }\href {https://doi.org/10.1080/13642810210142735} {\bibfield  {journal} {\bibinfo  {journal} {Philosophical magazine B}\ }\textbf {\bibinfo {volume} {82}},\ \bibinfo {pages} {1349} (\bibinfo {year} {2002})}\BibitemShut {NoStop}%
\bibitem [{\citenamefont {Bulut}\ and\ \citenamefont {Scalapino}(1996)}]{Bulut_PRB}%
  \BibitemOpen
  \bibfield  {author} {\bibinfo {author} {\bibfnamefont {N.}~\bibnamefont {Bulut}}\ and\ \bibinfo {author} {\bibfnamefont {D.~J.}\ \bibnamefont {Scalapino}},\ }\bibfield  {title} {\bibinfo {title} {{${d}_{{x}^{2}\ensuremath{-}{y}^{2}}$ symmetry and the pairing mechanism}},\ }\href {https://doi.org/10.1103/PhysRevB.54.14971} {\bibfield  {journal} {\bibinfo  {journal} {Phys. Rev. B}\ }\textbf {\bibinfo {volume} {54}},\ \bibinfo {pages} {14971} (\bibinfo {year} {1996})}\BibitemShut {NoStop}%
\bibitem [{\citenamefont {Sandvik}\ \emph {et~al.}(2004)\citenamefont {Sandvik}, \citenamefont {Scalapino},\ and\ \citenamefont {Bickers}}]{Sandvik_PRB}%
  \BibitemOpen
  \bibfield  {author} {\bibinfo {author} {\bibfnamefont {A.~W.}\ \bibnamefont {Sandvik}}, \bibinfo {author} {\bibfnamefont {D.~J.}\ \bibnamefont {Scalapino}},\ and\ \bibinfo {author} {\bibfnamefont {N.~E.}\ \bibnamefont {Bickers}},\ }\bibfield  {title} {\bibinfo {title} {{Effect of an electron-phonon interaction on the one-electron spectral weight of a d-wave superconductor}},\ }\href {https://doi.org/10.1103/PhysRevB.69.094523} {\bibfield  {journal} {\bibinfo  {journal} {Phys. Rev. B}\ }\textbf {\bibinfo {volume} {69}},\ \bibinfo {pages} {094523} (\bibinfo {year} {2004})}\BibitemShut {NoStop}%
\bibitem [{\citenamefont {Johnston}(2010)}]{JohnstonPhD}%
  \BibitemOpen
  \bibfield  {author} {\bibinfo {author} {\bibfnamefont {S.~S.}\ \bibnamefont {Johnston}},\ }\emph {\bibinfo {title} {{Electron-phonon Coupling in Quasi-Two-Dimensional Correlated Systems}}},\ \href {http://hdl.handle.net/10012/5274} {Ph.D. thesis},\ \bibinfo  {school} {University of Waterloo} (\bibinfo {year} {2010})\BibitemShut {NoStop}%
\bibitem [{\citenamefont {Zhang}\ \emph {et~al.}(2007)\citenamefont {Zhang}, \citenamefont {Louie},\ and\ \citenamefont {Cohen}}]{Zhang_PRB}%
  \BibitemOpen
  \bibfield  {author} {\bibinfo {author} {\bibfnamefont {P.}~\bibnamefont {Zhang}}, \bibinfo {author} {\bibfnamefont {S.~G.}\ \bibnamefont {Louie}},\ and\ \bibinfo {author} {\bibfnamefont {M.~L.}\ \bibnamefont {Cohen}},\ }\bibfield  {title} {\bibinfo {title} {{Electron-Phonon Renormalization in Cuprate Superconductors}},\ }\href {https://doi.org/10.1103/PhysRevLett.98.067005} {\bibfield  {journal} {\bibinfo  {journal} {Phys. Rev. Lett.}\ }\textbf {\bibinfo {volume} {98}},\ \bibinfo {pages} {067005} (\bibinfo {year} {2007})}\BibitemShut {NoStop}%
\bibitem [{\citenamefont {Sterling}\ and\ \citenamefont {Reznik}(2021{\natexlab{b}})}]{sterling2021}%
  \BibitemOpen
  \bibfield  {author} {\bibinfo {author} {\bibfnamefont {T.~C.}\ \bibnamefont {Sterling}}\ and\ \bibinfo {author} {\bibfnamefont {D.}~\bibnamefont {Reznik}},\ }\bibfield  {title} {\bibinfo {title} {{Effect of the electronic charge gap on \ch{LO} bond-stretching phonons in undoped \ch{La_2CuO_4} calculated using \ch{LDA}+\ch{U}}},\ }\href {https://doi.org/10.1103/PhysRevB.104.134311} {\bibfield  {journal} {\bibinfo  {journal} {Physical Review B}\ }\textbf {\bibinfo {volume} {104}},\ \bibinfo {pages} {134311} (\bibinfo {year} {2021}{\natexlab{b}})}\BibitemShut {NoStop}%
\bibitem [{\citenamefont {Reichardt}\ \emph {et~al.}(1989)\citenamefont {Reichardt}, \citenamefont {Pyka}, \citenamefont {Pintschovius}, \citenamefont {Hennion},\ and\ \citenamefont {Collin}}]{REICHARDT_PhC}%
  \BibitemOpen
  \bibfield  {author} {\bibinfo {author} {\bibfnamefont {W.}~\bibnamefont {Reichardt}}, \bibinfo {author} {\bibfnamefont {N.}~\bibnamefont {Pyka}}, \bibinfo {author} {\bibfnamefont {L.}~\bibnamefont {Pintschovius}}, \bibinfo {author} {\bibfnamefont {B.}~\bibnamefont {Hennion}},\ and\ \bibinfo {author} {\bibfnamefont {G.}~\bibnamefont {Collin}},\ }\bibfield  {title} {\bibinfo {title} {{Phonons in \ch{YBa_2Cu_3O_{7-$\delta$}}}},\ }\href {https://doi.org/https://doi.org/10.1016/0921-4534(89)91107-6} {\bibfield  {journal} {\bibinfo  {journal} {Physica C: Superconductivity and its Applications}\ }\textbf {\bibinfo {volume} {162-164}},\ \bibinfo {pages} {464} (\bibinfo {year} {1989})}\BibitemShut {NoStop}%
\bibitem [{\citenamefont {Pintschovius}\ \emph {et~al.}(1991)\citenamefont {Pintschovius}, \citenamefont {Pyka}, \citenamefont {Reichardt}, \citenamefont {Rumiantsev}, \citenamefont {Mitrofanov}, \citenamefont {Ivanov}, \citenamefont {Collin},\ and\ \citenamefont {Bourges}}]{PINTSCHOVIUS_PhC}%
  \BibitemOpen
  \bibfield  {author} {\bibinfo {author} {\bibfnamefont {L.}~\bibnamefont {Pintschovius}}, \bibinfo {author} {\bibfnamefont {N.}~\bibnamefont {Pyka}}, \bibinfo {author} {\bibfnamefont {W.}~\bibnamefont {Reichardt}}, \bibinfo {author} {\bibfnamefont {A.}~\bibnamefont {Rumiantsev}}, \bibinfo {author} {\bibfnamefont {N.}~\bibnamefont {Mitrofanov}}, \bibinfo {author} {\bibfnamefont {A.}~\bibnamefont {Ivanov}}, \bibinfo {author} {\bibfnamefont {G.}~\bibnamefont {Collin}},\ and\ \bibinfo {author} {\bibfnamefont {P.}~\bibnamefont {Bourges}},\ }\bibfield  {title} {\bibinfo {title} {{Lattice dynamical studies of HTSC materials}},\ }\href {https://doi.org/https://doi.org/10.1016/0921-4534(91)91965-7} {\bibfield  {journal} {\bibinfo  {journal} {Physica C: Superconductivity}\ }\textbf {\bibinfo {volume} {185-189}},\ \bibinfo {pages} {156} (\bibinfo {year} {1991})}\BibitemShut {NoStop}%
\bibitem [{\citenamefont {Pintschovius}\ \emph {et~al.}(2006)\citenamefont {Pintschovius}, \citenamefont {Reznik},\ and\ \citenamefont {Yamada}}]{Pintschovius_PRB}%
  \BibitemOpen
  \bibfield  {author} {\bibinfo {author} {\bibfnamefont {L.}~\bibnamefont {Pintschovius}}, \bibinfo {author} {\bibfnamefont {D.}~\bibnamefont {Reznik}},\ and\ \bibinfo {author} {\bibfnamefont {K.}~\bibnamefont {Yamada}},\ }\bibfield  {title} {\bibinfo {title} {{Oxygen phonon branches in overdoped \ch{La_{1.7}Sr_{0.3}Cu_3O_{4}}}},\ }\href {https://doi.org/10.1103/PhysRevB.74.174514} {\bibfield  {journal} {\bibinfo  {journal} {Phys. Rev. B}\ }\textbf {\bibinfo {volume} {74}},\ \bibinfo {pages} {174514} (\bibinfo {year} {2006})}\BibitemShut {NoStop}%
\bibitem [{\citenamefont {Braden}\ \emph {et~al.}(2005)\citenamefont {Braden}, \citenamefont {Pintschovius}, \citenamefont {Uefuji},\ and\ \citenamefont {Yamada}}]{Braden_PRB}%
  \BibitemOpen
  \bibfield  {author} {\bibinfo {author} {\bibfnamefont {M.}~\bibnamefont {Braden}}, \bibinfo {author} {\bibfnamefont {L.}~\bibnamefont {Pintschovius}}, \bibinfo {author} {\bibfnamefont {T.}~\bibnamefont {Uefuji}},\ and\ \bibinfo {author} {\bibfnamefont {K.}~\bibnamefont {Yamada}},\ }\bibfield  {title} {\bibinfo {title} {{Dispersion of the high-energy phonon modes in \ch{Nd_{1.85}Ce_{0.15}CuO_{4}}}},\ }\href {https://doi.org/10.1103/PhysRevB.72.184517} {\bibfield  {journal} {\bibinfo  {journal} {Phys. Rev. B}\ }\textbf {\bibinfo {volume} {72}},\ \bibinfo {pages} {184517} (\bibinfo {year} {2005})}\BibitemShut {NoStop}%
\bibitem [{\citenamefont {Lin}\ \emph {et~al.}(2020)\citenamefont {Lin}, \citenamefont {Miao}, \citenamefont {Mazzone}, \citenamefont {Gu}, \citenamefont {Nag}, \citenamefont {Walters}, \citenamefont {Garc\'{\i}a-Fern\'andez}, \citenamefont {Barbour}, \citenamefont {Pelliciari}, \citenamefont {Jarrige}, \citenamefont {Oda}, \citenamefont {Kurosawa}, \citenamefont {Momono}, \citenamefont {Zhou}, \citenamefont {Bisogni}, \citenamefont {Liu},\ and\ \citenamefont {Dean}}]{Lin_PRL}%
  \BibitemOpen
  \bibfield  {author} {\bibinfo {author} {\bibfnamefont {J.~Q.}\ \bibnamefont {Lin}}, \bibinfo {author} {\bibfnamefont {H.}~\bibnamefont {Miao}}, \bibinfo {author} {\bibfnamefont {D.~G.}\ \bibnamefont {Mazzone}}, \bibinfo {author} {\bibfnamefont {G.~D.}\ \bibnamefont {Gu}}, \bibinfo {author} {\bibfnamefont {A.}~\bibnamefont {Nag}}, \bibinfo {author} {\bibfnamefont {A.~C.}\ \bibnamefont {Walters}}, \bibinfo {author} {\bibfnamefont {M.}~\bibnamefont {Garc\'{\i}a-Fern\'andez}}, \bibinfo {author} {\bibfnamefont {A.}~\bibnamefont {Barbour}}, \bibinfo {author} {\bibfnamefont {J.}~\bibnamefont {Pelliciari}}, \bibinfo {author} {\bibfnamefont {I.}~\bibnamefont {Jarrige}}, \bibinfo {author} {\bibfnamefont {M.}~\bibnamefont {Oda}}, \bibinfo {author} {\bibfnamefont {K.}~\bibnamefont {Kurosawa}}, \bibinfo {author} {\bibfnamefont {N.}~\bibnamefont {Momono}}, \bibinfo {author} {\bibfnamefont {K.-J.}\ \bibnamefont {Zhou}}, \bibinfo {author} {\bibfnamefont {V.}~\bibnamefont {Bisogni}}, \bibinfo {author} {\bibfnamefont
  {X.}~\bibnamefont {Liu}},\ and\ \bibinfo {author} {\bibfnamefont {M.~P.~M.}\ \bibnamefont {Dean}},\ }\bibfield  {title} {\bibinfo {title} {{Strongly Correlated Charge Density Wave in \ch{La_{2-x}Sr_xCuO_4} Evidenced by Doping-Dependent Phonon Anomaly}},\ }\href {https://doi.org/10.1103/PhysRevLett.124.207005} {\bibfield  {journal} {\bibinfo  {journal} {Phys. Rev. Lett.}\ }\textbf {\bibinfo {volume} {124}},\ \bibinfo {pages} {207005} (\bibinfo {year} {2020})}\BibitemShut {NoStop}%
\bibitem [{\citenamefont {Martinelli}\ \emph {et~al.}(2024{\natexlab{b}})\citenamefont {Martinelli}, \citenamefont {Bia{\l}o}, \citenamefont {Hong}, \citenamefont {Oppliger}, \citenamefont {Lin}, \citenamefont {Schaller}, \citenamefont {K{\"u}spert}, \citenamefont {Fischer}, \citenamefont {Kurosawa}, \citenamefont {Momono} \emph {et~al.}}]{Martinelli2024}%
  \BibitemOpen
  \bibfield  {author} {\bibinfo {author} {\bibfnamefont {L.}~\bibnamefont {Martinelli}}, \bibinfo {author} {\bibfnamefont {I.}~\bibnamefont {Bia{\l}o}}, \bibinfo {author} {\bibfnamefont {X.}~\bibnamefont {Hong}}, \bibinfo {author} {\bibfnamefont {J.}~\bibnamefont {Oppliger}}, \bibinfo {author} {\bibfnamefont {C.}~\bibnamefont {Lin}}, \bibinfo {author} {\bibfnamefont {T.}~\bibnamefont {Schaller}}, \bibinfo {author} {\bibfnamefont {J.}~\bibnamefont {K{\"u}spert}}, \bibinfo {author} {\bibfnamefont {M.}~\bibnamefont {Fischer}}, \bibinfo {author} {\bibfnamefont {T.}~\bibnamefont {Kurosawa}}, \bibinfo {author} {\bibfnamefont {N.}~\bibnamefont {Momono}}, \emph {et~al.},\ }\bibfield  {title} {\bibinfo {title} {{Decoupled static and dynamical charge correlations in \ch{La_{2-x}Sr_xCuO_4}}},\ }\href {https://doi.org/10.48550/arXiv.2406.15062} {\bibfield  {journal} {\bibinfo  {journal} {arXiv preprint arXiv:2406.15062}\ } (\bibinfo {year} {2024}{\natexlab{b}})}\BibitemShut {NoStop}%
\bibitem [{\citenamefont {Lee}\ \emph {et~al.}(2021)\citenamefont {Lee}, \citenamefont {Zhou}, \citenamefont {Hepting}, \citenamefont {Li}, \citenamefont {Nag}, \citenamefont {Walters}, \citenamefont {Garcia-Fernandez}, \citenamefont {Robarts}, \citenamefont {Hashimoto}, \citenamefont {Lu} \emph {et~al.}}]{WSLee2021}%
  \BibitemOpen
  \bibfield  {author} {\bibinfo {author} {\bibfnamefont {W.-S.}\ \bibnamefont {Lee}}, \bibinfo {author} {\bibfnamefont {K.-J.}\ \bibnamefont {Zhou}}, \bibinfo {author} {\bibfnamefont {M.}~\bibnamefont {Hepting}}, \bibinfo {author} {\bibfnamefont {J.}~\bibnamefont {Li}}, \bibinfo {author} {\bibfnamefont {A.}~\bibnamefont {Nag}}, \bibinfo {author} {\bibfnamefont {A.}~\bibnamefont {Walters}}, \bibinfo {author} {\bibfnamefont {M.}~\bibnamefont {Garcia-Fernandez}}, \bibinfo {author} {\bibfnamefont {H.}~\bibnamefont {Robarts}}, \bibinfo {author} {\bibfnamefont {M.}~\bibnamefont {Hashimoto}}, \bibinfo {author} {\bibfnamefont {H.}~\bibnamefont {Lu}}, \emph {et~al.},\ }\bibfield  {title} {\bibinfo {title} {{Spectroscopic fingerprint of charge order melting driven by quantum fluctuations in a cuprate}},\ }\href {https://doi.org/10.1038/s41567-020-0993-7} {\bibfield  {journal} {\bibinfo  {journal} {Nature Physics}\ }\textbf {\bibinfo {volume} {17}},\ \bibinfo {pages} {53} (\bibinfo {year} {2021})}\BibitemShut {NoStop}%
\bibitem [{\citenamefont {Di~Castro}\ \emph {et~al.}(2014)\citenamefont {Di~Castro}, \citenamefont {Aruta}, \citenamefont {Tebano}, \citenamefont {Innocenti}, \citenamefont {Minola}, \citenamefont {Sala}, \citenamefont {Prellier}, \citenamefont {Lebedev},\ and\ \citenamefont {Balestrino}}]{diCastro_CCO_2014}%
  \BibitemOpen
  \bibfield  {author} {\bibinfo {author} {\bibfnamefont {D.}~\bibnamefont {Di~Castro}}, \bibinfo {author} {\bibfnamefont {C.}~\bibnamefont {Aruta}}, \bibinfo {author} {\bibfnamefont {A.}~\bibnamefont {Tebano}}, \bibinfo {author} {\bibfnamefont {D.}~\bibnamefont {Innocenti}}, \bibinfo {author} {\bibfnamefont {M.}~\bibnamefont {Minola}}, \bibinfo {author} {\bibfnamefont {M.~M.}\ \bibnamefont {Sala}}, \bibinfo {author} {\bibfnamefont {W.}~\bibnamefont {Prellier}}, \bibinfo {author} {\bibfnamefont {O.}~\bibnamefont {Lebedev}},\ and\ \bibinfo {author} {\bibfnamefont {G.}~\bibnamefont {Balestrino}},\ }\bibfield  {title} {\bibinfo {title} {{Tc up to 50 \ch{K} in superlattices of insulating oxides}},\ }\href {https://iopscience.iop.org/article/10.1088/0953-2048/27/4/044016} {\bibfield  {journal} {\bibinfo  {journal} {Superconductor Science and Technology}\ }\textbf {\bibinfo {volume} {27}},\ \bibinfo {pages} {044016} (\bibinfo {year} {2014})}\BibitemShut {NoStop}%
\bibitem [{\citenamefont {Castro}(2009)}]{dicastro_2009_CCO}%
  \BibitemOpen
  \bibfield  {author} {\bibinfo {author} {\bibfnamefont {D.}~\bibnamefont {Castro}},\ }\bibfield  {title} {\bibinfo {title} {{High-Tc superconductivity at the interface between the \ch{CaCuO_2} and \ch{SrTiO_3} insulating oxides}},\ }\href {https://journals.aps.org/prl/abstract/10.1103/PhysRevLett.115.147001} {\bibfield  {journal} {\bibinfo  {journal} {Phys. Rev. Lett.}\ }\textbf {\bibinfo {volume} {469}},\ \bibinfo {pages} {694} (\bibinfo {year} {2009})}\BibitemShut {NoStop}%
\bibitem [{\citenamefont {Bozovic}\ \emph {et~al.}(2002)\citenamefont {Bozovic}, \citenamefont {Logvenov}, \citenamefont {Belca}, \citenamefont {Narimbetov},\ and\ \citenamefont {Sveklo}}]{bozovic_2002_LSCO}%
  \BibitemOpen
  \bibfield  {author} {\bibinfo {author} {\bibfnamefont {I.}~\bibnamefont {Bozovic}}, \bibinfo {author} {\bibfnamefont {G.}~\bibnamefont {Logvenov}}, \bibinfo {author} {\bibfnamefont {I.}~\bibnamefont {Belca}}, \bibinfo {author} {\bibfnamefont {B.}~\bibnamefont {Narimbetov}},\ and\ \bibinfo {author} {\bibfnamefont {I.}~\bibnamefont {Sveklo}},\ }\bibfield  {title} {\bibinfo {title} {{Epitaxial Strain and Superconductivity in \ch{La_{2-x}Sr_xCuO_4} Thin Films}},\ }\href {https://journals.aps.org/prl/abstract/10.1103/PhysRevLett.89.107001} {\bibfield  {journal} {\bibinfo  {journal} {Physical review letters}\ }\textbf {\bibinfo {volume} {89}},\ \bibinfo {pages} {107001} (\bibinfo {year} {2002})}\BibitemShut {NoStop}%
\bibitem [{\citenamefont {Biagi}(2022)}]{BiagiPhD}%
  \BibitemOpen
  \bibfield  {author} {\bibinfo {author} {\bibfnamefont {M.}~\bibnamefont {Biagi}},\ }\emph {\bibinfo {title} {{{La}$_{2-x}${Sr}$_{x}${CuO}$_{4}$ thin films and nanostructures to study local ordering phenomena in a striped superconductor}}},\ \href {https://hdl.handle.net/10589/190124} {Ph.D. thesis},\ \bibinfo  {school} {Politecnico di Milano} (\bibinfo {year} {2022})\BibitemShut {NoStop}%
\bibitem [{\citenamefont {Arpaia}\ \emph {et~al.}(2018)\citenamefont {Arpaia}, \citenamefont {Andersson}, \citenamefont {Trabaldo}, \citenamefont {Bauch},\ and\ \citenamefont {Lombardi}}]{arpaia_2018_YBCO}%
  \BibitemOpen
  \bibfield  {author} {\bibinfo {author} {\bibfnamefont {R.}~\bibnamefont {Arpaia}}, \bibinfo {author} {\bibfnamefont {E.}~\bibnamefont {Andersson}}, \bibinfo {author} {\bibfnamefont {E.}~\bibnamefont {Trabaldo}}, \bibinfo {author} {\bibfnamefont {T.}~\bibnamefont {Bauch}},\ and\ \bibinfo {author} {\bibfnamefont {F.}~\bibnamefont {Lombardi}},\ }\bibfield  {title} {\bibinfo {title} {{Probing the phase diagram of cuprates with \ch{YBa_2Cu_3O_{7- $\delta$}} thin films and nanowires}},\ }\href {https://journals.aps.org/prmaterials/abstract/10.1103/PhysRevMaterials.2.024804} {\bibfield  {journal} {\bibinfo  {journal} {Physical Review Materials}\ }\textbf {\bibinfo {volume} {2}},\ \bibinfo {pages} {024804} (\bibinfo {year} {2018})}\BibitemShut {NoStop}%
\bibitem [{\citenamefont {Brookes}\ \emph {et~al.}(2018)\citenamefont {Brookes}, \citenamefont {Yakhou-Harris}, \citenamefont {Kummer}, \citenamefont {Fondacaro}, \citenamefont {Cezar}, \citenamefont {Betto}, \citenamefont {Velez-Fort}, \citenamefont {Amorese}, \citenamefont {Ghiringhelli}, \citenamefont {Braicovich} \emph {et~al.}}]{BROOKES_ID32}%
  \BibitemOpen
  \bibfield  {author} {\bibinfo {author} {\bibfnamefont {N.~B.}\ \bibnamefont {Brookes}}, \bibinfo {author} {\bibfnamefont {F.}~\bibnamefont {Yakhou-Harris}}, \bibinfo {author} {\bibfnamefont {K.}~\bibnamefont {Kummer}}, \bibinfo {author} {\bibfnamefont {A.}~\bibnamefont {Fondacaro}}, \bibinfo {author} {\bibfnamefont {J.}~\bibnamefont {Cezar}}, \bibinfo {author} {\bibfnamefont {D.}~\bibnamefont {Betto}}, \bibinfo {author} {\bibfnamefont {E.}~\bibnamefont {Velez-Fort}}, \bibinfo {author} {\bibfnamefont {A.}~\bibnamefont {Amorese}}, \bibinfo {author} {\bibfnamefont {G.}~\bibnamefont {Ghiringhelli}}, \bibinfo {author} {\bibfnamefont {L.}~\bibnamefont {Braicovich}}, \emph {et~al.},\ }\bibfield  {title} {\bibinfo {title} {The beamline id32 at the esrf for soft x-ray high energy resolution resonant inelastic x-ray scattering and polarisation dependent x-ray absorption spectroscopy},\ }\href {https://doi.org/https://doi.org/10.1016/j.nima.2018.07.001} {\bibfield  {journal} {\bibinfo  {journal} {Nuclear Instruments
  and Methods in Physics Research Section A: Accelerators, Spectrometers, Detectors and Associated Equipment}\ }\textbf {\bibinfo {volume} {903}},\ \bibinfo {pages} {175} (\bibinfo {year} {2018})}\BibitemShut {NoStop}%
\bibitem [{\citenamefont {Minola}\ \emph {et~al.}(2015)\citenamefont {Minola}, \citenamefont {Dellea}, \citenamefont {Gretarsson}, \citenamefont {Peng}, \citenamefont {Lu}, \citenamefont {Porras}, \citenamefont {Loew}, \citenamefont {Yakhou}, \citenamefont {Brookes}, \citenamefont {Huang}, \citenamefont {Pelliciari}, \citenamefont {Schmitt}, \citenamefont {Ghiringhelli}, \citenamefont {Keimer}, \citenamefont {Braicovich},\ and\ \citenamefont {Le~Tacon}}]{Minola_PRL}%
  \BibitemOpen
  \bibfield  {author} {\bibinfo {author} {\bibfnamefont {M.}~\bibnamefont {Minola}}, \bibinfo {author} {\bibfnamefont {G.}~\bibnamefont {Dellea}}, \bibinfo {author} {\bibfnamefont {H.}~\bibnamefont {Gretarsson}}, \bibinfo {author} {\bibfnamefont {Y.~Y.}\ \bibnamefont {Peng}}, \bibinfo {author} {\bibfnamefont {Y.}~\bibnamefont {Lu}}, \bibinfo {author} {\bibfnamefont {J.}~\bibnamefont {Porras}}, \bibinfo {author} {\bibfnamefont {T.}~\bibnamefont {Loew}}, \bibinfo {author} {\bibfnamefont {F.}~\bibnamefont {Yakhou}}, \bibinfo {author} {\bibfnamefont {N.~B.}\ \bibnamefont {Brookes}}, \bibinfo {author} {\bibfnamefont {Y.~B.}\ \bibnamefont {Huang}}, \bibinfo {author} {\bibfnamefont {J.}~\bibnamefont {Pelliciari}}, \bibinfo {author} {\bibfnamefont {T.}~\bibnamefont {Schmitt}}, \bibinfo {author} {\bibfnamefont {G.}~\bibnamefont {Ghiringhelli}}, \bibinfo {author} {\bibfnamefont {B.}~\bibnamefont {Keimer}}, \bibinfo {author} {\bibfnamefont {L.}~\bibnamefont {Braicovich}},\ and\ \bibinfo {author} {\bibfnamefont
  {M.}~\bibnamefont {Le~Tacon}},\ }\bibfield  {title} {\bibinfo {title} {{Collective Nature of Spin Excitations in Superconducting Cuprates Probed by Resonant Inelastic X-Ray Scattering}},\ }\href {https://doi.org/10.1103/PhysRevLett.114.217003} {\bibfield  {journal} {\bibinfo  {journal} {Phys. Rev. Lett.}\ }\textbf {\bibinfo {volume} {114}},\ \bibinfo {pages} {217003} (\bibinfo {year} {2015})}\BibitemShut {NoStop}%
\bibitem [{\citenamefont {Feng}\ \emph {et~al.}(2020)\citenamefont {Feng}, \citenamefont {Sallis}, \citenamefont {Shao}, \citenamefont {Qiao}, \citenamefont {Liu}, \citenamefont {Kao}, \citenamefont {Tremsin}, \citenamefont {Hussain}, \citenamefont {Yang}, \citenamefont {Guo},\ and\ \citenamefont {Chuang}}]{Feng_PRL}%
  \BibitemOpen
  \bibfield  {author} {\bibinfo {author} {\bibfnamefont {X.}~\bibnamefont {Feng}}, \bibinfo {author} {\bibfnamefont {S.}~\bibnamefont {Sallis}}, \bibinfo {author} {\bibfnamefont {Y.-C.}\ \bibnamefont {Shao}}, \bibinfo {author} {\bibfnamefont {R.}~\bibnamefont {Qiao}}, \bibinfo {author} {\bibfnamefont {Y.-S.}\ \bibnamefont {Liu}}, \bibinfo {author} {\bibfnamefont {L.~C.}\ \bibnamefont {Kao}}, \bibinfo {author} {\bibfnamefont {A.~S.}\ \bibnamefont {Tremsin}}, \bibinfo {author} {\bibfnamefont {Z.}~\bibnamefont {Hussain}}, \bibinfo {author} {\bibfnamefont {W.}~\bibnamefont {Yang}}, \bibinfo {author} {\bibfnamefont {J.}~\bibnamefont {Guo}},\ and\ \bibinfo {author} {\bibfnamefont {Y.-D.}\ \bibnamefont {Chuang}},\ }\bibfield  {title} {\bibinfo {title} {{Disparate Exciton-Phonon Couplings for Zone-Center and Boundary Phonons in Solid-State Graphite}},\ }\href {https://doi.org/10.1103/PhysRevLett.125.116401} {\bibfield  {journal} {\bibinfo  {journal} {Phys. Rev. Lett.}\ }\textbf {\bibinfo {volume} {125}},\ \bibinfo
  {pages} {116401} (\bibinfo {year} {2020})}\BibitemShut {NoStop}%
\bibitem [{\citenamefont {Bisogni}\ \emph {et~al.}(2014)\citenamefont {Bisogni}, \citenamefont {Kourtis}, \citenamefont {Monney}, \citenamefont {Zhou}, \citenamefont {Kraus}, \citenamefont {Sekar}, \citenamefont {Strocov}, \citenamefont {B\"uchner}, \citenamefont {van~den Brink}, \citenamefont {Braicovich}, \citenamefont {Schmitt}, \citenamefont {Daghofer},\ and\ \citenamefont {Geck}}]{Bisogni_PRL2014}%
  \BibitemOpen
  \bibfield  {author} {\bibinfo {author} {\bibfnamefont {V.}~\bibnamefont {Bisogni}}, \bibinfo {author} {\bibfnamefont {S.}~\bibnamefont {Kourtis}}, \bibinfo {author} {\bibfnamefont {C.}~\bibnamefont {Monney}}, \bibinfo {author} {\bibfnamefont {K.}~\bibnamefont {Zhou}}, \bibinfo {author} {\bibfnamefont {R.}~\bibnamefont {Kraus}}, \bibinfo {author} {\bibfnamefont {C.}~\bibnamefont {Sekar}}, \bibinfo {author} {\bibfnamefont {V.}~\bibnamefont {Strocov}}, \bibinfo {author} {\bibfnamefont {B.}~\bibnamefont {B\"uchner}}, \bibinfo {author} {\bibfnamefont {J.}~\bibnamefont {van~den Brink}}, \bibinfo {author} {\bibfnamefont {L.}~\bibnamefont {Braicovich}}, \bibinfo {author} {\bibfnamefont {T.}~\bibnamefont {Schmitt}}, \bibinfo {author} {\bibfnamefont {M.}~\bibnamefont {Daghofer}},\ and\ \bibinfo {author} {\bibfnamefont {J.}~\bibnamefont {Geck}},\ }\bibfield  {title} {\bibinfo {title} {{Femtosecond Dynamics of Momentum-Dependent Magnetic Excitations from Resonant Inelastic X-Ray Scattering in
  ${\mathrm{CaCu}}_{2}{\mathrm{O}}_{3}$}},\ }\href {https://doi.org/10.1103/PhysRevLett.112.147401} {\bibfield  {journal} {\bibinfo  {journal} {Phys. Rev. Lett.}\ }\textbf {\bibinfo {volume} {112}},\ \bibinfo {pages} {147401} (\bibinfo {year} {2014})}\BibitemShut {NoStop}%
\bibitem [{\citenamefont {Arpaia}\ \emph {et~al.}(2023)\citenamefont {Arpaia}, \citenamefont {Martinelli}, \citenamefont {Sala}, \citenamefont {Caprara}, \citenamefont {Nag}, \citenamefont {Brookes}, \citenamefont {Camisa}, \citenamefont {Li}, \citenamefont {Gao}, \citenamefont {Zhou} \emph {et~al.}}]{arpaia2023signature}%
  \BibitemOpen
  \bibfield  {author} {\bibinfo {author} {\bibfnamefont {R.}~\bibnamefont {Arpaia}}, \bibinfo {author} {\bibfnamefont {L.}~\bibnamefont {Martinelli}}, \bibinfo {author} {\bibfnamefont {M.~M.}\ \bibnamefont {Sala}}, \bibinfo {author} {\bibfnamefont {S.}~\bibnamefont {Caprara}}, \bibinfo {author} {\bibfnamefont {A.}~\bibnamefont {Nag}}, \bibinfo {author} {\bibfnamefont {N.~B.}\ \bibnamefont {Brookes}}, \bibinfo {author} {\bibfnamefont {P.}~\bibnamefont {Camisa}}, \bibinfo {author} {\bibfnamefont {Q.}~\bibnamefont {Li}}, \bibinfo {author} {\bibfnamefont {Q.}~\bibnamefont {Gao}}, \bibinfo {author} {\bibfnamefont {X.}~\bibnamefont {Zhou}}, \emph {et~al.},\ }\bibfield  {title} {\bibinfo {title} {{Signature of quantum criticality in cuprates by charge density fluctuations}},\ }\href@noop {} {\bibfield  {journal} {\bibinfo  {journal} {Nature Communications}\ }\textbf {\bibinfo {volume} {14}},\ \bibinfo {pages} {7198} (\bibinfo {year} {2023})}\BibitemShut {NoStop}%
\bibitem [{\citenamefont {Baroni}\ \emph {et~al.}(2001)\citenamefont {Baroni}, \citenamefont {de~Gironcoli}, \citenamefont {Dal~Corso},\ and\ \citenamefont {Giannozzi}}]{Baroni2001}%
  \BibitemOpen
  \bibfield  {author} {\bibinfo {author} {\bibfnamefont {S.}~\bibnamefont {Baroni}}, \bibinfo {author} {\bibfnamefont {S.}~\bibnamefont {de~Gironcoli}}, \bibinfo {author} {\bibfnamefont {A.}~\bibnamefont {Dal~Corso}},\ and\ \bibinfo {author} {\bibfnamefont {P.}~\bibnamefont {Giannozzi}},\ }\bibfield  {title} {\bibinfo {title} {{Phonons and related crystal properties from density-functional perturbation theory}},\ }\href {https://doi.org/10.1103/RevModPhys.73.515} {\bibfield  {journal} {\bibinfo  {journal} {Rev. Mod. Phys.}\ }\textbf {\bibinfo {volume} {73}},\ \bibinfo {pages} {515} (\bibinfo {year} {2001})}\BibitemShut {NoStop}%
\bibitem [{\citenamefont {Meyer}\ \emph {et~al.}(shed)\citenamefont {Meyer}, \citenamefont {Els\"asser}, \citenamefont {Lechermann},\ and\ \citenamefont {F\"ahnle}}]{Meyer}%
  \BibitemOpen
  \bibfield  {author} {\bibinfo {author} {\bibfnamefont {B.}~\bibnamefont {Meyer}}, \bibinfo {author} {\bibfnamefont {C.}~\bibnamefont {Els\"asser}}, \bibinfo {author} {\bibfnamefont {F.}~\bibnamefont {Lechermann}},\ and\ \bibinfo {author} {\bibfnamefont {M.}~\bibnamefont {F\"ahnle}},\ }\bibfield  {title} {\bibinfo {title} {{FORTRAN90 Program for Mixed-Basis Pseudopotential Calculations for Crystals}},\ }\href@noop {} {\bibfield  {journal} {\bibinfo  {journal} {Max-Planck-Institut f\"ur Metallforschung, Stuttgart}\ } (\bibinfo {year} {unpublished})}\BibitemShut {NoStop}%
\bibitem [{\citenamefont {Heid}\ and\ \citenamefont {Bohnen}(1999)}]{Heid1999}%
  \BibitemOpen
  \bibfield  {author} {\bibinfo {author} {\bibfnamefont {R.}~\bibnamefont {Heid}}\ and\ \bibinfo {author} {\bibfnamefont {K.~P.}\ \bibnamefont {Bohnen}},\ }\bibfield  {title} {\bibinfo {title} {{Linear response in a density-functional mixed-basis approach}},\ }\href {https://doi.org/10.1103/PhysRevB.60.R3709} {\bibfield  {journal} {\bibinfo  {journal} {Phys. Rev. B}\ }\textbf {\bibinfo {volume} {60}},\ \bibinfo {pages} {R3709} (\bibinfo {year} {1999})}\BibitemShut {NoStop}%
\bibitem [{\citenamefont {Perdew}\ and\ \citenamefont {Wang}(1992)}]{Perdew1992}%
  \BibitemOpen
  \bibfield  {author} {\bibinfo {author} {\bibfnamefont {J.~P.}\ \bibnamefont {Perdew}}\ and\ \bibinfo {author} {\bibfnamefont {Y.}~\bibnamefont {Wang}},\ }\bibfield  {title} {\bibinfo {title} {{Accurate and simple analytic representation of the electron-gas correlation energy}},\ }\href {https://doi.org/10.1103/PhysRevB.45.13244} {\bibfield  {journal} {\bibinfo  {journal} {Phys. Rev. B}\ }\textbf {\bibinfo {volume} {45}},\ \bibinfo {pages} {13244} (\bibinfo {year} {1992})}\BibitemShut {NoStop}%
\bibitem [{Las()}]{LastComment}%
  \BibitemOpen
  \href@noop {} {\bibinfo {title} {{We note that in our simplified RFFM model the intensity of the buckling phonons ($c$-axis motion of in-plane oxygens) is expected to be significantly lower with respect to the intensity of the bond-stretching modes. The reason is that the transverse motion of the oxygens does not produce at the first order a variation of the Cu-O bond length along the $\textbf{q}$ direction. The treatmet of the transverse phonons requires further order terms in Eq.\ref{eq: XX8} }}}\BibitemShut {NoStop}%
\bibitem [{\citenamefont {HC4412}()}]{HC4412}%
  \BibitemOpen
  \bibfield  {author} {\bibinfo {author} {\bibnamefont {HC4412}},\ }\href@noop {} {}\bibinfo {howpublished} {\doi {10.15151/ESRF-ES-433978160}}\BibitemShut {NoStop}%
\bibitem [{\citenamefont {HC4607}()}]{HC4607}%
  \BibitemOpen
  \bibfield  {author} {\bibinfo {author} {\bibnamefont {HC4607}},\ }\href@noop {} {}\bibinfo {howpublished} {\doi {10.15151/ESRF-ES-641834549}}\BibitemShut {NoStop}%
\bibitem [{\citenamefont {HC5029}()}]{HC5029}%
  \BibitemOpen
  \bibfield  {author} {\bibinfo {author} {\bibnamefont {HC5029}},\ }\href@noop {} {}\bibinfo {howpublished} {\doi {10.15151/ESRF-ES-1095669447}}\BibitemShut {NoStop}%
\bibitem [{\citenamefont {HC5221}()}]{HC5221}%
  \BibitemOpen
  \bibfield  {author} {\bibinfo {author} {\bibnamefont {HC5221}},\ }\href@noop {} {}\bibinfo {howpublished} {\doi {10.15151/ESRF-ES-1106943771}}\BibitemShut {NoStop}%
\bibitem [{\citenamefont {HC5627}()}]{HC5627}%
  \BibitemOpen
  \bibfield  {author} {\bibinfo {author} {\bibnamefont {HC5627}},\ }\href@noop {} {}\bibinfo {howpublished} {\doi {10.15151/ESRF-ES-1556145667}}\BibitemShut {NoStop}%
\end{thebibliography}%

\section*{\label{sec:Methods} Methods}
\subsection*{Derivation of the 1-phonon structure factor}

We begin with the well-known expression for the inelastic scattering cross section for phonons, which should be valid regardless of the probe:
\begin{align}
\begin{split}
    \frac{\partial^2\sigma}{\partial\Omega\partial E_f} = \frac{k_f}{k_i}S(\boldsymbol{\mathrm{Q}},\omega)
     \end{split}\label{eq: ApAeq1}
    \end{align}
where
\begin{align}
\begin{split}
   S(\boldsymbol{\mathrm{Q}},\omega_{\boldsymbol{\mathrm{q}}}) = \delta(E_f-E_i-\hbar\omega_{\boldsymbol{\mathrm{q}}})\left|F(\boldsymbol{\mathrm{Q}},q)\right|^2
     \end{split}\label{eq: ApAeq2}
    \end{align}
and
\begin{align}
\begin{split}
  F(\boldsymbol{\mathrm{Q}},q) = \sum_n\sum_if_i(n)e^{i\boldsymbol{\mathrm{Q}}\cdot(\boldsymbol{\mathrm{R}}_n+\boldsymbol{\mathrm{r}}_i+\boldsymbol{\delta}_i(n))}-F_{\mathrm{el}}(\boldsymbol{\mathrm{Q}})
     \end{split}\label{eq: ApAeq3}
    \end{align}
is the phonon structure factor, $F_{\mathrm{el}}(\boldsymbol{\mathrm{Q}})$ is the elastic-scattering structure factor, the $n$-sum runs over unit cells and the $i$-sum runs over sites within the unit cell. In the
case of neutrons, $f_i(n)$ is the Fermi length independent of $n$, while for x-rays we can generally write:
\begin{align}
\begin{split}
 f_i(n) = \boldsymbol{\varepsilon}_i\cdot\boldsymbol{F}_i(n,E)\cdot\boldsymbol{\varepsilon}_i
     \end{split}\label{eq: ApAeq4}
    \end{align}
where $\boldsymbol{\varepsilon}_i$, $\boldsymbol{\varepsilon}_f$ are the incident and scattered photon polarization and $\boldsymbol{F}_i(n, E)$ rank-2 complex tensor (is strongly an energy-dependent for resonant processes) that can depend on the unit cell (see below). This dependence comes about because the amplitude of the distortion due to phonons is different in different unit cells. To first order, we can write:
    \begin{align}
\begin{split}
 f_i(n) \simeq f_i^0+\frac{\partial f_i}{\partial \Qc}\Qc(n)
     \end{split}\label{eq: ApAeq5}
    \end{align}
and for a cos-wave phonon with propagation vector $\boldsymbol{\mathrm{q}}$    
    \begin{align}
\begin{split}
 \Qc(n) =\Qc^0\frac{1}{2}(e^{\boldsymbol{\mathrm{q}}\cdot\boldsymbol{\mathrm{R}}_n}+c.c.)
     \end{split}\label{eq: ApAeq6}
    \end{align}
where $\Qc^0$ is the one-phonon amplitude. \par
Similarly, we can write:
    \begin{align}
\begin{split}
 \boldsymbol{\delta}_i(n) = \Qc^0\varepsilon_i\frac{1}{2}(e^{i\boldsymbol{\mathrm{q}}\cdot\boldsymbol{\mathrm{R}}_n+\varphi_i}+c.c.)
     \end{split}\label{eq: ApAeq7}
    \end{align}
where the $\varphi_i$ are phase shifts for individual atoms.\par
In the limit of small $\Qc^0$, we can do the following approximation:
    \begin{align}
\begin{split}
e^{i\boldsymbol{\mathrm{Q}}\cdot\boldsymbol{\delta}_i(n)}\simeq 1 +  \Qc^0i(\boldsymbol{\mathrm{Q}}\cdot\boldsymbol{\varepsilon}_i)\frac{1}{2}(e^{i\boldsymbol{\mathrm{q}}\cdot\boldsymbol{\mathrm{R}}_n+\varphi_i}+c.c.)
     \end{split}\label{eq: ApAeq8}
    \end{align}
Inserting the previous equations into Eq.\,\ref{eq: ApAeq3} and to linear order in $\Qc^0$, we obtain:
 \begin{align}
\begin{split}
\sum_n\sum_if_i(n)e^{i\boldsymbol{\mathrm{Q}}\cdot(\boldsymbol{\mathrm{R}}_n+\boldsymbol{\mathrm{r}}_i+\boldsymbol{\delta}_i(n))} = \sum_n e^{i\boldsymbol{\mathrm{Q}}\cdot\boldsymbol{\mathrm{R}}_n}F^0(\boldsymbol{\mathrm{Q}})\\
+ \Qc^0\sum_n e^{i(\boldsymbol{\mathrm{Q}}+\boldsymbol{\mathrm{q}})\cdot\boldsymbol{\mathrm{R}}_n}F^-(\boldsymbol{\mathrm{Q}})\\
+ \Qc^0\sum_n e^{i(\boldsymbol{\mathrm{Q}}-\boldsymbol{\mathrm{q}})\cdot\boldsymbol{\mathrm{R}}_n}F^+(\boldsymbol{\mathrm{Q}})
     \end{split}\label{eq: ApAeq9}
    \end{align}
with
 \begin{align}
\begin{split}
F^0(\boldsymbol{\mathrm{q}}) &= \sum_i f_i^0 e^{i\boldsymbol{\mathrm{Q}}\cdot\boldsymbol{\mathrm{r}}_i}\\
F^-(\boldsymbol{\mathrm{Q}}) &= \sum_i \left( \frac{\partial f_i}{\partial \Qc}+f_i^0(\boldsymbol{\varepsilon}_i\cdot\boldsymbol{\mathrm{Q}})e^{i(\varphi+\pi/2)})\right)e^{i\boldsymbol{\mathrm{Q}}\cdot\boldsymbol{\mathrm{r}}_i}\\
F^+(\boldsymbol{\mathrm{Q}}) &= \sum_i \left( \frac{\partial f_i}{\partial \Qc}+f_i^0(\boldsymbol{\varepsilon}_i\cdot\boldsymbol{\mathrm{Q}})e^{i(-\varphi+\pi/2)})\right)e^{i\boldsymbol{\mathrm{Q}}\cdot\boldsymbol{\mathrm{r}}_i}
     \end{split}\label{eq: ApAeq10}
    \end{align}
In the approximation of an infinite lattice, the lattice sums in $n$ in Eq.\,\ref{eq: ApAeq9} can be converted into $\delta$ functions as usual, yielding:
 \begin{align}
\begin{split}
\sum_n\sum_if_i(n)e^{i\boldsymbol{\mathrm{Q}}\cdot(\boldsymbol{\mathrm{R}}_n+\boldsymbol{\mathrm{r}}_i+\boldsymbol{\delta}_i(n))} =\delta(\boldsymbol{\mathrm{G}}-\boldsymbol{\mathrm{Q}})F^0(\boldsymbol{\mathrm{Q}})\\
+\Qc^0\delta(\boldsymbol{\mathrm{G}}-(\boldsymbol{\mathrm{Q}}+\boldsymbol{\mathrm{q}}))F^-(\boldsymbol{\mathrm{Q}})\\
+\Qc^0\delta(\boldsymbol{\mathrm{G}}-(\boldsymbol{\mathrm{Q}}-\boldsymbol{\mathrm{q}}))F^+(\boldsymbol{\mathrm{Q}})
     \end{split}\label{eq: ApAeq11}
    \end{align}
$\boldsymbol{\mathrm{G}}$ being a reciprocal lattice vector. The first term correspond to the Bragg (elastic)
cross section and is subtracted off in Eq.\,\ref{eq: ApAeq3}, while the other two yield peaks at $\boldsymbol{\mathrm{Q}} =
\boldsymbol{\mathrm{G}} - \boldsymbol{\mathrm{q}}$ and $\boldsymbol{\mathrm{Q}} = \boldsymbol{\mathrm{G}} + \boldsymbol{\mathrm{q}}$, respectively.\par

\iffalse %begin omitted section
\subsection*{Mode coordinate for 1 phonon}
We still need to evaluate the parameter $\Qc^0$ in Eq.\,\ref{eq: ApAeq11}, which is usually taken to correspond to the mean squared displacement of the phonon. We have used the following definitions, which are rather common to solve the normal mode problem:
 \begin{align}
\begin{split}
\zeta_i &= \sqrt{m_i}x_i = Q_{ci}e^{i\omega t}\\
Q_{ci} &= \sqrt{m_i}\Qc^0\mu_i
     \end{split}\label{eq: ApAeq13}
    \end{align}
where the index $i$ runs over atoms and components and
 \begin{align}
\begin{split}
\mu_i = \frac{\partial x_i}{\partial \Qc}
     \end{split}\label{eq: ApAeq14}
    \end{align}
and $\Qc$ is dimensionless. Assuming:
 \begin{align}
\begin{split}
\sqrt{\left < x^2\right >} = \sqrt{\frac{2\hbar}{\omega m_{eff}}}
     \end{split}\label{eq: ApAeq15}
    \end{align}
 $m_{eff}$ being a mode ‘effective mass’, we obtain:
     \begin{align}
\begin{split}
\Qc^0 = \frac{1}{\mu_{eff}}\sqrt{\frac{2\hbar}{\omega m_{eff}}}
     \end{split}\label{eq: ApAeq16}
    \end{align}
where $\mu_{eff}$ is a weighted average of the parameters $\mu_i$.
\fi %end omitted section

\subsection*{Simplified RFFM model}

In the present case, there is only one Cu atom located at the origin, surrounded by four oxygen atoms, two located in the same unit cell and two in adjacent unit cells.  In the presence of a bond-stretching phonon, their normalized O displacements are summarized in Table\,\ref{tab: displ}, while Cu displacements can be omitted to first order.

\begin{table}[h]
\begin{center}
\tabcolsep 1.5 mm
\caption{Normalized oxygen displacements for our simplified model of bond-stretching phonons, which are all degenerate at the $\Gamma$ point.  The half-breathing mode (HB) and the BS$_{10}$ mode disperse along the $(\zeta,0)$ direction, while the full-breathing mode and the BS$_{11}$ mode (also known as the \emph{quadrupolar mode}), disperse along the $(\zeta,\zeta)$ direction. }

\label{tab: displ}
\begin{tabular} {c c c c c}
%{p{1.5cm} p{5.0cm} p{1.5cm} p{1.5cm}p{1.5cm} p{1.5cm}p{1.5cm}}%{llllllll}
\hline \hline
O site & HB   & FB &  BS$_{10}$  &  BS$_{11}$    \\
\hline
\vspace{-8pt} \\
$(\frac{1}{2},0)$ & $(e^{ \pi i \zeta},0)$ & $(e^{ \pi i \zeta},0)$ & $(0,0)$ & $(e^{ \pi i \zeta},0)$ \\
$(0,\frac{1}{2})$ & $(0,0)$ & $(0,e^{ \pi i \zeta})$& $(0,1)$& $(0,-e^{ \pi i \zeta})$ \\
$(-\frac{1}{2},0)$ & $(e^{- \pi i \zeta},0)$ & $( e^{- \pi i \zeta},0)$& $(0,0)$& $( e^{- \pi i \zeta},0)$ \\
$(0,-\frac{1}{2})$ & $(0,0)$ & $(0, e^{- \pi i \zeta})$& $(0,1)$& $(0, -e^{- \pi i \zeta})$ \\
\vspace{-8pt} \\
\hline \hline
\end{tabular}
\vspace{-0.6cm}
\end{center}
\end{table}

Time-dependent displacements for the unit cell labeled as $[n_x, n_y]$ are obtained as
\begin{align}
    \begin{split}
(\delta_x, \delta_y)(t)=\Qc\frac{(\epsilon_x,\epsilon_y)}{2}\left( e^{i( 2 \pi (q_xn_x+q_yn_y)-\omega t)} + c.c.\right)
\end{split}\label{eq: XX1}
    \end{align}

where $(\epsilon_x,\epsilon_y)$ are from Table\,\ref{tab: displ}.\par

In our simplified model, we assume that the change of the Cu form factor is proportional to the mean squared amplitude of the coherent breathing displacement of the oxygens towards (or away from) Cu.

\begin{align}
    \begin{split}
\frac{\partial f}{\partial \Qc}=\frac{1}{\Qc}\sqrt{\left<\Delta d_B^2\right>}
     \end{split}\label{eq: XX8}
\end{align}  

In other words, we ignore distortions of the local environment that do not affect the average Cu-O bond lengths.  We can immediately see that modes BS$_{10}$ and BS$_{11}$ produce no coherent breathing displacement, hence we expect that their RIXS intensity will be very small (this is entirely consistent with other theoretical models based on the EPC coupling).

\iffalse %%%%%%Alternative table -Section to be omitted
\begin{table}[h]
\begin{center}
\tabcolsep 1.5 mm
\caption{Alternative expression for the mode}
\label{tab: alt_displ}
\begin{tabular} {c c c c c}
%{p{1.5cm} p{5.0cm} p{1.5cm} p{1.5cm}p{1.5cm} p{1.5cm}p{1.5cm}}%{llllllll}
\hline \hline
O site & HB   & FB &  BS$_{10}$  &  BS$_{11}$    \\
\hline
\vspace{-8pt} \\
$(\frac{1}{2},0)$ & $(1,0)$ & $(1,0)$ & $(0,0)$ & $(1,0)$ \\
$(0,\frac{1}{2})$ & $(0,0)$ & $(0,1)$& $(0,1)$& $(0,-1)$ \\
$(-\frac{1}{2},0)$ & $( e^{-2 \pi i \zeta},0)$ & $( e^{-2 \pi i \zeta},0)$& $(0,0)$& $( e^{-2 \pi i \zeta},0)$ \\
$(0,-\frac{1}{2})$ & $(0,0)$ & $(0, e^{-2 \pi i \zeta})$& $(0,1)$& $(0, -e^{-2 \pi i \zeta})$ \\
\vspace{-8pt} \\
\hline \hline
\end{tabular}
\vspace{-0.6cm}
\end{center}
\end{table}
\fi %%%%%%end section to be omitted

%%%%%% end section to be omitted

For the HB and FB modes, the average change in bond lengths over a pair of co-aligned Cu-O bonds meeting at the origin is:

\begin{align}
    \begin{split}
\Qc \sin (\pi \zeta) \; \sin (\omega_{\boldsymbol{\mathrm{q}}} t)+O(\Qc^2)
   \end{split}\label{eq: XX5b}
    \end{align}  
which has amplitude $\Delta d_B=\Qc \sin (\pi \zeta)$.\par

Taking the root mean square over the four bonds, one obtains:

\begin{align}
    \begin{split}
\left[\frac{\partial f}{\partial \Qc}\right]_{HB}&=\frac{1}{\sqrt{2}}\sin (\pi \zeta)\\
\left[\frac{\partial f}{\partial \Qc}\right]_{FB}&=\sin (\pi \zeta)\\
     \end{split}\label{eq: XX10}
\end{align}  

\par

From this, one can calculate the expected one-phonon RIXS intensity, which is taken to be proportional to $|\partial f_i/\partial \Qc|^2$

\begin{align}
    \begin{split}
I_{HB}&\propto\sin^2 (\pi \zeta)=\sin^2 (\pi q_\parallel)\\
I_{FB}&\propto 2\sin^2 (\pi \zeta)=2\sin^2 (\pi q_\parallel/\sqrt{2})\\
     \end{split}\label{eq: approx_RFFM}
\end{align}

Eq.\,\ref{eq: approx_RFFM} should be compared with the TB expression (Eq.\,\ref{eq:M_br_fermi}): 

\begin{align}
    \begin{split}
I_{TB}&\propto \sin^2 (\pi q_\parallel \cos \varphi)+\sin^2 (\pi q_\parallel  \sin \varphi)
     \end{split}\label{eq: TB_Int_Scaled}
\end{align}  

\par

Remarkably, the TB model (Eq.\,\ref{eq:M_br_fermi}) and \ref{eq: TB_Int_Scaled}~ produce an identical dependence on $q_\parallel$ along the two high-symmetry directions.
\par

\subsection*{\label{sec:Appendixx}DFT-RFFM}

For the DFT-RFFM model, breathing displacement amplitudes are calculated using the O and Cu eigenmodes obtained from DFT.  The average Cu-O breathing amplitudes along $x$ and $y$ are:

\begin{align}
    \begin{split}
\Delta_{Bx}=\Qc\;Q_x \sin (\pi q_x)\\
\Delta_{By}=\Qc\; Q_y \sin (\pi q_y)
     \end{split}\label{eq: Br_x_y}
\end{align}  

and the mean squared displacement is 

\begin{align}
    \begin{split}
\Delta_{RMS}=\frac{\Qc}{\sqrt{2}}\sqrt{Q_x^2 \sin^2 (\pi q_x)+ Q_y^2 \sin^2 (\pi q_y)}\\
     \end{split}\label{eq: Br_RMS}
\end{align}

Hence, for the intensity

\begin{align}
    \begin{split}
I \propto Q_x^2 \sin^2 (\pi q_x)+ Q_y^2 \sin^2 (\pi q_y)\\
     \end{split}\label{eq: Br_Int}
\end{align}

Eqs.\,\ref{eq:M_br_fermi}, \ref{eq: TB_Int_Scaled} and \ref{eq: Br_Int} yield the same result with $Q_x=Q_y=1$.  The DFT-RMMF model is obtained by replacing $Q_x$ and $Q_y$ with the eigenmodes obtained from DFT.
\appendix
\section{\label{sec:Appendix A} Detuning data on LCO-UD22}
\begin{figure*}
    %\centering
    \textbf{Detuning method and phonon features in LCO-UD22.}\\[0.5em] 
    \includegraphics[width=0.75\textwidth]{ 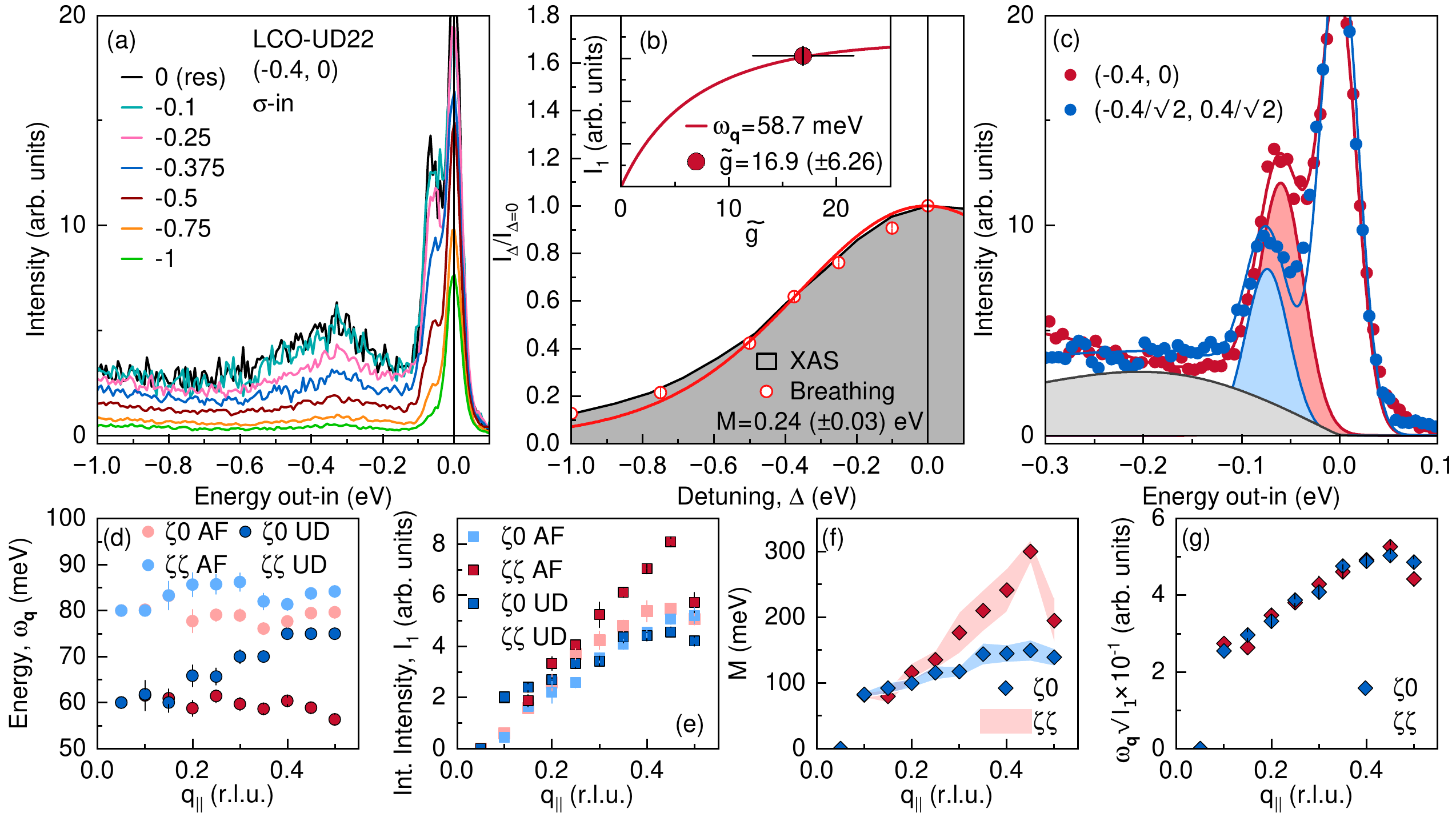}
    \caption{(a) RIXS spectra of LCO-UD22 collected at $\mathbf{q_{\parallel}}= (-0.4, 0)$\,r.l.u.  as a function of the detuning energy $\Delta$. (b) The evolution of the integrated phonon intensity (symbols) upon detuning and its fitting with Eq.\,\ref{eq:RIXS_Int_1} (solid lines) on top of corresponding XAS plot (gray area). The inset indicates the position of the reference point on the Ament curve which is used for the calibration between $I_1$ and $\tilde{g}$. (c) An example of fit (solid lines) of RIXS data (markers) along ($\zeta$,0) (red) and ($\zeta$,$\zeta$) (blue). The phonon Gaussians are highlighted with red and blue area. Gray area indicates the broad paramagnon along ($\zeta$,$\zeta$) excitation which overlaps with phonon signal. Other fitting features are omitted for clarity. (d) Experimental breathing phonon energy (symbols) with $\mathbf{q_{\parallel}}$ along the high-symmetry lines. The dark symbols refer to LCO-UD22 while light symbols to LCO-AF for comparison. (e) Experimental breathing phonon intensity with $\mathbf{q_{\parallel}}$ for LCO-UD22 and LCO-AF. (f) Evolution of the EPC strength $M(\mathbf{q_{\parallel}})$ for LCO-UD22 determined with the complete Ament model. (g) Evolution of the product $\omega_{\mathbf{q}}\sqrt{I_1(\mathbf{q_{\parallel}})}$ for LCO-UD22, so-called simplified Ament model.}
    \label{fig:Detuning_LCOUD}
\end{figure*}

Here, we discuss the detuning and ($\zeta$,0)/($\zeta$,$\zeta$) data of superconducting underdoped La$_2$CuO$_4$ (LCO-UD22, $T_{\mathrm{c}} = 22$\,K, $t\approx30$\,nm, $a=b=3.76$\,\si{\angstrom} and $c\approx3.18$\,\si{\angstrom}), collected under the same experimental conditions as for LCO-AF. Figure\,\ref{fig:Detuning_LCOUD}\,(a) shows RIXS spectra measured at several values of $\Delta$. From their fitting, we find that the breathing phonon energy is significantly reduced compared to the LCO-AF case-decreasing from approximately 80\,meV to 60\,meV (see Fig.\,\ref{fig:Detuning_LCOUD}\,(d)). This shift implies that the low-energy phonon mode near 30\,meV can no longer be resolved with the current energy resolution of $\sim 45$\,meV. As a result, part of its spectral weight is likely attributed to the breathing phonon, leading to a partial contamination of the extracted results. It is the main reason why these data are shown in the Appendix.\par
By applying the Ament model, we estimate the EPC in LCO-UD22 at $\mathbf{q_{\parallel}} = (-0.4, 0)$\,r.l.u. to be 0.24\,($\pm 0.03$)\,eV (see Fig.,\ref{fig:Detuning_LCOUD}\,(b)), which is significantly higher than in LCO-AF. As discussed in the main text, one possible reason for this increase is the broad paramagnon excitation, whose intensity may be incorrectly attributed to the phonon.\par 
Figure\,\ref{fig:Detuning_LCOUD}\,(f) and (g) displays the $\mathbf{q}$-dependence of the EPC evaluated using the compete Ament approach and its simplified version, respectively. In the former, we observe a pronounced enhancement of the EPC at $\mathbf{q_{\parallel}} = (-0.45, 0)$\,r.l.u., which is completely absent in the latter. Currently, we cannot confirm whether the feature at $\mathbf{q_{\parallel}} = (-0.45, 0)$\,r.l.u. is real, and higher-resolution measurements are necessary for validation.\par 

\section{\label{sec:Appendix B} Details of the detuning method}
\begin{figure}
    %\centering
        \textbf{Detuning curves for various values of $\tilde{g}$.}\\[0.5em] 
    \includegraphics[width=0.75\columnwidth]{ 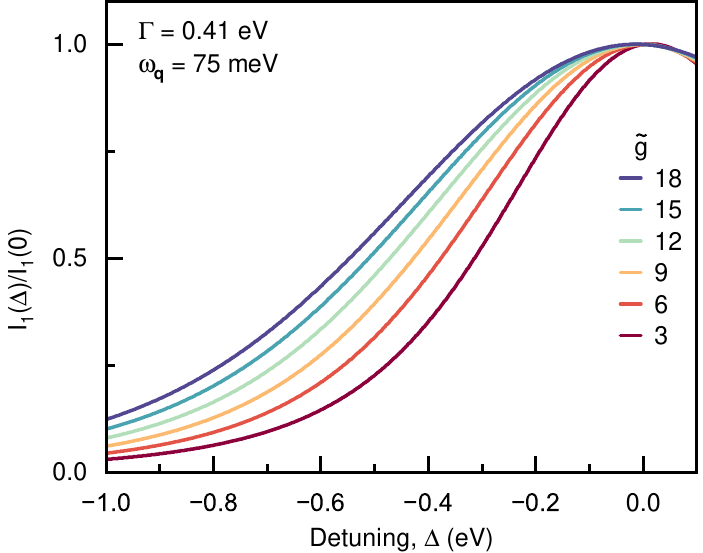}
    \caption{The behavior of the one-phonon intensity under the detuning $\Delta$ for various EPC strength $\tilde{g}$. All curves are normalized to their value at resonance. $\Gamma$ and $\omega_{\mathbf{q}}$ are fixed: $\Gamma$ = 0.41\,eV, $\omega_{\mathbf{q}}$ = 75\,meV.}
    \label{fig:Detuning_details}
\end{figure}

To extract the EPC from the detuning data, we analyze the so-called detuning curves, which describe the evolution of $I_1$ from Eq.\,\ref{eq:RIXS_Int_1} as a function of $\Delta$ for a fixed coupling constant $\tilde{g}$. For each value of $\tilde{g}$, the calculated $I_1(\Delta)$ is normalized to its on-resonance intensity, at $\Delta = 0$. The experimental normalized $I_1(\Delta)$ values are then compared with these theoretical detuning curves, and the corresponding value of $\tilde{g}$ is identified from the best matching curve, see Fig.\,\ref{fig:Detuning_details}.\par

\par
   \end{document}